\documentclass[prb, twocolumn, superscriptaddress]{revtex4-1}
\usepackage{amsmath,amsfonts, amssymb, amsthm, dsfont}
\usepackage{yfonts}
\usepackage{bm}
\usepackage{mathrsfs}
\usepackage{graphicx}
\usepackage{verbatim}
\usepackage{hyperref}
\usepackage{color}
\usepackage{tikz}
\usepackage{pgffor}
\usepackage{enumitem}

\newcommand{\comments}[1]{}
\newcommand{\mb}[1]{\mathbf{#1}}
\newcommand{\ftj}{\ensuremath{\mathsf{T}}} 
\newcommand{\Z}{\mathbb Z}
\newcommand{\U}{\mathrm{U}}

\newcommand{\Ref}[1]{Ref.~\onlinecite{#1}}
\newcommand{\Refs}[1]{Refs.~\onlinecite{#1}}

\newcommand{\spd}[1]{#1D}

\usetikzlibrary{decorations.pathreplacing,decorations.markings}
\tikzset{middlearrow/.style={
        decoration={markings,
            mark= at position 0.5 with {\arrow{#1}} ,
        },
        postaction={decorate}
    }
}

\begin{document}

\title{Loop Braiding Statistics and Interacting Fermionic Symmetry-Protected Topological Phases in Three Dimensions}
\author{Meng Cheng}
\affiliation{Department of Physics, Yale University, New Haven, CT 06511-8499, USA}
\author{Nathanan Tantivasadakarn}
\affiliation{Perimeter Institute for Theoretical Physics, Waterloo, ON N2L 2Y5, Canada}
\author{Chenjie Wang}
\affiliation{Perimeter Institute for Theoretical Physics, Waterloo, ON N2L 2Y5, Canada}
\date{\today}

\begin{abstract}
	We study Abelian braiding statistics of loop excitations in three-dimensional (\spd{3}) gauge theories with fermionic particles  and the closely related problem of classifying \spd{3} fermionic symmetry-protected topological (FSPT) phases with unitary symmetries. It is known that the two problems are related by turning FSPT phases into gauge theories through gauging the global symmetry of the former.  We show that there exist certain types of Abelian loop braiding statistics that are allowed only in the the presence of fermionic particles, which correspond to 3D ``intrinsic''  FSPT phases, i.e., those that do not stem from bosonic SPT phases. While such intrinsic FSPT phases are ubiquitous in \spd{2} systems and in \spd{3} systems with anti-unitary symmetries, their existence in 3D systems with unitary symmetries was not confirmed previously due to the fact that strong interaction is necessary to realize them. We show that the simplest unitary symmetry to support \spd{3} intrinsic FSPT phases is $\mathbb{Z}_2\times\mathbb{Z}_4$. To establish the results, we first derive a complete set of physical constraints on Abelian loop braiding statistics. Solving the constraints, we obtain all possible Abelian loop braiding statistics in 3D gauge theories, including those that correspond to intrinsic FSPT phases. Then, we construct exactly soluble state-sum models to realize the loop braiding statistics. These state-sum models generalize the well-known Crane-Yetter and Dijkgraaf-Witten models.
\end{abstract}

\maketitle

\tableofcontents

\section{Introduction}
\label{sec:intro}
Topological phases in three spatial dimensions (3D) can support particle and loop excitations~\footnote{We do not discuss gapped phases in \spd{3} which do not have a well-defined continuum limit, such as the so-called ``fracton'' topological orders}. While we learn in undergraduate quantum mechanics that there are only bosonic and fermionic exchange statistics for particles in \spd{3}, the rich statistical properties of loop excitations only begin to be uncovered recently~\cite{Wang_PRL2014, Ran_PRX2014}, in conjunction with the study of bosonic symmetry-protected topological (BSPT) phases in \spd{3}~\cite{Chen_science, ChenPRB2013, senthil_3D}. In particular, it was found that to characterize BSPT phases protected by finite unitary symmetries, it is necessary to consider the so-called ``three-loop'' braiding statistics in gauge theories, a new type of braiding statistics in \spd{3}~\cite{Wang_PRL2014,WangLevinPRB, Ran_PRX2014, Wang_DW, Chen2014}, where the gauge theories are obtained by gauging the global symmetries of BSPT phases. 

So far, all studies of loop braiding statistics have focused on gauge theories where all particles are bosons. Loop braiding statistics in the presence of fermionic particles are much less explored. Perhaps the most important question to address is: Does the presence of fermions allow new types of loop braiding statistics that are not possible otherwise?

This question is closely related to the problem of classifying interacting fermionic symmetry-protected topological phases (FSPT) in \spd{3}, in which the braiding statistics of vortex loops serves as a topological invariant for the bulk phase. To put it into context, we briefly review the classification of FSPT phases with unitary symmetries in \spd{3}. For non-interacting fermionic systems, it is well known that there are no nontrivial FSPT phases protected by on-site unitary symmetries~\cite{schnyder2008, kitaev2009, Ludwig}
\footnote{A quick derivation is the following: one can first block diagonalize the single-particle Hamiltonian according to the irreducible representations of the unitary symmetry group. Within each block, there are no further constraints on the single-particle Hamiltonian matrix, so the classification becomes that of the class D. It is known that the class D has a $\mathbb{Z}$ classification in \spd{2}, but is completely trivial in \spd{3}}. 
On the other hand, we can create interacting FSPT phases by effectively turning fermions into bosons with the help of strong interactions (i.e. fermions forming spins or molecular bound states), and letting the bosons form SPT states. An interesting question then arises: are there ``intrinsic'' FSPT phases in \spd{3} protected by unitary symmetries? Here, by ``intrinsic'', we mean those FSPT phases that do not stem from BSPT phases. If there were any, we know that they must be strongly interacting states (since we know there are no nontrivial non-interacting SPT phases). Then, the loop braiding statistics obtained by gauging symmetries of these ``intrinsic'' FSPT phases are the ones that are only allowed in the presence of fermionic particles.


\begin{figure}[b]

\begin{tikzpicture}[>=stealth, scale=0.9]

\draw [gray!60](0,0)--(1.5,1.5)--(4.5,1.5)--(3,0)--cycle;
\draw [gray!60] (1.5,1.5)--(1.5,4.5)--(4.5,4.5)--(4.5,1.5);
\draw [gray] (0,3)--(1.5,4.5)--(4.5,4.5)--(3,3);
\draw [fill=gray!40,opacity=0.7,very thick,rotate=3] (1.9,2) ellipse (1 and 0.35);
\draw [gray] (0,0)--(0,3)--(3,3)--(3,0);

\draw [->,red!70] (1.2,1.89)..controls (1.4,1.8) and (1.9,1.8)..(2.4,1.95);
\draw [<-,blue!50] (1.2,1.7)..controls (1.4,1.65) and (1.9,1.63)..(2.4,1.75);

\end{tikzpicture}
\caption{A $\mathbb Z_4$ symmetry defect in the $\mathbb Z_2^f\times\mathbb Z_2\times\mathbb Z_4$ intrinsic FSPT phase. There lives a 1D helical Dirac fermion (denoted by red and blue arrows) on the defect. The shaded region represents a branch surface associated with the defect.}
\label{fig_defect}
\end{figure}
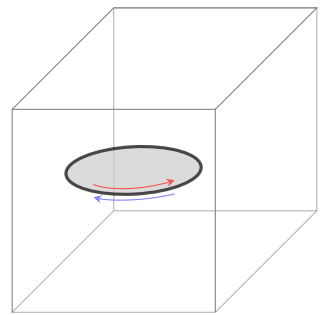

Due to the non-existence of non-interacting phases, one has to confront the complexity of interacting systems from the very beginning to realize FSPT phases. A fruitful approach is to construct and study exactly-solvable lattice models. A systematic construction of fermionic SPT phases has been proposed in \Ref{GuWen}, although no explicit examples in \spd{3} for unitary symmetries (except the bosonic ones) were given. In addition, it is not clear what kind of physical properties characterize the constructed FSPT states in \Ref{GuWen}. We should mention that other classification schemes have been proposed, such as spin cobordism group in \Ref{KapustinFSPT} and the invertible TQFTs in Refs. \onlinecite{Freed2014, Freed2016}.

In this work, we study \emph{Abelian} loop braiding statistics in \spd{3} gauge theories with fermionic particles. As we will see, the presence of fermionic particles indeed enables new types of three-loop braiding statistics, forbidden when the particles are bosonic. Because of the well-established correspondence between SPT phases and gauge theories~\cite{LevinGu2012, Wang_PRL2014, WangLevinPRB, ChengPRL2014, Gu_2013}, our results also imply existence of intrinsic FSPT phases. We derive these results through a combination of physical arguments and exactly-solvable models. Moreover, we derive a complete classification of Abelian three-loop braiding statistics in Abelian gauge theories (i.e., the gauge group is Abelian) in the presence of fermionic particles. 

We show that the simplest symmetry group which allows for intrinsic interacting FSPT phases is $\mathbb{Z}_2^f\times\mathbb{Z}_2\times\mathbb{Z}_4$. Here $\mathbb{Z}_2^f$ is the fermion parity conservation. It turns out that this example captures the essence of all \spd{3} ``Abelian'' FSPT phases discussed in this work. Hence, we give an intuitive picture for one of the $\mathbb{Z}_2^f\times\mathbb{Z}_2\times\mathbb{Z}_4$ intrinsic FSPT phases --- namely, the ``root'' phase --- in terms of decorated domain walls. In this description, a symmetric state can be obtained by proliferating domain walls of the global symmetry. If domain walls themselves are ``decorated'' by lower-dimensional SPT phases, then the wavefunction of proliferated domain walls may represent a nontrivial SPT state~\cite{Chen_NC2014}. In the $\mathbb{Z}_2^f\times\mathbb{Z}_2\times\mathbb{Z}_4$ FSPT phase, a $\mathbb{Z}_4$ domain wall is decorated by a \spd{2} FSPT phase protected by the $\mathbb{Z}_2^f\times\mathbb{Z}_2$ symmetry. Using the terminology of \Ref{Cheng_fSPT}, the one that we use for decoration is the ``root'' Abelian FSPT phase in \spd{2}, which has a $\mathbb{Z}_4$ classification (note that the full classification of 2D $\mathbb{Z}_2^f\times\mathbb Z_2$ FSPT phases is $\mathbb Z_8$; however, the $\mathbb Z_8$ root phase is ``non-Abelian''). In fact, because this Abelian root phase has a $\mathbb{Z}_4$ classification, it can only exist on a $\mathbb{Z}_4$ domain walls~\cite{Ryu_arxiv2012, Gu_2013, GuWen, Cheng_fSPT}. This \spd{2} phase can be easily realized with non-interacting fermions, and a simple example is the following: the system consists of two layers, a Chern insulator with Chern number $C=1$ (or equivalently, two copies of $p_x+ip_y$ superconductors), and its time-reversal image with $C=-1$. This non-interacting FSPT has helical Dirac fermions on the edge, and since the two chiral modes carry opposite $\mathbb{Z}_2$ charges, they can not backscatter. Interactions on the edge can cause spontaneous breaking of the $\mathbb{Z}_2$ symmetry, but the edge can not be both symmetric and non-degenerate~\cite{Lu_arxiv2012, Gu_2013}.

To expose the physics of the 3D FSPT phase constructed above, we imagine inserting a $\mathbb{Z}_4$ symmetry defect loop into the FSPT phase (Fig.~\ref{fig_defect}). Since the defect loop can be thought of as the effective boundary of a domain wall, it must carry similar gapless modes as the edge of the 2D FSPT phase decorated on the domain wall, assuming no spontaneous symmetry breaking along the loop. As we already mentioned, one possible effective low-energy theory for the gapless modes is a \spd{1} helical Dirac fermion. This 1D hilical Dirac fermion on $\mathbb{Z}_4$ symmetry defects is an important property of the FSPT phase.

To explicitly show that the above 3D FSPT phase is intrinsic, we need to gauge the $\mathbb{Z}_2^f\times\mathbb{Z}_2\times\mathbb{Z}_4$ symmetry. Then, symmetry defect loops turn into dynamical vortex loops in the gauged system. There are two ``three-loop'' braiding process (see Fig.~\ref{fig_braiding}) which can reveal the intrinsic nature of the FSPT phase: first, consider braiding a $\mathbb{Z}_2^f$ (i.e. fermion parity) vortex loop around a $\mathbb Z_2$ vortex loop, while both are linked to a unit $\mathbb{Z}_4$ vortex loop. We find that this three-loop braiding statistics is either semionic or anti-semionic ($\pm \pi/2$). On the other hand, using a result that will be established in Sec.~\ref{sec:parityloop}, if this FSPT phase stems from a BSPT phase, this three-loop braiding phase can only be $0$ or $\pi$. The essence in this difference is that  fermion parity vortex loops play a nontrivial role in the three-loop braiding statistics in the FSPT phase constructed from decorated domain walls. Hence, this FSPT must be intrinsically fermionic and the corresponding loop braiding statistics can exist only in the presence of fermionic particles. The other process is to exchange two identical $\mathbb{Z}_2$ vortex loops linked to the unit $\mathbb{Z}_4$ vortex. The resulting Berry phase turns out to be $\pi/4$ (up to a $\pi$ ambiguity), which as we will see is not allowed in systems with bosonic charges.

Due to the length of the paper, we briefly outline the strategy underlying our approach: in Sec. \ref{sec:constraint} we define a set of topological invariants to unambiguously characterize Abelian three-loop braiding statistics in the presence of fermionic particles, and derive physical constraints satisfied by the topological invariants. We then solve the constraints to obtain possible solutions corresponding to gauged intrinsic FSPT phases.  Next, to show that the constraints are complete and the solutions we found are indeed physical, we consider an exactly-solvable lattice model of topological twisted gauge theories which realize all solutions we have found in Sec. \ref{sec_state-sum} and \ref{sec:braiding}.

\section{Preliminaries}

\subsection{Symmetry}
Any fermionic system has a fundamental unbreakable symmetry, namely the conservation of the total fermion parity: $P_f=(-1)^{N_f}$, where $N_f$ is the number of fermions. The two operators $\{\mathds{1}, P_f\}$ form a symmetry group, which we denote as $\mathbb Z_2^f$. 

In addition, the system may be symmetric under other global symmetry transformations. Together with the fermion parity, all symmetry transformations form a symmetry group $\mathcal G$. It is generally required that $P_f$ commutes with all elements in $\mathcal G$. Accordingly, $\mathbb Z_2^f$ is a normal subgroup of $\mathcal G$. The quotient group $G=\mathcal{G}/\mathbb{Z}_2^f$  in a sense contains all the ``physical'' symmetries (i.e. those can be broken by physical perturbations). Mathematically, $\mathcal G$ is a central extension of $G$ by $\mathbb Z_2^f$. Given $G$, such an extension is not unique. Possible extensions are mathematically classified by the second group cohomology $\mathcal H^2[G, \mathbb Z_2^f]$. For example,  it is well known that fermionic systems with time-reversal symmetry $\mathbb{Z}_2^\ftj=\{\mathds{1}, T\}$ come in two types: one with $T^2=\mathds{1}$, and the other with $T^2 = P_f$, corresponding to the two elements in $\mathcal H^2[\mathbb Z_2^\ftj, \mathbb Z_2^f]=\mathbb Z_2$.

In this work, we consider fermionic systems with an Abelian unitary symmetry group $\mathcal{G}$. Without loss of generality, we can represent $\mathcal{G}$ as follows:\cite{WangFSPT}
\begin{equation}
	\mathcal{G}=\mathbb{Z}_{N_0}^f\times\prod_{i=1}^K \mathbb{Z}_{N_i}.
	\label{eqn:symmetry}
\end{equation}
where $N_0$ is an even number.  
We use integer vectors $a=(a_0, a_1,\dots, a_K)$ to denote the group elements, with $a_i=0,1,\dots, (N_i-1)$, and use additive convention for group multiplication. Generators of the group are denoted as
\begin{equation}
e_i=(0,\dots,1,\dots,0) \nonumber
\end{equation}
where the $i$th entry equals 1 and other entries equals 0.

The group $\mathbb{Z}_{N_0}^f$ is singled out because the fermion parity element is given by 
\begin{equation}
P_f = \frac{N_0}{2}e_0. \nonumber
\end{equation}
Equivalently, it means that the unit charge under $\mathbb Z_{N_0}^f$ is a fermion, while the unit charge under other $\mathbb{Z}_{N_i}$ subgroups  are all bosons. If we consider the more ``physical'' global symmetry group $G=\mathbb{Z}_{N_0/2}\times\prod_{i=1}^K \mathbb{Z}_{N_i}$, we find that all fermions carry ``half charges'' under $\mathbb{Z}_{N_0/2}$, forming the so-called ``projective representations'' of $\mathbb Z_{N_0/2}$. (If $N_0/2$ is odd, this actually does not give a true projective representation, due to the familiar isomorphism that $\mathbb{Z}^f_{N_0}=\mathbb{Z}_2^f\times \mathbb{Z}_{N_0/2}$).

\subsection{FSPT phases and gauging symmetry}

As discussed in the introduction,  fermionic systems with a symmetry $\mathcal G$ may form various SPT phases, i.e., gapped symmetric short-range entangled states.  In this work we study FSPT phases protected by Abelian unitary symmetries  in Eq.~\eqref{eqn:symmetry}. 

One way to characterize FSPT phases is to ``gauge'' the global symmetry $\mathcal G$ in lattice Hamiltonians. That is, we minimally couple the system to a lattice gauge field of a (discrete) gauge group $\mathcal G$. There is a well-defined procedure to do so; see e.g. \Refs{LevinGu2012, WangLevinPRB}. The resulting model, a gauge theory coupled to fermionic matter, is actually \emph{topologically ordered}, in the sense that it is gapped and it hosts topologically nontrivial excitations. In a gauge theory, nontrivial excitations carry either gauge charge or gauge flux. It has been proposed and verified in various systems that braiding statistics of charge and flux excitations in the gauged models are able to distinguish the original SPT phases. In this work, we study braiding statistics in \spd{3} gauged FSPT systems, extending previous works on \spd{2}/\spd{3} BSPT phases~\cite{Wang_PRL2014, WangLevinPRB} and \spd{2} FSPT phases~\cite{Cheng_fSPT, WangFSPT}. 

It is sometimes useful to only gauge the fermion parity symmetry subgroup $\mathbb{Z}_{N_0}^f$. In this way, the fermionic system is turned into a bosonic one, in the sense that there are no local fermionic excitations. Because of the direct product structure in Eq. \eqref{eqn:symmetry}, the other global symmetries remain unaffected by the gauging procedure, so we end up with a $\mathbb{Z}_{N_0}$ gauge theory enriched by a symmetry group $\mathcal{G}/\mathbb{Z}_{N_0}=\prod_{i=1}^K \mathbb{Z}_{N_i}$~\cite{Ying_2012, Lu_arxiv2013, Chen2014, BBCW, Tarantino_SET, TeoSET}.

\subsection{Basics of \spd{3} braiding statistics}

We now discuss the basics of braiding statistics between excitations in a gauged \spd{3} FSPT system, i.e., a $\mathcal G$ gauge theory coupled to fermionic matter. 

There are two kinds of excitations in the system: particle excitations that carry gauge charges, and vortex loop excitations that carry gauge fluxes. For an Abelian group $\mathcal G$, we use integer vectors $q=(q_0, q_1, \dots, q_K)$, with $q_i=0, \dots, (N_i-1)$, to denote the charge excitations. The self statistics associated with exchanging two identical charges is given by
\begin{equation}
\theta_{q} = \pi q_0 \label{exchange}.
\end{equation}
That is, it is a fermion/boson if $q_0$ is odd/even. 

Vortex excitations are string-like and must form closed loops inside the bulk of the system. They carry gauge flux. We use vectors $$\phi = \left(\frac{2\pi}{N_0} a_0, \frac{2\pi}{N_1} a_1, \dots, \frac{2\pi}{N_K}a_K\right)$$ to label gauge fluxes, where $a_i=0,1,\dots, (N_i-1)$ is an integer.  There is a well-known correspondence between gauge fluxes and group elements: one may regard the vector $a=(a_0,\dots, a_K)$ that labels $\phi$ as an group element of $\mathcal G$. Accordingly, the fermion parity group element corresponds to the fermion parity flux $(\pi, 0,\dots,0)$.

Unlike charge excitations, vortex excitations cannot be uniquely labeled by their gauge fluxes. Two distinct vortices $\alpha$ and $\alpha'$ may carry the same gauge flux, i.e., $\phi_\alpha= \phi_{\alpha'}$. It can be shown that two vortices carrying the same gauge flux can be transformed to one another by attaching gauge charges.  The mutual braiding statistics between a charge excitation $q$ and a vortex loop $\alpha$ follows the Aharonov-Bohm law
\begin{equation}
\theta_{q\alpha} = q\cdot\phi_\alpha 
\end{equation}
where ``$\cdot$'' is the vector inner product. In particular, the Aharonov-Bohm phase between $q$ and a fermion parity vortex is given by $\pi q_0$.

\begin{figure}
\begin{tikzpicture}[>=stealth, scale=1.5]

\def \lw{0.7 pt};
\begin{scope}
\draw [line width = \lw] (1.7, -0.5) .. controls (2, -0.5) and (2, 0.5) ..(1.7,0.5 );
\draw [line width = \lw] (2.3, -0.5) .. controls (2.6, -0.5) and (2.6, 0.5) ..(2.3,0.5 );

\begin{scope}[xshift=0.7cm]
\draw [thick, blue!40](1, 0.5)..controls(1.3,0.4) and (1.5,0.2)..(1.8,0.4)..controls (1.9, 0.467) and (1.9, 0.6) .. (1.8, 0.65)..controls (1.7,0.7) and (1.3, 0.65)..(1.0, 0.5);
\draw [thick, blue!40, -stealth](1.4, 0.636)--(1.35, 0.6255);
\end{scope}
\begin{scope}[xshift=0.7cm, yscale=-1]
\draw [thick, blue!40](1, 0.5)..controls(1.3,0.4) and (1.5,0.2)..(1.8,0.4)..controls (1.9, 0.467) and (1.9, 0.6) .. (1.8, 0.65)..controls (1.7,0.7) and (1.3, 0.65)..(1.0, 0.5);
\draw [thick, blue!40, -stealth](1.4, 0.636)--(1.35, 0.6255);
\end{scope}

\draw [white, line width = 3pt ] (2, 0) .. controls (3.6,0) and (3.6, 0.8) ..(2, 0.8);
\draw [white, line width = 3pt ] (2, 0) .. controls (0.4,0) and (0.4, 0.8) ..(2, 0.8);
\draw [line width = \lw ] (2, 0) .. controls (3.6,0) and (3.6, 0.8) ..(2, 0.8);
\draw [line width = \lw ] (2, 0) .. controls (0.4,0) and (0.4, 0.8) ..(2, 0.8);

\draw [white, line width = 3pt] (1.7, -0.5) .. controls (1.4, -0.5) and (1.4, 0.5) ..(1.7,0.5 );
\draw [white, line width = 3pt] (2.3, -0.5) .. controls (2, -0.5) and (2, 0.5) ..(2.3,0.5 );
\draw [line width = \lw] (1.7, -0.5) .. controls (1.4, -0.5) and (1.4, 0.5) ..(1.7,0.5 );
\draw [line width = \lw] (2.3, -0.5) .. controls (2, -0.5) and (2, 0.5) ..(2.3,0.5 );

\draw [ ->](2, 0.8) --(1.93, 0.8);
\draw [ -> ] (2.089, 0.2) --(2.075, 0.13);
\draw [ -> ] (1.489, 0.2) --(1.475, 0.13);

\node at (1.35, -0.05)[anchor=north, scale=1]{$\alpha$};
\node at (2.62, -0.05)[anchor=north, scale=1]{$\beta$};
\node at (2,0.8)[anchor=south, scale=1]{$\gamma$};

\end{scope}
\end{tikzpicture}
\caption{Three-loop braiding process involving vortex loops $\alpha$, $\beta$ and $\gamma$. The blue lines are trajectories swept out by two points on $\alpha$}
\label{fig_braiding}
\end{figure}
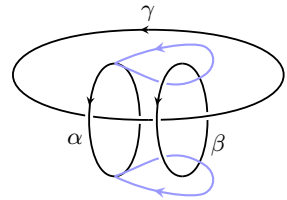

The most interesting part of 3D braiding statistics is between vortex loops. It was shown~\cite{Wang_PRL2014,Ran_PRX2014}  that the fundamental loop braiding process involves three loops (Fig.~\ref{fig_braiding}): a loop $\alpha$ is braided around a loop $\beta$ while both are linked to a third ``base'' loop $\gamma$. This three-loop braiding process has been used to characterize various \spd{3} topological phases.\cite{WangLevinPRB} Following the notation in \Ref{Wang_PRL2014}, we denote the three-loop braiding phase by $\theta_{\alpha\beta,c}$, where $c$ is the integer vector that labels the gauge fluxes carried by $\gamma$: $c\equiv\phi_\gamma = (\frac{2\pi}{N_1}c_1, \dots, \frac{2\pi}{N_K}c_K)$. We use $\theta_{\alpha\beta,c}$ instead of $\theta_{\alpha\beta,\gamma}$, because the three-loop braiding statistical phase depends only on the flux $\phi_\gamma$ parametrized by $c$ and is totally insensitive to the amount of charges attached to $\gamma$. We will also consider an exchange or half-braiding process: two identical loops $\alpha$, both linked to the base loop $\gamma$, exchange their positions. We denote this three-loop exchange statistics by $\theta_{\alpha,c}$. 

In most part of the paper, we will only consider Abelian braiding statistics.  Non-Abelian loop braiding statistics can also appear in gauge theories with Abelian gauge group. We will briefly touch upon non-Abelian loop braiding at the end of the paper.

\section{Physical Constraints on Abelian Loop Braiding Statistics}
\label{sec:constraint}

In this section, we study general properties of loop braiding statistics in gauged \spd{3} FSPT systems. For simplicity,  we only consider Abelian loop statistics, i.e., every Berry phase associated with braiding excitations is Abelian. We will discuss physical constraints on Abelian loop braiding statistics, and will discuss which types can result from ``intrinsic'' FSPT phases.

\subsection{Topological invariants}

To begin, we define a set of topological invariants $\{\Theta_{ij,k}, \Theta_{i,k}\}$ for loop braiding statistics. These topological invariants are a subset of the full braiding statistics data, and hence are simpler to deal with compared to the latter. Nevertheless, they are equivalent to the full set of loop braiding statistics, since the latter can be reconstructed out of the former. Similar topological invariants have been introduced in \spd{2}/\spd{3} gauged BSPT phases~\cite{Wang_PRL2014, WangLevinPRB}, as well as 2D gauged FSPT phases.\cite{WangFSPT}

Let $\alpha, \beta$ and $\gamma$ be vortex loops, carrying unit flux $\frac{2\pi}{N_i}e_i, \frac{2\pi}{N_j}e_j$ and $\frac{2\pi}{N_k}e_k$, respectively. Here, $e_i$ is an integer vector $(0,\dots,1,\dots,0)$ where the $i$th entry is 1 and all other entries are 0, with $i=0,1,\dots,K$. We define the topological invariant $\Theta_{ij,k}$ as follows:
\begin{equation}
	\Theta_{ij,k}=N^{ij}\theta_{\alpha\beta,e_k},
	\label{}
\end{equation}
where $\theta_{\alpha\beta,e_k}$ is the mutual braiding statistics between $\alpha$ and $\beta$ while both are linked to the base loop $\gamma$ (Fig.~\ref{fig_braiding}). Here, we use $N^{ij}$ to denote the least common multiple of $N_i$ and $N_j$. 

Similarly, we define a topological invariant $\Theta_{i,k}$ for the self statistics associated with exchanging two identical $\alpha$'s, both of which are linked to the base loop $\gamma$. It is defined as follows
\begin{equation}
\Theta_{i,k} = \tilde N_i \theta_{\alpha,e_k}
\end{equation}
where 
\begin{equation}
	\begin{split}
	\tilde N_0 &=
	\begin{cases}
		\displaystyle N_0,  &  \frac{N_0}{2}\equiv 0\, (\text{mod } 2)\\
		\frac{N_0}{2},  &  \frac{N_0}{2}\equiv 1\, (\text{mod } 2)\\
	\end{cases}	
	\end{split}
	\label{tildeN0} 
\end{equation}
and for $i\ge 1$,
\begin{equation}
	\begin{split}
	\tilde N_i &=
	\begin{cases}
		N_i,  &  N_i\equiv 0\, (\text{mod } 2)\\
		2N_i,  &  N_i\equiv 1\, (\text{mod } 2)\\
	\end{cases}	
	\end{split}
	\label{tildeNi} 
\end{equation}

The above topological invariants $\{\Theta_{ij,k}, \Theta_{i,k}\}$ are defined in a way such that (i) they only depend on the flux of $\alpha$, $\beta$ and $\gamma$ and (ii) the full set of braiding statistics can be reconstructed out of $\{\Theta_{ij,k}, \Theta_{i,k}\}$. One can check the property (i) by replacing $\alpha,\beta,\gamma$ with $\alpha',\beta',\gamma'$ respectively. It is easy to show that the topological invariants do not change if $\phi_{\alpha'}=\phi_{\alpha}$, $\phi_{\beta'}=\phi_{\beta}$ and $\phi_{\gamma'} = \phi_\gamma$. The proof of property (ii) is more involved, so we give the proof in Appendix \ref{app:reconstruction}. 

\subsection{Physical constraints}

The topological invariants $\{\Theta_{ij,k}, \Theta_{i,k}\}$ cannot take arbitrary values. We argue that the topological invariants have to satisfy the following constraints:
\begin{subequations}
	\begin{align}
		&\Theta_{ii,k}=\frac{2N_i}{\tilde{N}_i}\Theta_{i,k}, \label{eqn:const1}\\
		&\Theta_{ij,k}=\Theta_{ji,k},\label{eqn:const2}\\
		&N_{ijk}\Theta_{ij,k}=0,\label{eqn:const3}\\
		&N_{k}\Theta_{i,k}=0,\label{eqn:const4}\\
		&N_i \Theta_{i,k}=	\frac{N_{0i}}{2}\Theta_{0i,k},\quad\quad \ \ (\text{$N_i$ even, $i\ge 1$})\label{eqn:const5}\\
	&\Theta_{i,k}\frac{N^{ik}}{\tilde{N}_i}+\Theta_{ik,i}=0, \quad\quad (\text{$N^{ik}$ even}) \label{eqn:const6}\\
	&\frac{N^{ijk}}{N^{ij}}\Theta_{ij,k}+\frac{N^{ijk}}{N^{jk}}\Theta_{jk,i} + \frac{N^{ijk}}{N^{ki}}\Theta_{ki,j}=0,\label{eqn:const7}\\
		&\Theta_{i,i}=0, \: \ \ \quad\quad \ \ \ \ \ \quad\quad\quad (\text{conjectured})  \label{eqn:const8}\\
		&\frac{N_0^2}{4\tilde{N}_0}\Theta_{0,i}=0, \ \ \ \ \  \label{eqn:const9}
	\end{align}
\end{subequations}
where $N_{i\dots k}$ denotes the greatest common divisor of $N_i, \dots, N_k$, and $N^{i\dots k}$ denotes the least common multiple of $N_i, \dots, N_k$. These constraints can be proved by checking various consistency conditions on the three-loop braiding statistics. The proofs are given separately in Appendix \ref{appendix_proof}.

These constraints are {\it necessary} conditions that physical Abelian three-loop braiding must satisfy (except Eq.~\eqref{eqn:const8} which remains a conjecture at this stage). On the other hand, we do not know at this point whether these constraints are also {\it sufficient}, in the sense that every solution to these constraints can be realized in physical systems. To verify the completeness of the constraints, we will present a family of exactly solvable lattice models in Sec.~\ref{sec_state-sum} and show that indeed every solution to the constraints are physical.

Several comments are in order. First, similar constraints were obtained for gauge theories coupled to BSPT systems in \Ref{Wang_PRL2014} and \Ref{WangLevinPRB}. Most of the constraints here are just variants of those for BSPT phases, and some are even the same, e.g., Eqs.~\eqref{eqn:const3}, \eqref{eqn:const4} and \eqref{eqn:const7}. However, Eq.~\eqref{eqn:const5} is more ``fermionic'' than others, since it has no bosonic analog. It replaces the stronger condition $N_i\Theta_{i,k}=0$ in bosonic theories(see Ref.~\onlinecite{WangLevinPRB}). Nevertheless, in a sense it is a ``2D'' constraint~\cite{WangFSPT}, since the base loop does not enter the constraint in any nontrivial way. Similarly, Eq. \eqref{eqn:const9} has no analog in bosonic systems.

Second, the constraint Eq. \eqref{eqn:const8} remains a conjecture at this stage. We are not able to give a general proof. A weaker constraint can be derived from Eqs. \eqref{eqn:const1}, \eqref{eqn:const6} and \eqref{eqn:const7}:
\begin{equation}
	\frac{3N_0}{\tilde N_0}\Theta_{0,0}=0; \quad \text{gcd}(3, N_i)\Theta_{i,i}=0 \quad (i\ge 1). 
\end{equation}
This weaker result provides some evidence for the conjecture Eq. \eqref{eqn:const8}. In fact, it remains a conjecture in gauged BSPT systems too\cite{WangLevinPRB} (though, see \Ref{WangPRX2016} for a derivation of Eq. \eqref{eqn:const8} by exploiting the bulk-boundary correspondence under certain assumptions).

Third, Eq.~\eqref{eqn:const5} holds only for even $N_i$ with $i\ge 1$. There are no analogous constraints for odd $N_i$. In addition, Eq.~\eqref{eqn:const6} holds only when $N^{ik}$ is even.

Lastly, we derive several useful corollaries. The first one follows from \eqref{eqn:const1} and \eqref{eqn:const3}:
\begin{equation}
	\frac{2N_i}{\tilde{N}_i}N_{ik}\Theta_{i,k}=0.
	\label{eqn:corollary1}
\end{equation}
Note that for BSPT theories, we have a stronger condition $N_{ik}\Theta_{i,k}=0$ (see \Ref{WangLevinPRB}). Another corollary follows from \eqref{eqn:const4} and \eqref{eqn:const5}. Setting $k=i$ in Eq.~\eqref{eqn:const5} and using Eq.~\eqref{eqn:const4}, we immediately obtain
\begin{equation}
	\frac{N_{0i}}{2}\Theta_{0i,i}=0, \quad (\text{$N_i$ even, $i\ge 1$})
	\label{eqn:corollary2}
\end{equation}
Finally, combining Eq.~\eqref{eqn:corollary2} with Eq. \eqref{eqn:const6}, we have
\begin{equation}
	\frac{N_0}{2}\Theta_{i,0}=0
	\label{eqn:corollary3}
\end{equation}
Even though Eq.~\eqref{eqn:corollary3} follows from Eq.~\eqref{eqn:corollary2} which holds only for even $N_i$ with $i\ge1$, one can easily check that it holds in general.

\subsection{Solutions from BSPT phases}
\label{sec:bspt}

Later on, we will solve the constraints Eqs.~\eqref{eqn:const1}-\eqref{eqn:const9}, where each solution leads to a consistent set of loop braiding statistics and corresponds to an FSPT phase. BSPT phases form a subset of FSPT phases, so we first write down a class of solutions that stem from BSPT phases.

Physically, we can imagine the following construction: first, we form ``molecules'' from pairs of fermions, where each molecule is a boson. The bosonic molecules are neutral under $\mathbb{Z}_2^f$, and thus only sense the quotient symmetry group $G=\mathcal G/\mathbb Z_2^f=\mathbb Z_m \times \prod_i \mathbb Z_{N_i}$. We then put the molecules into a BSPT state protected by the symmetry $G$. It is now generally believed that the \spd{3} BSPT phases with unitary symmetry $G$ are classified by the cohomology group $\mathcal{H}^4[G, \U(1)]$.\cite{Chen_arxiv2012} The loop braiding statistics in gauged BSPT models were studied in \Ref{Wang_PRL2014, WangLevinPRB}. 

Now we can adapt the loop braiding statistics of BSPT phases from Ref.~\onlinecite{WangLevinPRB} into $\Theta_{i,k}$ and $\Theta_{ij,k}$ and find the following expressions:
\begin{align}
\Theta_{ij,k} & = \frac{2\pi N^{ij}}{\bar N_{ik}\bar N_j}(M_{ikj}-M_{kij}) +\frac{2\pi  N^{ij}}{\bar N_{jk}\bar N_i}(M_{jki}-M_{kji}) \label{bspt1}
\end{align}
 and 
\begin{align}
\Theta_{i,k} & =  \frac{\tilde N_i}{\bar N_i} \frac{2\pi}{\bar N_{ik}}(M_{iki}-M_{kii}).
\label{bspt2}
\end{align}
Here $M_{ijk}$ is an arbitrary three-index integer tensor, $\bar N_{ik}=\gcd(\bar N_i, \bar N_k)$, and
\begin{equation}
	\begin{split}
	\bar N_i &=
	\begin{cases}
		\frac{N_0}{2},  & \quad i=0\\[3pt]
		N_i,  &\quad  i\ge1\\
	\end{cases}	
	\end{split}
\end{equation}
One can easily check that the above expressions satisfy all the constraints \eqref{eqn:const1}-\eqref{eqn:const9}. Note that different values of $M_{ijk}$ can lead to the same values of the topological invariants.


\subsection{Braiding statistics of fermion parity loops}
\label{sec:parityloop}

\setitemize[0]{leftmargin=10pt,rightmargin=10pt}

We are mainly interested in loop braiding statistics beyond those given by Eqs.~\eqref{bspt1} and \eqref{bspt2}, i.e., those resulting from gauging  ``instrinsic'' FSPT phases. Such loop braiding statistics will be explicitly discussed in the next subsection. Before doing that, we would like to answer this question:  Given a solution to the constraints Eqs.~\eqref{eqn:const1}-\eqref{eqn:const9}, i.e., a set of three-loop braiding statistics, how do we know whether it is ``intrinsically fermionic'', and not just a gauged BSPT phase in disguise?

We claim that the following criterion holds: 
\begin{itemize}
\item[]  \textbf{Criterion}: {\it A set of three-loop braiding statistics results from gauging an intrinsic FSPT phase, if and only if some of the three-loop braiding statistics involving fermion parity loops are ``nontrivial''. }
\end{itemize}
Those three-loop braiding statistics that involve fermion parity loops include $\theta_{\alpha,\mathbf f}$, $\theta_{\alpha\beta,\mathbf f}$, $\theta_{\mathbf f, \gamma}$ and $\theta_{\alpha\mathbf f,\gamma}$, where $\alpha,\beta,\gamma$ are arbitrary vortex loops and $\mathbf f$ stands for a fermion parity loop.

We need to clarify what we mean by ``nontrivial'' three-loop braiding statistics. Just as any other vortex excitations, there are many fermion parity loops which  differ by charge attachments. Attaching charges to loops shifts three-loop braiding statistics by Aharonov-Bohm phases. Accordingly, we call a three-loop braiding statistical phase ``trivial'' if it can be tuned to $0$ by attaching charges to the loops involved in the braiding process.  To remove the ambiguity due to charge attachment and Aharonov-Bohm phases, we define the following quantities for Abelian loop braiding statistics:  
\begin{align}
\Theta_{i,\mathbf{f}}  & =\tilde{N_i}\theta_{\alpha,\mathbf{f}}, \nonumber\\
\Theta_{ij, \mathbf{f}}& = N^{ij}\theta_{\alpha\beta,\mathbf{f}}, \nonumber\\
\Theta_{\mathbf{f},k}  & = \theta_{\mathbf{f},e_k}, \nonumber\\
\Theta_{\mathbf{f}i, k}& = \text{lcm}(2,N_k)\theta_{\mathbf{f}\alpha,e_k}, \label{fp_def}
\end{align}
where ``$\text{lcm}$'' stands for least common multiple, $\theta_{\alpha,\mathbf{f}}$ is the exchange statistics of two identical $\alpha$'s linked to a fermion parity loop $\mathbf{f}$, $\theta_{\alpha\beta,\mathbf{f}}$ is the mutual braiding statistics between $\alpha$ and $\beta$ both linked to $\mathbf{f}$, $\theta_{\mathbf{f},e_k}$ is the exchange statistics of two identical $\mathbf{f}$'s linked to a base loop $\gamma$, and $\theta_{\mathbf{f}\alpha,e_k}$ is the mutual statistics between $\alpha$ and $\mathbf{f}$ while both are linked to $\gamma$. Here, $\alpha,\beta,\gamma$ are vortex loops carrying gauge flux $\frac{2\pi}{N_i}e_i$,$\frac{2\pi}{N_j}e_j$, and $\frac{2\pi}{N_k}e_k$ respectively. These quantities are defined in a way  similar to the topological invariants $\Theta_{i,k}$ and $\Theta_{ij,k}$. One can easily show that if $\Theta_{i,\mathbf{f}}$, $\Theta_{ij,\mathbf{f}}$, $\Theta_{\mathbf{f},k}$ and $\Theta_{\mathbf{f}j,k}$ vanish, all three-loop braiding statistics involving fermion parity loops are ``trivial''.  Therefore to see if a set of three-loop braiding statistics corresponds to an intrinsic FSPT phase, we only need to check if any of the quantities $\Theta_{i,\mathbf{f}}$, $\Theta_{ij,\mathbf{f}}$, $\Theta_{\mathbf{f},k}$ and $\Theta_{\mathbf{f} j,k}$ is nonvanishing.

This criterion can be proven by explicitly solving the constraints Eqs.~\eqref{eqn:const1}-\eqref{eqn:const9} and checking if all the solutions with vanishing $\Theta_{i,\mathbf{f}}$, $\Theta_{ij,\mathbf{f}}$, $\Theta_{\mathbf{f},k}$ and $\Theta_{\mathbf{f} j,k}$ are in the form of Eqs.~\eqref{bspt1} and \eqref{bspt2} (which will be discussed in the next subsection). On the other hand, Eqs.~\eqref{bspt1} and \eqref{bspt2} indeed lead to vanishing  $\Theta_{i,\mathbf{f}}$, $\Theta_{ij,\mathbf{f}}$, $\Theta_{\mathbf{f},k}$ and $\Theta_{\mathbf{f} j,k}$. 

Here, we would like to give a more intuitive argument. Although heuristic, it provides a physical interpretation for the criterion and can be applied more generally to non-Abelian loop braiding. Let us first show the ``if'' direction in the criterion. Consider those FSPT phases formed by bosonic pairs of fermions. Since the pairs do not transform under the fermion parity symmetry, they only need the symmetries in $\mathcal G/\mathbb Z_2^f$ for protection. Under the assumption that FSPT phases have a one-to-one correspondence to equivalence classes of three-loop braiding statistics (i.e. up to charge attachment), it is reasonable to expect that the fermion parity loops should not play a nontrivial role in three-loop braiding statistics beyond Aharonov-Bohm phases after gauging the symmetry $\mathcal{G}$. Hence, the ``if'' direction holds. 

To see the ``only if'' direction, let us assume that fermion parity loops do not play any role beyond Aharonov-Bohm phases. Then, for any three-loop structure involving a fermion parity $\mathbf{f}$, we can always attach charges to the loops such that (1) the new fermion parity loop $\mathbf{f}'$ has a bosonic exchange statistics and (2) for any given gauge flux, there always exists a vortex such that its mutual braiding statistics with respect to $\mathbf{f}'$ is trivial, while both are linked to the same base loop.  Accordingly, we can condense $\mathbf{f}'$, confining the fermionic gauge charges. This condensation leaves behind a gauge theory with purely bosonic charges, without affecting the other gauge symmetries, i.e., the resulting theory has a gauge group $\mathcal{G}/\mathbb{Z}_2^f$. As a consequence, the corresponding  FSPT phases are always equivalent to those formed by bosonic pairs. This argument is reasonable but not quite rigorous, because a complete theory of loop condensation in topological phases is not available yet. We notice that similar arguments have been applied in two dimensions~\cite{You2015}.

Let us now combine this criterion with the constraints Eqs.~\eqref{eqn:const1}-\eqref{eqn:const9} of topological invariants. We will see that the quantities $\Theta_{i,\mathbf{f}}, \Theta_{ij,\mathbf{f}}, \Theta_{\mathbf{f},k}$ are forced to vanish due to the constraints; only $\Theta_{\mathbf{f}i,k}$ may possibly be nonzero. Accordingly, for Abelian loop braiding statistics, we only need to compute $\Theta_{\mathbf{f}i,k}$ to see if a given set of Abelian three-loop braiding statistics corresponds to an intrinsic FSPT phase.

To derive these results, we first relate the quantities defined in Eq.~\eqref{fp_def} to the topological invariants as follows: 
\begin{subequations}
\begin{align}
\Theta_{i,\mathbf{f}}  & =\frac{N_0}{2}\Theta_{i,0}, \label{fp1}\\
\Theta_{ij, \mathbf{f}}& =\frac{N_0}{2}\Theta_{ij,0}, \label{fp2}\\
\Theta_{\mathbf{f},k}  & =\frac{N_0^2}{4\tilde{N}_0}\Theta_{0,k}, \label{fp3}\\
\Theta_{\mathbf{f}i, k}& = \frac{N_{0i}}{\gcd(2, N_i)}\Theta_{0i,k}. \label{fp4}
\end{align}
\end{subequations}
These relations follow straightforwardly from the definitions of related quantities. 

The fact that $\Theta_{i,\mathbf f} =0$ follows immediately from Eq.~\eqref{eqn:corollary3}, and $\Theta_{\mathbf{f},k}=0$ follows immediately form Eq.~\eqref{eqn:const9}. To see $\Theta_{ij,\mathbf{f}}=0$, we consider three cases. 
\begin{enumerate}
	\item $i=j=0$, it follows from Eqs.~\eqref{eqn:const1} and (\ref{eqn:corollary3}).
	\item $i=0, j\ge{1}$. If $N_j$ is odd, then $N_{0j}=\gcd(m,N_j)$. Then, $\Theta_{0j,\mathbf{f}}=0$ follows from the constraint Eq.~\eqref{eqn:const3}. If $N_j$ is even, we have
\begin{equation}
	\Theta_{0j,\mathbf{f}}=\frac{N_0}{2}\Theta_{0j,0}=\frac{N_0}{N_{0j}}N_j \Theta_{j,0}=N^{0j}\Theta_{j,0}\equiv 0,
\end{equation}
where the first equality follows Eq.~\eqref{fp1}, the second follows Eq.~\eqref{eqn:const5}, and the last follows Eq.~\eqref{eqn:const4}.
\item $i,j\ge 1$. Let us denote $N_0=2^{r_0}t_0$, $N_i= 2^{r_i}t_i$ and $N_j=2^{r_j}t_j$, where $t_0, t_i,t_j$ are odd numbers. Without loss of generality, we assume $r_i\le r_j$. It is not hard to see that $\Theta_{ij,\mathbf{f}}$ can only be $0$ or $\pi$. Using Eqs.~\eqref{eqn:const3} and \eqref{fp2}, we find that $\Theta_{ij,\mathbf{f}}$ may be nonzero only if $r_0\le r_i\le r_j$. Assuming this is the case, we write Eq.~\eqref{eqn:const7} in the following form:
\begin{equation}
\frac{t^{0ij}}{t^{ij}}\Theta_{ij,0} = -\frac{2^{r_j-r_i}t^{0ij}}{t^{0i}}\Theta_{0i,j} -\frac{t^{0ij}}{t^{0j}}\Theta_{j0,i} \label{fp5}.
\end{equation}
With this equation, we then have
\begin{align}
\Theta_{ij,\mathbf{f}} & =  \frac{N_0}{2}\Theta_{ij,0} \cdot t_{0i}t_{0j}\frac{t^{0ij}}{t^{ij}} \nonumber \\
  &  = 2^{r_0-1}t_0t_{0i}t_{0j}\left(-\frac{2^{r_j-r_i}t^{0ij}}{t^{0i}}\Theta_{0i,j} -\frac{t^{0ij}}{t^{0j}}\Theta_{j0,i}\right) \nonumber\\
& =  -t_0t_{0j}\frac{2^{r_j-r_i}t^{0ij}}{t^{0i}} N_i\Theta_{i,j}  -  t_0t_{0i}\frac{t^{0ij}}{t^{0j}} N_j \Theta_{j,i} \nonumber \\
& =  -t_{0i}t_{0j}t^{0ij} 2^{r_j} (\Theta_{i,j}+\Theta_{j,i})
\end{align} 
where the first line uses the facts that $\Theta_{ij,\mathbf{f}}$ can only be $0$ or $\pi$ and that $ t_{0i}t_{0j}\frac{t^{0ij}}{t^{ij}}$ is odd, the second line uses Eq.~\eqref{fp5}, the third line uses the constraint Eq.~\eqref{eqn:const5}, and the last line is a simplification. It is easy to see that both $N_i$ and $N_j$ divide the coefficient  $t_{0i}t_{0j}t^{0ij} 2^{r_j}$. Then, using the constraint Eq.~\eqref{eqn:const4}, we prove that $\Theta_{ij,\mathbf{f}}=0$. 
\end{enumerate}

Therefore, only $\Theta_{\mathbf{f}i,k}$ is potentially nonzero. Accordingly to Eqs.~\eqref{eqn:const3} and \eqref{fp4}, one can see that it can only take values $0$ or $\pi$. Moreover, if either $N_i$ or $N_k$ is odd, Eq.~\eqref{eqn:const3} is enough to guarantee $\Theta_{\mathbf{f}i,k}=0$. If both $N_i$ and $N_k$ are even, from Eq.~\eqref{eqn:const5}, we have $\Theta_{\mathbf{f}i,k}=N_i\Theta_{i,k}$ for $i\ge 1$, which further lead to $\Theta_{\mathbf{f}i,i}=0$ using Eq.~\eqref{eqn:const8}. For $i=0$, we have $\Theta_{\mathbf{f}0,k} = \frac{N_0}{2}\Theta_{00,k} = 0$ according to Eq.~\eqref{fp3}.  

To summarize, we have shown that to check whether a set of topological invariants corresponds to an intrinsic FSPT phase or not, we only need to check if $\Theta_{\mathbf{f}i,k}$ is nonzero for $i\neq k$, $i\ge 1$, and $N_i$ and $N_k$ are both even.

\subsection{Solving the constraints}
\label{sec_solving}

We now explicitly solve the constraints Eqs.~\eqref{eqn:const1}-\eqref{eqn:const9}. Mathematically speaking, the constraints are linear equations of the tensors $\Theta_{i,k}$ and $\Theta_{ij,k}$. Solving them is straightforward, though tedious due to the fact that the equations are defined modulo $2\pi$. 

We first notice the following structure of the solutions: Given two sets of topological invariants $\Theta^{(1)}$ and $\Theta^{(2)}$, if all intrinsic FSPT indicators $\Theta_{\mathbf{f}i,k}$ are identical (for $i\neq k, i\ge 1$), we can define $\Theta'=\Theta^{(1)}-\Theta^{(2)}$. Due to the linearity of the constraints, $\Theta'$ also satisfy all constraints. In fact, we see that
\begin{equation}
	N_i\Theta'_{i,k}=\Theta_{\mathbf{f}i,k}^{(1)}-\Theta_{\mathbf{f}i,k}^{(2)}=0.
	\label{}
\end{equation}
Combined with $N_i\Theta'_{i,k}=0$ we obtain $N_{ik}\Theta'_{i,k}=0$. In fact, $\Theta'$ satisfies essentially the stronger constraints for BSPT phases, whose solutions have been given in Sec. \ref{sec:bspt}. Therefore, once we know the solutions corresponding to intrinsic FSPT phases, all others can be obtained by adding BSPT solutions.

To solve the constraints, a useful observation is that the constraints only relate those components of tensors whose indices differ at most by one index 0. Accordingly, we can divide the components of the tensors into four groups:
\begin{align}
(a): &\ \Theta_{0,0}, \Theta_{00,0}\nonumber\\
(b): &\ \Theta_{i,0}, \Theta_{0,i}, \Theta_{i,i}, \Theta_{i0,0},\Theta_{00,i}, \Theta_{ii,0},  \Theta_{i0,i}, \Theta_{ii,i} \nonumber\\
(c): &\ \Theta_{i,j}, \Theta_{j,i}, \Theta_{ij,0}, \Theta_{j0,i}, \Theta_{i0,j}, \Theta_{jj,i}, \Theta_{ii,j},\Theta_{ij,i}, \Theta_{ij,j}\nonumber\\
(d): &\ \Theta_{ij,k}, \Theta_{jk,i}, \Theta_{ki,j} \nonumber
\end{align}
where $i\neq j\neq k \neq 0$. Since $\Theta_{ij,k}$ is symmetric in the first two indices, we do not list other components that are related by this symmetry above. 

In the group $(a)$, only the trivial solution is allowed: $\Theta_{00,0}=\Theta_{0,0}=0$. It follows directly from the constraints Eq.~\eqref{eqn:const1} and \eqref{eqn:const8}. Also, invariants in the group $(d)$ satisfy the same equations with those of BSPT phases. Hence, solutions for the group $(d)$ are exactly the same as in BSPT phases, i.e., all can be written in the form Eqs.~\eqref{bspt1}. 

Below we solve the constraints for cases $(b)$ and $(c)$. Without loss of generality, we consider $\mathcal G=\mathbb Z_{2m}^f\times \mathbb Z_{N_1}$ for case $(b)$ and consider $\mathcal G = \mathbb Z_{2m}^f\times \mathbb Z_{N_1}\times\mathbb Z_{N_2}$ for case $(c)$.

\subsubsection{$\mathcal G=\mathbb{Z}_{2m}^f\times\mathbb{Z}_{N_1}$}

Consider the symmetry group $\mathcal{G} =\mathbb{Z}_{N_0}^f\times\mathbb{Z}_{N_1} $ with $N_0=2m$.  We solve the constraints for topological invariants in the group $(b)$ with $i=1$. Among the eight components, we find that $\Theta_{0,1}$, $\Theta_{1,0}$ and $\Theta_{1,1}$ completely determine the rest. More explicitly,
\begin{align}
\Theta_{11,1}& =\frac{2N_1}{\tilde N_1}\Theta_{1,1} \nonumber\\
\Theta_{00,1}&= \frac{2N_0}{\tilde N_0}\Theta_{0,1}, \quad  \Theta_{11,0}= \frac{2N_1}{\tilde N_1}\Theta_{1,0} \nonumber\\
\Theta_{10,0}&= -\frac{N^{01}}{\tilde N_0}\Theta_{0,1}, \quad  \Theta_{10,1} =-\frac{N^{01}}{\tilde N_1}\Theta_{1,0} \label{case_b1}
\end{align} 
where  the first and second lines  follow Eq.~\eqref{eqn:const1} and the third line follows Eq.~\eqref{eqn:const6}. 

Since $\Theta_{1,1}=0$ following Eq.~\eqref{eqn:const8}, we only need to find possible values for $\Theta_{0,1}$ and $\Theta_{1,0}$. Let us first consider odd $N_1$. In this case, using the constraints Eqs.~\eqref{eqn:const1},\eqref{eqn:const3}, \eqref{eqn:const4}, and \eqref{eqn:const9}, we find that 
\begin{align}
\Theta_{0,1} = \frac{2\pi}{\gcd(m, N_1)}x,
\Theta_{1,0} &=\frac{2\pi}{\gcd(m,N_1)} y.
\end{align}
Here $x,y$ are integers. It is not hard to see that this solution is in the form of Eq. \eqref{bspt2}. This agrees with the criterion discussed in Sec.~\ref{sec:parityloop}.

When $N_1$ is even, using Eqs.~\eqref{eqn:const4} and \eqref{eqn:const9}, we find that $\Theta_{0,1}$ can take the following values
\begin{align}
\Theta_{0,1} = \left\{
\begin{array}{ll}
\displaystyle
\frac{2\pi}{\gcd(m, N_1)} x, & \text{if $m$ is odd}\\[17pt]
\displaystyle
\frac{2\pi}{\gcd(m/2, N_1)} x, & \text{if $m$ is even} \\
\end{array},
\right. \label{case_b2}
\end{align}
where $x$ is an integer. According to the corollary Eq.~\eqref{eqn:corollary3}, we have $m \Theta_{1,0}=0$. In addition, multiplying Eq.~\eqref{eqn:const5} by 2 and using Eq.~\eqref{eqn:const3}, we have $2N_1\Theta_{1,0}=0$. Together we find
\begin{align}
\Theta_{1,0} = \left\{
\begin{array}{ll}
\displaystyle
\frac{2\pi}{\gcd(m, N_1)} y, & \text{if $m$ is odd} \\[17pt]
\displaystyle
\frac{2\pi}{\gcd(m, 2N_1)}y, & \text{if $m$ is even}
\end{array}
\right. \label{case_b3}
\end{align}
For odd $m$, the parameters $x$ and $y$ are independent. For even $m$, there exists a relation between $x$ and $y$:  Taking $i=1$ and $k=0$ in Eq.~\eqref{eqn:const5} and using the expression of $\Theta_{01,0}=\Theta_{10,0}$ in Eq.~\eqref{case_b1}, we find that 
\begin{equation}
\frac{N_1}{\gcd(m/2, N_1)}y\pi = \frac{N_1}{\gcd(m/2, N_1)}x\pi \label{case_b4}
\end{equation}
This relation puts an constraint $x\equiv y\,(\text{mod }2)$ on $x$ and $y$ only if $N_1/\gcd(m/2, N_1)$ is odd. 

Let us see which solutions correspond to ``intrinsic'' FSPT phases. According to the criterion discussed in Sec.~\ref{sec:parityloop}, we only need to check the quantity $\Theta_{\mathbf{f}1,0}$. It is non-vanishing only when $m$ is even and $N_1$ is even, in which case we find that 
\begin{equation}
\Theta_{\mathbf{f}1,0} = \frac{N_1}{\gcd(m/2, N_1) }x \pi.
\end{equation}
Let $N_0=2^{r_0} t_0$ and $N_1 = 2^{r_1}t_1$, where $r_0\ge2$, $r_1\ge 1$ and $t_0,t_1$ are odd numbers. Then, it is easy to see that $\Theta_{\mathbf{f}1,0}=\pi$ only if
\begin{equation}
r_0\ge r_1+2 \ge 3
\end{equation}
and $x$ is an odd number. Therefore, the simplest symmetry group to support intrinsic FSPT phases is $\mathcal{G} = \mathbb{Z}^f_8\times\mathbb{Z}_2$.

\subsubsection{$\mathcal G=\mathbb{Z}_{2m}^f\times\mathbb{Z}_{N_1}\times\mathbb{Z}_{N_2}$}
In this case, one will find that if either $N_1$ or $N_2$ is odd, all solutions to the constraints correspond to BSPT phases, given by \eqref{bspt1} and \eqref{bspt2}. Hence, below we focus on the more interesting case where both $N_1$ and $N_2$ are even. One can show that any odd factors of $N_0,N_1, N_2$ cannot add solutions that correspond to intrinsic FSPT phases. Hence we assume that $N_i=2^{r_i}$ for $i=0,1,2$, with $r_0,r_1,r_2\ge 1$ for simplicity. Without loss of generality, we further take $r_1\le r_2$. 

There are nine topological invariants in the group $(c)$ with $i=1$ and $j=2$. $\Theta_{11,2}, \Theta_{22,1}, \Theta_{12,1}, \Theta_{12,2}$ are determined by $\Theta_{1,2}$ or $\Theta_{2,1}$ through Eqs.~\eqref{eqn:const1} and \eqref{eqn:const6}:
\begin{align}
\Theta_{11,2}& =2\Theta_{1,2}, \quad \Theta_{22,1} = 2\Theta_{2,1} \nonumber\\
\Theta_{12,1} & = -\frac{N^{12}}{N_1} \Theta_{1,2}, \quad \Theta_{12,2} = -\frac{N^{12}}{N_2}\Theta_{2,1}\label{case_c1}
\end{align}
Multiplying Eq.~\eqref{eqn:const5} by 2 and using Eq.~\eqref{eqn:const3}, we have $2N_1\Theta_{1,2}=2N_2\Theta_{2,1}=0$. Combining this with Eq.~\eqref{eqn:const4}, we have
\begin{align}
\displaystyle \Theta_{1,2} & = \frac{2\pi}{2^{\min(r_1+1,r_2)}} a_1, \nonumber\\
\displaystyle \Theta_{2,1} & = \frac{2\pi}{2^{r_1}} a_2 \label{case_c2}.
\end{align}
Solving the constraints Eq.~\eqref{eqn:const2} and \eqref{eqn:const7}, we find that
\begin{align}
\Theta_{01,2} & = \frac{2\pi N^{01}}{N_{02}N_1}b_1 - \frac{2\pi N^{01}}{N_{12}N_0}b_2,\nonumber\\
\Theta_{12,0} & = \frac{2\pi N^{12}}{N_{01}N_2} b_3 - \frac{2\pi N^{12}}{N_{02}N_1} b_1, \nonumber\\
\Theta_{20,1} & = \frac{2\pi N^{02}}{N_{12}N_0}b_2 -\frac{2\pi N^{02}}{N_{01}N_2}b_3. \label{case_c3}
\end{align}
The parameters $a_1,b_1,b_2,b_3$ are not arbitrary. Using Eq.~\eqref{eqn:const5}, they should satisfy the following relations
\begin{align}
\frac{\pi N_2}{N_{12}} b_2 = \frac{\pi N_0}{N_{01}}b_3, \nonumber\\
\frac{2\pi}{2^{\min(r_1+1, r_2)-r_1}} a_1 + \pi b_2&  = \frac{\pi N_0}{N_{02}}b_1.
\end{align}

Let us see which of the solutions correspond to intrinsic FSPT phases. Accordingly to the criterion in Sec.~\ref{sec:parityloop}, we need to evaluate $\Theta_{\mathbf{f}1,2}$ and $\Theta_{\mathbf{f}2,1}$. We find that $\Theta_{\mathbf{f}2,1}=0$, and 
\begin{equation}
\Theta_{\mathbf{f}1,2} = N_1\Theta_{1,2} = \frac{2\pi}{2^{\min(r_1+1, r_2)-r_1}} a_1.
\end{equation} 
Therefore intrinsic FSPT phases with $\Theta_{\mathbf{f}1,2}=\pi$ can only occur for
\begin{equation}
r_2 \ge r_1+1 \ge 2,
\end{equation}
and $a_1$ is an odd number. The simplest example is then the group $\mathcal{G}=\mathbb Z_2^f\times\mathbb Z_2\times\mathbb Z_4$.

\subsection{Examples}
In this subsection, we discuss two examples whose three-loop braiding statistics correspond to intrinsic FSPT phases.

\subsubsection{$\mathcal G = \mathbb Z_8^f\times \mathbb Z_2$}
One of the simplest groups that support intrinsic FSPT phases is $\mathcal{G} = \mathbb{Z}_8^f\times\mathbb{Z}_2$. It supports the following three-loop braiding statistics
\begin{align}
\Theta_{0,1} = \pi, \quad  \Theta_{1,0}=\frac{\pi}{2}, \quad \Theta_{11,0}= \pi, \quad \Theta_{01,0}=\pi
\end{align}
and all other invariants are $0$. It is a solution obtained in Sec.~\ref{sec_solving} by taking $x=y=1$ in Eqs.~\eqref{case_b2} and \eqref{case_b3}.

\subsubsection{$\mathcal G = \mathbb Z_2^f\times\mathbb Z_2\times \mathbb Z_4$}

Another simple group that supports intrinsic FSPT phases is  $\mathcal{G} = \mathbb{Z}_2^f\times\mathbb{Z}_2\times\mathbb{Z}_4$. The ``root'' intrinsic phase is characterized by the following topological invariants:
\begin{equation}
	\Theta_{1,2}=\frac{\pi}{2}, \quad \Theta_{11,2}=\pi, \quad \Theta_{01,2}=\pi, \quad \Theta_{12,1}=\pi.
	\label{eqn:z2z4}
\end{equation}
And all other invariants are $0$. It is obtained by setting $a_1=b_1=1$, $a_2=b_2=b_3=0$ in Eqs. \eqref{case_c1}-\eqref{case_c3}. 

\subsubsection{Physical picture}
\label{sec:picture}

Let us understand the two examples in more physical terms. Although the two examples have seemingly different symmetry groups, they are in fact closely related. Both symmetry groups can be regarded as central extensions of $\mathbb{Z}_2\times\mathbb{Z}_4$: namely, we can take the physical symmetries to be $G=\mathbb{Z}_2\times\mathbb{Z}_4$ in both cases, and for $\mathbb{Z}_8^f\times\mathbb{Z}_2$ the fermions carry half charges under the $\mathbb{Z}_4$ subgroup.

Consider a $\frac{2\pi}{4}e_2$ base loop. By dimensional reduction we obtain a \spd{2} fermionic SPT protected by the $\mathbb{Z}_2\times\mathbb{Z}_4$ symmetry. In fact, because of $\Theta_{1,2}=\frac{\pi}{2}$, the protecting symmetry is just the $\mathbb{Z}_2$ subgroup. As we have already discussed in Sec. \ref{sec:intro}, this is the ``root'' Abelian $\mathbb{Z}_2$ FSPT phase in 2D, which has a $\mathbb{Z}_4$ classification (thus can only exist on a $\mathbb{Z}_4$ base loop)~\cite{Gu_2013, GuWen, Cheng_fSPT}. Besides the non-interacting realization mentioned in Sec. \ref{sec:intro},  commuting-projector Hamiltonians for such 2D phases have also been found in \Ref{GuWen, WarePRB2016, TarantinoPRB2016}. $\Theta_{1,2}=\frac{\pi}{2}$ translates into the fractional exchange statistics of symmetry fluxes in the 2D FSPT phase, which as proven in \Ref{Gu_2013} implies its edge (i.e. the $\mathbb{Z}_4$ vortex loop) has to be degenerate.

\section{Topological State-Sum Models}
\label{sec_state-sum}

In this section we introduce a class of lattice models to realize the fermionic gauge theories found in the previous section. We define these lattice models with a path integral representation of the partition function in discretized Euclidean space-time.  More specifically, we define a partition function for any closed oriented manifolds with a triangulation. The partition function however is a topological invariant of the space-time manifold (i.e. independent of the choice of the triangulation). Hence, it is a type of lattice topological quantum field theory (TQFT). It is generally believed that such topological state-sum models can be cast into commuting-projector Hamiltonians~\cite{WilliamsonAP}.

We will first recall a few useful facts regarding triangulations of $n$-dimensional manifolds.  We will work with simplicial triangulations for simplicity~\footnote{In fact, only a $\Delta$-complex triangulation is needed} and denote the set of $k$-simplices ($0\leq k \leq n$) in the triangulation as $\Delta_k$. For a given triangulation, we pick an arbitrary ordering of the vertices as $0,1,2,\dots$. The restriction of the ordering on each $k$-simplex $\sigma_k$ induces a relative ordering of the vertices of $\sigma_k$. Under this relative ordering, we write $\sigma_k$ as $[i_0 i_1 \cdots i_k]$, where $i_0<i_1 <\dots<i_k$ are the vertices of $\sigma_k$. 

On an oriented manifold, all simplices can be equipped with orientations, induced from the orientation of the manifold $M$.  For each $\sigma_n$, we define $\varepsilon(\sigma)$ to be $1$ if the orientation on $\sigma$ induced from that of $M$ coincides with the one coming from the ordering of its vertices;  otherwise if they are opposite then $\varepsilon(\sigma)=-1$.


\subsection{Twisted Crane-Yetter TQFT}
We now present models for fermionic gauge theories. The construction was first introduced by Kapustin and Thorngren recently in the context of higher-form gauge theories~\cite{KapustinHigherGauge}. We will call these models the \emph{twisted Crane-Yetter models}. The general input of the twisted Crane-Yetter TQFT involves $(i)$ a braided fusion category (BFC), $(ii)$ a finite group $G$ and $(iii)$ certain cohomological data $(\beta,\lambda,\omega)$ associated with $G$ and the BFC. 

For simplicity, we present the construction for an Abelian BFC $\mathcal{A}$. The anyon labels in $\mathcal{A}$ are denoted by $a, b, c, \dots$. The identity (i.e., the trivial anyon) is denoted by $0$. The BFC $\mathcal{A}$ can be viewed as an Abelian group, with the group multiplication given by the fusion rules.  As a BFC, $\mathcal{A}$ is equipped with further topological data, in particular the $F$ and $R$ symbols.
We will further assume that the $F$ symbols of $\mathcal{A}$ can be chosen to be trivial. In this case, the hexagon equations simplify to 
\begin{equation}
	R^{a,c}R^{b,c}=R^{a+b,c},R^{c,a}R^{c,b}=R^{c,a+b}.
	\label{}
\end{equation}
Notice that because $\mathcal{A}$ is Abelian, we denote the multiplication additively. In other words, $R^{a,b}$ defines a bi-character on the Abelian group $\mathcal{A}$. We define $\mathcal{T}$ as the following subset of $\mathcal{A}$:
\begin{equation}
	\mathcal{T}=\{x\in\mathcal{A}\,|\, R^{a,x}R^{x,a}=1, \forall a\in \mathcal{A}\}.
	\label{}
\end{equation}
We refer to $\mathcal{T}$ as the subset of transparent particles. In many cases, we will actually take a BCF such that $\mathcal{A}=\mathcal{T}$.

The other pieces of the input data are a finite group $G$ and two group cocycles: a $3$-cocycle $[\beta]\in \mathcal{H}^3[G, \mathcal{T}]$, and a $2$-cocycle $[\lambda] \in \mathcal{H}^2[G, \mathcal{A}]$ ($[\cdot]$ denotes the cohomology class). A 3-cocycle $\beta$ is a $3$-cochain (i.e., a function) $G^3\rightarrow \mathcal{T}$ that satisfies the 3-cocycle condition:
\begin{align}
\beta(\mathbf{h},\mathbf{k},\mathbf{l}) & - \beta(\mathbf{g}\mathbf{h},\mathbf{k},\mathbf{l}) + \beta(\mathbf{g},\mathbf{h}\mathbf{k},\mathbf{l}) \nonumber\\ 
&- \beta(\mathbf{g},\mathbf{h},\mathbf{k}\mathbf{l})+ \beta(\mathbf{g},\mathbf{h},\mathbf{k})= 0
\end{align}
Similarly, a 2-cocycle $\lambda$ is a function $G^2\rightarrow \mathcal{A}$ that satisfies the 2-cocycle condition:
\begin{equation}
\lambda(\mathbf{h},\mathbf{k}) - \lambda(\mathbf{g}\mathbf{h},\mathbf{k}) + \lambda(\mathbf{g},\mathbf{h}\mathbf{k})- \lambda(\mathbf{g},\mathbf{h})=0
\end{equation}
The final piece of data $\omega$ is a group 4-cochain $\omega:G^4\rightarrow \U(1)$. It is generally speaking not a 4-cocycle, however it does satisfy a similar condition which will be determined later. We will frequently use the following short-hand notation for a group $n$-cochain $\nu$: 
\begin{equation}
	\nu_{i_0i_1\cdots i_n}\equiv \nu(\mb{g}_{i_0i_1}, \mb{g}_{i_1i_2}, \cdots, \mb{g}_{i_{n-1}i_n}).
	\label{}
\end{equation}

With this understanding of the input data, we now describe the model. We start with an ordered triangulation of an oriented $4$-manifold. Each $1$-simplex $[ij]$ ($i<j$) is assigned a group element $\mb{g}_{ij}\in G$, which can be thought as $G$-connections of the gauge field. As a topological gauge theory, we require that there is no $G$ flux in every face, i.e. the connection is flat. So for each $2$-simplex $[ijk]$ ($i<j<k$) the flatness condition is imposed:
\begin{equation}
	\mb{g}_{ij}\cdot\mb{g}_{jk}=\mb{g}_{ik}.
	\label{gflat}
\end{equation}
To each $2$-simplex $[ijk]$ ($i<j<k$), we assign a simple object $f_{ijk}$ from $\mathcal{A}$. For each $3$-simplex $[ijkl]$, we demand that the following ``flatness condition'' holds:
\begin{equation}
	f_{jkl}-f_{ikl} +f_{ijl}-f_{ijk} =\beta_{ijkl}.
	\label{fflat}
\end{equation}

Let us now write down the partition function. To each $4$-simplex, say $\sigma_4 = (01234)$, we assign a phase factor:
\begin{equation}
	\begin{split}
		\ftj^+({\sigma_4})=&R^{f_{012},f_{234}}\big(R^{f_{034},\beta_{0123}}R^{f_{014},\beta_{1234}}R^{\lambda_{012}, f_{234}}\big)^{-1}\\
	&\cdot\omega(\mb{g}_{01},\mb{g}_{12},\mb{g}_{23},\mb{g}_{34}).
	\end{split}
		\label{eqn:4-simplex}
\end{equation}
The phase factor $\ftj^+(\sigma_4)$ is assigned to $\sigma_4$ assuming its local orientation coincides with the global orientation of $M$. If they have the opposite orientations, we instead assign $\ftj^-({\sigma_4})=\big[\ftj^+({\sigma_4})\big]^*$ to  $\sigma_4$. The partition function is then defined as
\begin{equation}
	\mathcal{Z}(M)=\frac{1}{|G|^{|\Delta_0|} |\mathcal{A}|^{|\Delta_1|-|\Delta_0|}}\sum_{\{\mb{g}_{ij},f_{ijk}\}}\prod_{\sigma_4\in\Delta_4}\ftj^{\epsilon(\sigma_4)}({\sigma_4}).
	\label{}
\end{equation}

We require that $\mathcal{Z}(M)$ defines a topological quantum field theory. Namely, $\mathcal{Z}(M)$ should yield a topological invariant of the manifold $M$, which means that 1) it must be independent of the specific choice of triangulation and 2) independent of the ordering of the vertices. It is known that all triangulations can be related to each other via a finite series of elementary moves, known as Pachner moves~\cite{Pachner}. 
To this end, we define the following ``obstruction class''
\begin{equation}
	\begin{split}
		\mathscr{O}(\mb{g}_1,\mb{g}_2,\mb{g}_3,\mb{g}_4,\mb{g}_5)=&	
	R^{\beta(\mb{g}_1,\mb{g}_2,\mb{g}_3\mb{g}_4\mb{g}_5),\beta(\mb{g}_3,\mb{g}_4,\mb{g}_5)}\\
	&R^{\beta(\mb{g}_1,\mb{g}_2\mb{g}_3\mb{g}_4,\mb{g}_5),\beta(\mb{g}_2,\mb{g}_3,\mb{g}_4)}\\
	&R^{\beta(\mb{g}_1\mb{g}_2\mb{g}_3,\mb{g}_4,\mb{g}_5),\beta(\mb{g}_1,\mb{g}_2,\mb{g}_3)}\\
	&R^{\lambda(\mb{g}_1, \mb{g}_2), \beta(\mb{g}_3, \mb{g}_4, \mb{g}_5)}.
	\end{split}
		\label{obstruction-class}
\end{equation}
One can show that $\mathscr{O}$ is actually a $5$-cocycle in $\mathcal{H}^5[G, \U(1)]$. We show in Appendix \ref{appendix_pachner} that invariance under Pachner moves requires that
		\begin{equation}
			\mathscr{O}(\mb{g}_1,\mb{g}_2,\mb{g}_3,\mb{g}_4,\mb{g}_5)=\frac{\omega_{12345}\omega_{01345}\omega_{01235}}{\omega_{01234}\omega_{01245}\omega_{02345}}.
			\label{eqn:obstruction-vanishing}
		\end{equation}
 We observe that the right-hand side of Eq. \eqref{eqn:obstruction-vanishing} is the coboundary of the 4-cochain $\omega$. Hence, it implies that the obstruction class must be cohomologically trivial in order for the twisted Crane-Yetter model to be well defined. Otherwise, we say the model is ``obstructed''. For obstruction-free models, Eq.~\eqref{eqn:obstruction-vanishing} becomes a ``twisted'' 4-cocycle condition on $\omega$ (compared to the regular 4-cocycle condition in which the left-hand side is 1).

The twisted Crane-Yetter models reduce to known models in two special limits:
\begin{enumerate}
	\item $G$ is trivial. In this case, the state sum reduces to the well-known Crane-Yetter theories~\cite{CY1, CY2} (the Hamiltonian version of the TQFT is known as the Walker-Wang model~\cite{WW} in the condensed matter literature).  Excitations in the model can be understood as a $\mathcal{T}$ gauge theory, however with an interesting twist: particle excitations are labeled by elements of $\mathcal{T}$. A particle $a$ then has topological spin $\theta_a=R^{a,a}=\pm 1$. In fact, the characterization of particle excitations hold generally, not just for the Abelian BFCs discussed here. Therefore, in general Crane-Yetter models also produce topological gauge theories. Recently it is shown that with non-Abelian BFC as the input the Crane-Yetter model can also realize twisted gauge theory.~\cite{WangChen}
	\item $\mathcal{A}$ is trivial. In this case, the theory reduces to the Dijkgraaf-Witten topological gauge theory~\cite{Dijkgraaf90}.
\end{enumerate}

\subsection{Gauge-theoretical interpretation}
The state-sum model can be understood as a topological gauge theory, for a $2$-form gauge field $f$ and a $1$-form gauge field $\mb{g}$. In other words, the theory embodies two kinds of gauge symmetries: the 1-form gauge transformations on $f$
\begin{equation}
	f_{ijk}\rightarrow f_{ijk}+\xi_{jk}-\xi_{ik}+\xi_{ij}, \xi\in \mathcal{A},
	\label{}
\end{equation}
and $0$-form gauge transformations on $\mb{g}$:
\begin{equation}
	\mb{g}_{ij}\rightarrow \mb{h}_i^{-1}\mb{g}_{ij}\mb{h}_j, \mb{h}\in G.
	\label{}
\end{equation}

We explicitly prove that the partition function is invariant under the two types of gauge transformations, with two simplifying assumptions: (a) $G$ is Abelian and (b) The entire $\mathcal{A}$ is transparent (i.e. $\mathcal{T}=\mathcal{A}$). While the proof is rather technical and the details can be found in Appendix \ref{sec:gauge-invariance}, we introduce the reformulation of the TQFT as a topological gauge theory following \Ref{KapustinHigherGauge}, using the notations of simplicial calculus (see Appendix \ref{sec:review} for a review of relevant mathematical concepts). In the following the multiplication in $G$ will also be denoted additively. We define the discretized ``action'' $\mathcal{S}$ by $\ftj(\sigma_4)=e^{i\mathcal{S}(\sigma_4)}$. $f$ can be viewed as a $2$-cochain valued in $\mathcal{A}$, and $\mb{g}$ is a $1$-cochain valued in $G$. The flatness conditions Eqs.~\eqref{gflat} and \eqref{fflat} then can be written as $\delta \mathbf{g}=0$ and $\delta f=\beta$. The latter implies that $f$ is not closed.

In the partition function, the product of three $R$ symbols closely resembles the ``Pontryagin square'' in \Ref{KapustinHigherGauge}.  Roughly speaking, if $f$ is a closed $2$-cochain, the Pontryagin square is just the cup product $f\cup f$. However, if $f$ is not quite closed, the cup product is no longer closed and we need to amend it by an additional term: $f\cup f - f\cup_1 \delta f$ to get a closed cochain.  In this notation, the action can be written as
\begin{equation}
	\mathcal{S}=2\pi  [f\cup f - f\cup_1 \beta + \lambda\cup f +\eta].
	\label{}
\end{equation}
Here $\eta=\frac{\ln \omega}{2\pi i}$ is the linearized form of $\omega$.

With this notation, we can now give a quick derivation of the obstruction-vanishing condition. In order for the action $\mathcal{S}$ to be a topological invariant, all we need to show is that $\mathcal{S}$ is a closed $4$-cochain:
\begin{equation}
	\begin{split}
	\delta(f\cup f&- f\cup_1 \beta + \lambda \cup f)\\
	&=\delta f\cup f + f\cup \delta f - \delta(f\cup_1\delta f) - \lambda\cup \delta f\\
	&=-\delta f\cup_1\delta f - f\cup_1\delta^2f - \lambda\cup \beta\\
	&= -\beta\cup_1\beta - \lambda\cup \beta.
	\end{split}
	\label{}
\end{equation}
We thus require $\delta \eta =\beta\cup_1\beta + \lambda \cup \beta$, which is the obstruction-vanishing condition Eq.~\eqref{eqn:obstruction-vanishing}.

\subsection{Relation to symmetry-enriched topological phases}
\label{sec:dualitySET}
We now define a variant of the Crane-Yetter TQFT. Instead of having $G$ labels on the 1-simplices, we dualize them to 0-simplices, i.e. vertices. Namely, the actual labels are $G$ group elements on vertices, and $\mb{g}_{ij}=\mb{g}_i^{-1}\mb{g}_j$. The flatness condition for $f$'s is the same as before, as well as the partition function
$\tilde{\ftj}^+(\sigma_4)$:
\begin{equation}
	\tilde{\ftj}^+(\sigma_4)=\ftj^+(\sigma_4).
	\label{}
\end{equation}
Here $\ftj^+(\sigma_4)$ is the partition function defined in Eq. \eqref{eqn:4-simplex}, where the $G$ label on $[ij]$ is given by $\mb{g}_i^{-1}\mb{g}_j$. 

This ``dual'' state-sum model can be viewed as a model of symmetry-enriched topological phase. For each $\mb{h}\in G$, a global symmetry transformation is defined as 
\begin{equation}
	\mb{h}: \mb{g}_i\rightarrow \mb{hg}_i.
	\label{}
\end{equation}
Apparently the partition function is invariant under such global symmetry transformations since it only depends on $\mb{g}_{i}^{-1}\mb{g}_j$.

As we have argued, the $G$ fields can be understood as connections of a discrete $G$ gauge field. As a result, gauge-equivalent configurations of $G$ fields yield the same partition function. In the SET model, the connections $\{\mb{g}_{ij}\}$ are by definition ``pure gauge''.  Thus the partition function on any oriented closed manifold is identical to $\mb{g}_i=1$, i.e. the Crane-Yetter TQFT of $\mathcal{A}$. The original state-sum model can then be viewed as ``gauging'' the SET model. This relation was first considered between DW gauge theories and group-cohomology models of bosonic SPT phases~\cite{LevinGu2012}.

Therefore, we can understand that the bulk excitations of the SET model are described by the Crane-Yetter TQFT, i.e. a $\mathcal{T}$ gauge theory (with possibly fermionic gauge charges). An important question is then how excitations, including both particles and loops, transform under the symmetry group $G$. In the next section, using dimensional reduction, we will argue that the particles transform as projective representations of $G$ essentially determined by $\lambda$, while the loop excitations exhibit nontrivial symmetry actions corresponding to nontrivial three-loop braiding statistics after gauging.

\section{Braiding Statistics in the State-Sum Models}
\label{sec:braiding}

In this section, we analyze the braiding statistics between particle and loop excitations in the twisted Crane-Yetter state-sum models. For simplicity, we assume that the whole BFC $\mathcal{A}$ is transparent, i.e., $\mathcal{A}=\mathcal{T}$, throughout the section.

\subsection{Particle excitations}
\label{sec:particles}
First let us consider the properties of particle excitations. For this purpose, it is useful to view the state-sum model as a $G$-symmetry enriched gauge theory. As we have argued earlier, the particle excitations are charged under the emergent gauge group $\mathcal{A}$. We need to understand how they transform under $G$. We will present some evidence that the symmetry transformations on particles are determined by $\lambda$.

To begin, we consider a simpler theory defined by
\begin{equation}
	\ftj^+(\sigma_4)=R^{\lambda_{012}, f_{234}},
	\label{eqn:simpler}
\end{equation}
together with the twisted flatness condition $\delta f=\beta$.  Basically we drop the term in the action which contributes to the exchange statistics of the particle excitations (so now they are all bosons). \Ref{KapustinHigherGauge} showed that this theory can be analyzed in a dual representation, where the $2$-form is dualized to a $1$-form gauge field $a$. To simplify the discussion, we use the simplicial calculus. First the constraint $\delta f-\beta=0$ can be implemented by introducing a 1-cochain Lagrange multiplier $a$ and the following modification to the action:
\begin{equation}
	\begin{split}
	\mathcal{S}&=-\lambda\cup f + a\cup (\delta f - \beta)\\
	&=(\delta a-\lambda)\cup f - a\cup \beta.
	\end{split}
	\label{}
\end{equation} 
We should notice that this applies only to non-degenerate $R$, namely $\frac{1}{|\mathcal{A}|}\sum_{a\in\mathcal{A}}R^{a,b}=\delta(b)$.
Since $f$ is no longer constrained, we can sum over $f$ to get $\delta a-\lambda=0$, and the action becomes $\mathcal{S}=-a\cup \beta$. This is a symmetry-enriched $\mathcal{A}$ gauge theory, where the gauge charges, labeled by characters $\chi$ of $\mathcal{A}$, transform as projective representations of $G$. More specifically, the projective representation $U_\chi$ on the gauge charge $\chi$ is given by
\begin{equation}
U_\chi(\mathbf{g})U_{\chi}(\mathbf{h})=\eta_\chi(\mathbf{g},\mathbf{h})U_\chi(\mathbf{g}\mathbf{h}), \label{eqn:u}
\end{equation}
where the projective phase factor is given by 
\begin{equation}
	\eta_\chi(\mb{g,h})=\chi\big(\lambda(\mb{g,h})\big).
	\label{eqn:chi}
\end{equation}

In general, with the $f\cup f$ term in $\ftj^+(\sigma_4)$, we can not apply the above duality transformation. However for a special case $\mathcal{A}=\mathbb{Z}_2^f$ and with a trivial $\beta$, we can ``linearize'' the term $f\cup f$ using the following relation:
\begin{equation}
	f\cup f = w_2\cup f,
	\label{}
\end{equation}
where $w_2$ is the second Stiefel-Whitney class of the manifold. This relation only holds when $f$ is a 2-cocycle. Then a similar duality transformation leads to $\delta a= w_2 + \lambda$.  The $w_2$ term accounts for the fermionic statistics of particles~\cite{ThorngrenSPT}, and the symmetry transformation under $G$ is again given by Eqs. \eqref{eqn:u} and \eqref{eqn:chi}.

Extrapolating from these two special cases, we conjecture that Eqs. \eqref{eqn:u} and \eqref{eqn:chi} hold more generally in the full theory (when the $R$ symbol is nondegenerate). We do not have a proof of this statement at the moment, but will show that it is also consistent with the dimensional reduction result.


\subsection{Dimensional reduction}
To understand properties of the loop excitations, we consider the theory on a $4$-manifold $M_4=M_3\times S^1$, where $M_3$ is a $3$-manifold and $S^1$ is the circle.  Since these models have zero correlation length,  we can analyze the theory in the limit where there is only one cell in the $S^1$ direction, and view  it as a \spd{2} theory on $M_3$. We will fix the $G$ flux through the ``hole'' of $S^1$ to be $\mb{h}$. Some of the particle excitations in the \spd{2} theory correspond to the loop excitations that are linked to a $\mathbf{h}$ gauge flux in the \spd{3} theory. Accordingly, if we can extract the braiding statistics in the \spd{2} theory, three-loop braiding statistics in the \spd{3} theory can be inferred from there. 

Let us understand the fields in the \spd{2} theory after dimensional reduction. All the 1-form $G$ gauge fields in $M_3$ remain, so do the $2$-form $\mathcal{A}$ gauge fields in $M_3$. Both of them satisfy the same (twisted) flatness conditions as before. However, there are additional dynamical $1$-form fields, denoted by $m$, coming from the dimensional reduction of the full $2$-form gauge fields in $M_4$. They obey the following flatness conditions:
\begin{equation}
	m_{ik}=m_{ij}+m_{jk}+n(\mb{g}_{ij}, \mb{g}_{jk}).
\end{equation}
Here $n$ is the slant product of $\beta$: $n=i_\mb{h}\beta$. The explicit expression of $n$ in terms of $\beta$ reads:
\begin{equation}
	n(\mb{k,l})=\beta(\mb{k,l,h})-\beta(\mb{k,h,l})+\beta(\mb{h,k,l}).
	\label{}
\end{equation}

The partition function with a fixed holonomy $\mb{h}$ around $S^1$ will be denoted by $\mathcal{Z}_\mb{h}(M_3\times S^1)$. After a straightforward but lengthy calculation, we find
\begin{widetext}
	\begin{equation}
		\mathcal{Z}_\mb{h}(M_3\times S^1)=\frac{1}{|G|^{|\Delta_0|}|\mathcal{A}|^{|\Delta_1|}} \sum_{\{\mb{g}\}\in l(\Delta_0)}\bigg(\sum_{\{f\} \in l(\Delta_2)}\delta_{\delta f=\beta}\prod_{\sigma_3\in \Delta_3}\mathsf{S}_{2+1}^{\varepsilon(\sigma_3)}(\sigma_3)\bigg) \bigg(  \sum_{\{m\}\in l(\Delta_1)}\delta_{\delta m=-n}\prod_{\sigma_3\in\Delta_3} \ftj_{2+1}^{\varepsilon(\sigma_3)}(\sigma_3)\bigg)
	\label{eq:Zh}
\end{equation}
\end{widetext}
In this expression $\Delta_k\equiv \Delta_k(M_3)$, and the weight on a tetrahedron is given by 
\begin{equation}
	\begin{split}
		\mathsf{S}_{2+1}^+([0123])&= R^{\xi_{01}, f_{123}},\\
		\ftj_{2+1}^+([0123])
		&=R^{m_{23},n_{012}}R^{\lambda_{012}, m_{23}}\alpha(\mb{g}_{01},\mb{g}_{12},\mb{g}_{23}).
	\end{split}
		\label{st}
\end{equation}
Here the 1-cocycle $\xi$ is the slant product of $\lambda$:
\begin{equation}
	\xi(\mb{k})=\lambda(\mb{k,h})-\lambda(\mb{h,k}).
	\label{xi}
\end{equation}
and $\alpha$ is a 3-cochain that depends on  $\beta$ and $\omega$. The obstruction vanishing condition Eq.\eqref{eqn:obstruction-vanishing} reduces to the following equation 
\begin{equation}
R^{n_{012},n_{234}}R^{\lambda_{012},n_{234}}=\frac{\alpha_{1234}\alpha_{0134}\alpha_{0123}}{\alpha_{0234}\alpha_{0124}} \label{2d-obstruction}
\end{equation}
We refer the readers to Appendix \ref{sec:dim-reduc} for the derivation of the dimensional reduction.

\subsection{Braiding statistics in the \spd{2} theory}

After dimensional reduction, we now analyze the \spd{2} theory.  To simply the analysis, we assume that $G$ is Abelian, and $\xi$ in Eq.~\eqref{xi} is cohomologically trivial. With the latter assumption, we see that in the 2D partition function $\mathcal{Z}_\mb{h}$ the sum over $f$ becomes completely independent of $\mb{h}$, and the $\mb{h}$ dependence only enters through the sum over $m$. Therefore for the purpose of extracting loop braiding statistics on the $\mb{h}$ base loop, we only need to focus on the sum over $m$.  More discussions on the physical meaning of the sum over $f$ can be found in Appendix \ref{sec:dim-reduc}. It turns out that to realize those three-loop braiding statistics studied in Sec.~\ref{sec:constraint}, it is sufficient to assume $\xi$ is cohomologically trivial.

To infer the loop braiding statistics in the original \spd{3} theory, our analysis of the \spd{2} theory should proceed in two steps.  First, we need to establish the correspondence between the excitations in the \spd{2} and \spd{3} theories.  More precisely, we need to identify which of the \spd{2} anyons correspond to the \spd{3} particle excitations and which correspond to the \spd{3} loop excitations. Second, we need to extract the braiding statistics of the \spd{2} anyons. 

All properties of the \spd{2} theory $\mathcal{Z}_\mathbf{h}$ should depend on $\mathbf{h}$. For notational brevity, we suppress this $\mathbf{h}$ dependence below. It is easy to recover this dependence later.

\subsubsection{Correspondence between \spd{2} and \spd{3} excitations}
\label{sec:correspondence}

As discussed above, the \spd{2} theory has two kinds of dynamical variables, $m_{ij}\in \mathcal{A}$ and $\mb{g}_{ij} \in G$ living on each link $[ij]$. On each 2-simplex $[ijk]$, they satisfy the twisted and untwisted flatness conditions respectively:
\begin{align}
m_{ij}+m_{jk}+n(\mb{g}_{ij},\mb{g}_{jk}) & = m_{ik}, \nonumber\\
\mb{g}_{ij}\cdot\mb{g}_{jk} & = \mb{g}_{ik}. \label{2d-flatness}
\end{align}
Instead of viewing $m_{ij}$ and $\mb{g}_{ij}$ as two independent degrees of freedom, we can combine them into one and denote it as $(m_{\mb{g}})_{ij}$. In fact, $\{m_{\mb{g}}\}|_{m\in\mathcal A, \mb{g}\in G}$ form a group $\tilde G$ under the following definition of group multiplication:
\begin{equation}
	m_\mb{g}\times m'_\mb{k}=[m+m'+n(\mb{g,k})]_{\mb{gk}}.
	\label{}
\end{equation}
The group $\tilde{G}$ is known as a central extension of $G$ by $\mathcal A$, associated with the $2$-cocycle $n(\mb{g}, \mb{k})\in \mathcal{H}^2[G, \mathcal A]$.

With this notation, we claim that the partition function $\ftj_{\text{2+1}}$ actually represents a \spd{2} Dijkgraaf-Witten gauge theory of group $\tilde G$ associated with the following 3-cocycle
\begin{equation}
	\omega_{2+1}(a_\mb{g},b_\mb{k},c_\mb{l})=R^{c, n(\mb{g,k})}R^{\lambda(\mb{g,k}),c}\alpha(\mb{g,k,l}).
	\label{2d-3cocycle}
\end{equation}
To see this, one can check that (i) $\omega_{2+1}$ is indeed a 3-cocycle in $\mathcal H^3[\tilde G, \U(1)]$, as long as $n,\lambda, \alpha$ satisfy Eq.~\eqref{2d-obstruction} and (ii) the conditions (\ref{2d-flatness}) leads to
\begin{equation}
(m_{\mb{g}})_{ij} \times (m_{\mb{g}})_{jk} = (m_{\mb{g}})_{ik}
\end{equation}
i.e., every 2-simplex has a flat $\tilde G$ connection.

The above mapping is most convenient for general computations of braiding statistics of excitations in the \spd{2} theory, since braiding statistics in Dijkgraaf-Witten theory is known (see e.g. Ref. \onlinecite{twistedQD}). Below we will take a slightly different approach. We will dualize the $G$ gauge fields in the $\tilde{G}$ Dijkgraaf-Witten theory, similarly as discussed in Sec. \ref{sec:dualitySET}, and view the result as an $\mathcal{A}$ gauge theory, enriched by the symmetry group $G$.

The Hamiltonian version of this symmetry-enriched gauge theory was recently considered in Ref. \onlinecite{ChengSET2016}. The anyons in an Abelian $\mathcal{A}$ gauge theory can be labeled as dyons $(a, \chi)$ where $a\in \mathcal{A}$ is the gauge flux, and $\chi:\mathcal{A}\rightarrow \mathrm{U}(1)$ denotes a character of $\mathcal{A}$, labeling a gauge charge. Since the symmetry group $G$ does not permute any anyons, each anyon carries a projective representation of $G$.
The twisted flatness condition Eq.~\eqref{2d-flatness} is interpreted as $\mathcal{A}$ gauge charges carrying projective representations of $G$.  As shown in Ref. \onlinecite{ChengSET2016}, the projective phases on a pure charge $({1}, \chi)$ is
\begin{equation}
	\eta_{(1,\chi)}(\mb{g,k})=\chi\big(n(\mb{g,k})\big).
	\label{eta1}
\end{equation}
For a gauge flux $(a, \mathds{1})$,
\begin{equation}
	\eta_{(a,\mathds{1})}(\mb{g,k})=R^{a, n(\mb{g,k})}R^{\lambda(\mb{g,k}),a}
	\label{eta2}
\end{equation}
More generally, the projective phase of the dyon $(a,\chi)$ is given by $\eta_{(a,\chi)}=\eta_{(a,\mathds{1})}\eta_{(1,\chi)}$.

We now identify the correspondence between the \spd{2} and \spd{3} excitations. For each $a$, we define  $\chi_a(x) = R^{x,a}$ with $x\in\mathcal{A}$. It is clear that $\chi_a$ is a character of $\mathcal{A}$. Then, the dyon $(a,\chi_a)$ transforms under $G$ with a projective phase
\begin{align}
	\eta_{(a,\chi_a)}(\mb{g,k})& =R^{a, n(\mb{g,k})}R^{\lambda(\mb{g,k}),a}R^{n(\mb{g,k}),a} \nonumber\\
	& = R^{\lambda(\mb{g,k}),a} \nonumber\\
	& = \chi_a\big(\lambda(\mb{g,k})\big)
	\label{}
\end{align}
where we used the fact that $\beta$, and therefore $n$, is transparent to obtain the second line. It is easy to see that the anyons $\{(a,\chi_a)\}_{a\in\mathcal{A}}$ form a fusion group $\mathcal{A}$, and we identify them as the descendants of the 3D quasiparticles. Indeed, the topological twist of $(a,\chi_a)$ is $\chi_a(a)=R^{a,a}=\theta_a$, and the mutual braiding between $(a,\chi_a)$ and $(b, \chi_b)$ is $\chi_a(b)\chi_b(a)=R^{a,b}R^{b,a}=1$, as expected. In fact, it is not difficult to see that the set $\{(a, \chi_a)\}$ forms the maximal subset of transparent anyons in the \spd{2} theory, i.e. any other particles not in this set have nontrivial braiding with at least one of the anyons in the set. As a result, all other particles should be understood as the descendants of gauge flux loops in the \spd{3} theory.

\subsubsection{Braiding statistics with $\mathcal{A}=\mathbb Z_{N_0}^f$}

With the above understanding, we now specialize to the case $\mathcal A = \mathbb Z_{N_0}$ and $G = \prod_{i=1}^K {\mathbb Z}_{N_i}$. For the rest of the section we assume the even number $N_0=2m$, and the R symbols are given by $R^{a,b}=(-1)^{ab}$, which corresponds to $\mathbb{Z}_{N_0}^f$.  We will also work with the followng parametrization of $\lambda$:
\begin{equation}
\lambda(\mb{g},\mb{k})=\sum_{i=1}^K \frac{q_i}{N_i}\left(\mb{g}_i+\mb{k}_i - [\mb{g}_i+\mb{k}_i] \right) \quad ({\rm mod}\ N_0 ),
	\label{eqn:lambdaexpr}
\end{equation}
where $q_i$ are integers, and $[\mb{g}_i+\mb{k}_i]$ equals $\mb{g}_i+\mb{k}_i$ modulo $N_i$. We have used integer vector $\mathbf{g}=(\mathbf{g}_1,\dots,\mathbf{g}_K)$ to deonote group elements of $G$. In fact, this parameterization exhausts all cohomology classes in $H^2[G,\mathbb{Z}_{N_0}]$ which satisfy the assumption that $\xi$ in Eq.~\eqref{xi} is cohomologically trivial.

We would like to extract a part of the braiding statistics data from given $n(\mb{g},\mb{k})$ and $\alpha(\mb{g}, \mb{k}, \mb{l})$, focusing on those that will lead to Abelian statistics. To start, let us give an explicit parametrization of the inputs $n(\mb{g}, \mb{k})$ and $\alpha(\mb{g},\mb{k}, \mb{l})$. Let us take the following class of 2-cocycles $n\in \mathcal{H}^2[G,\mathbb Z_{N_0}]$:
\begin{equation}
	\begin{gathered}
	n(\mb{g},\mb{k}) = \sum_{i=1}^K \frac{p_i}{N_i}\left(\mb{g}_i+\mb{k}_i - [\mb{g}_i+\mb{k}_i] \right) \quad ({\rm mod}\ N_0 ),\\
		\end{gathered}
	\label{eq:abelian-n}
\end{equation}
where $p_i$ are integer parameters. Here, we have used additive convention for group multiplication in both $G$ and $\mathcal A$. It is worth to mention that this class of $n$ satisfies $n(\mb{g}, \mb{k}) = n(\mb{k}, \mb{g})$. This property is a necessary and sufficient condition for the braiding statistics to be Abelian.  At the same time, we can choose
\begin{align}
	\alpha(\mb{g},\mb{k},\mb{l}) & = e^{i\pi\sum_{ij}\frac{p_ip_j}{N_iN_j} \mb{g}_i(\mb{k}_j+\mb{l}_j-[\mb{k}_j+\mb{l}_j])} \nonumber\\
	& \quad \times e^{i2\pi\sum_{ij}\frac{t_{ij}}{N_iN_j} \mb{g}_i(\mb{k}_j+\mb{l}_j-[\mb{k}_j+\mb{l}_j])}
\end{align}
where $t_{ij}$ are integers. One can check that the above $\lambda$, $n$, $\alpha$ indeed make $\omega_{2+1}$ a $3$-cocycle. The fact the existence of $\alpha$ for given $\lambda$ and $n$ (not just those parametrized by Eq. \eqref{eq:abelian-n}) shows that there is no obstruction in the \spd{2} theory. Notice that if we modify the $4$-cochain $\omega$ by a $4$-cocycle, $\alpha$ will be modified by a $3$-cocycle correspondingly.

One can in principle compute the braiding statistics using the understanding from previous subsection and using the general results for Dijkgraaf-Witten models~\cite{twistedQD, lin2014}, but in our case we will take a short cut.  Notice that in our parametrization of $n(\mb{g,k})$, different $\mathbb{Z}_{N_i}$ subgroups are decoupled, so let us focus on an individual $\mathbb{Z}_{N_i}$ for now. In the \spd{2} $\mathbb{Z}_{N_0}$ gauge theory, following the discussion of dual SET phases in the previous subsection we denote the unit gauge flux by $v$, and the unit gauge charge $e$ corresponding to the character $\lambda_e(x)=\exp(\frac{2\pi i}{N_0}x)$, with $x\in\mathcal{A}$. The (bulk) fermion in this notation is represented by $e^{N_0/2} v$.

With a $\mathbb{Z}_{N_i}$ global symmetry, anyons can carry fractional charges under $\mathbb{Z}_{N_i}$.  Denote the projective phase on an anyon $a$ by $\eta_a(\mathbf{g},\mathbf{k})$. We can extract the fractional charge $Q_a$ as follows:
\begin{equation}
	e^{2\pi i Q_a}=\prod_{j=1}^{N^{0i}}\eta_a([j], [1]),
	\label{}
\end{equation}
where $[j]=j \ (\text{mod } N_i)$ is a group element of $\mathbb{Z}_{N_i}$. Using this and Eqs.~\eqref{eta1} and \eqref{eta2}, we find
\begin{equation}
	 e^{i2\pi Q_e}= e^{\frac{2\pi ip_i}{N_{0i}}}, \quad  e^{i2\pi Q_v}=e^{\frac{2\pi i N_0(p_i+q_i)}{2N_{0i}}}.
	\label{}
\end{equation}
We can say that $e$ anyon carries a $\frac{p_i}{N_{0i}}$ fractional charge of the $\mathbb{Z}_{N_i}$ symmetry, while $v$ anyon carries $\frac{N_0(p_i+q_i)}{2N_{0i}}$ fractional charge. The fermion $e^{N_0/2}v$ carries a fractional charge $\frac{N_0q_i}{2N_{0i}}$, which can only be $0$ or $\frac{1}{2}$. 

Such an SET can be easily described by an Abelian Chern-Simon theory with the following K matrix and charge vector $t_i$~\cite{WenZee1992, Wen_AdvPhys1995, Lu_arxiv2013}:
\begin{equation}
	K=
	\begin{pmatrix}
		0 & N_0\\
		N_0 & 0
	\end{pmatrix},
	 \ \ \ t_i=
	\begin{pmatrix}
		\frac{N_0^2(p_i+q_i)}{2N_{0i}}\\[5pt]
	 \frac{N_0p_i}{N_{0i}}
	\end{pmatrix}.
	\label{}
\end{equation}
In this formalism, anyons are labelled by integer vectors. The exchange statistics $\theta_l$ of an anyon $l$ and mutual braiding statistics between anyons $l$ and $l'$ are given by
\begin{equation}
\theta_l = \pi l^T K^{-1}l, \quad \theta_{ll'}=2\pi l^T K^{-1} l' \label{Kmatrix_braiding}.
\end{equation} 
In addition, the $\mathbb Z_{N_i}$ charge carried by $l$ is given by
\begin{equation}
Q_l = l^T K^{-1} t_i.
\end{equation}
We denote the gauge charge by $e=(1,0)$ and the gauge flux $v=(0,1)$. One can easily check that such $K$ and $t_i$ indeed descirbe the above $\mathbb Z_{N_0}$ gauge theory enriched by $\mathbb{Z}_{N_i}$ symmetry.

So far we have focused on the symmetry fractionalization in the \spd{2} theory, which is completely determined by $\beta$ and $\lambda$. We have not accounted for the possibility of adding \spd{2} BSPT layers (i.e. different 3-cocycle $\alpha$). While the details of braiding statistics surely depend on the BSPT layer, this subtlety does not affect the indicator $\Theta_{\mb{f}i,k}$ for intrinsic FSPT phases, discussed in Sec.~\ref{sec:parityloop}, which is of central interest to us. So, we ignore the BSPT ambiguity (i.e., $\alpha$ dependence) for the moment.

We now gauge the symmetry $G$ and switch back to the twisted Crane-Yetter model. After gauging, a vortex carrying the unit $\mathbb{Z}_{N_i}$ flux can be represented by the fractional vectors $t_i/N_i$. The braiding statistics between these vortices can be calculated again using Eq.~\eqref{Kmatrix_braiding}. Accordingly, we obtain the exchange and mutual braiding phases:
\begin{equation}
	\begin{gathered}
	\theta_i=\frac{\pi}{N_i^2}t_i^\mathsf{T}K^{-1} t_i=\frac{\pi N_0^2 p_i(p_i+q_i)}{N_i^2N_{0i}^2},\\
	\theta_{ij}=\frac{2\pi}{N_iN_j}t_i^\mathsf{T}K^{-1}t_j=\frac{\pi N_0^2 (2p_ip_j+p_iq_j+q_ip_j)}{N_iN_jN_{0i}N_{0j}}.
	\end{gathered}
	\label{2dtheta1}
\end{equation}
The mutual braiding between the $\mathbb{Z}_{N_i}$ unit flux and the $e$ anyon, as well as the $v$ anyon, is given by 
\begin{align}
	\theta_{ei} & = \frac{2\pi p_i}{N_iN_{0i}},
	\theta_{vi}  = \frac{\pi N_0(p_i+q_i)}{N_iN_{0i}}. \label{2dtheta2}
\end{align}
These braiding statistics are only part of the full set of braiding statistics. Note that there exist other $\mathbb{Z}_{N_i}$ vortices differing by charge attachments. However, knowing the braiding statistics between the vortices $\{t_i/N_i\}_{i=1,\dots,K}$, as well as the anyons $e$ and $v$, is enough to extract the topological invariants for gauged intrinsic FSPT phases defined in Sec.~\eqref{sec:constraint}. 


\subsection{Braiding statistics in the \spd{3} theory}

Now we come back to the \spd{3} theory. Again we consider $G = \prod_{i=1}^K {\mathbb Z}_{N_i}$ and $\mathcal A = \mathbb Z_{N_0}^f$, with $R^{a,b}=(-1)^{ab}$. The 2-cocycle $\lambda$ is given in Eq.~\eqref{eqn:lambdaexpr}. We will ignore the $\omega$ dependence of the braiding statistics, which is not relevant for intrinsic FSPT indicator $\Theta_{\mb{f}i,k}$. 

We use the following explicit expression for representative cocyclces $\beta \in \mathcal{H}^3[G,\mathbb{Z}_{N_0}]$:
\begin{equation}
	\begin{split}
		\beta(\mb{g,h,k})=\sum_{ij}&\frac{N_0 p_{ij}}{N_{i0}N_{j}}\mb{g}_i(\mb{h}_j+\mb{k}_j-[\mb{h}_j+\mb{k}_j]) \\
		&+\sum_{ijk}\frac{{N_0}p_{ijk}}{N_{ijk0}}\mb{g}_i\mb{h}_j\mb{k}_k,
	\end{split}
	\label{eqn:betaexpr}
\end{equation}
where $p_{ij}$ and $p_{ijk}$ are integer parameters, and $[\mathbf{h}_j+\mathbf{k}_j]$ equals $\mathbf{h}_j+\mathbf{k}_j$ modulo $N_j$. It turns out that the first term in the explicit expression describes Abelian braiding statistics for loops, while the second term leads to non-Abelian loop statistics. For this reason we will refer to the two types of cocycles as ``Abelian'' and ``non-Abelian''. In this subsection, we only consider the Abelian part of the cocycle, leaving a discussion on non-Abelian loop braiding statistics in Sec.~\ref{sec:discussion}. One may calculate the cohomology group $\mathcal{H}^3[G, \mathbb{Z}_{N_0}]$ using the K\"unneth decomposition:
\begin{equation}
	\mathcal{H}^3[G, \mathbb{Z}_{N_0}]=\prod_{i}\mathbb{Z}_{N_{i0}}\prod_{i<j}\mathbb{Z}_{N_{ij0}}^2 \prod_{i<j<k}\mathbb{Z}_{N_{ijk0}}.
	\label{}
\end{equation}
We believe that the parametrization in Eq.~\eqref{eqn:betaexpr} exhausts all cohomology classes in $\mathcal{H}^3[G, \mathbb{Z}_{N_0}]$. 


We now focus on the Abelian part of $\beta$. As discussed previously, dimension reduction of the 3D state-sum model leads to a 2D model described by the slant product $n=i_\mb{h}\beta$. For $\beta$ in Eq.~\eqref{eqn:betaexpr}, the slant product is given by
\begin{equation}
	n(\mb{k},\mb{l})=\sum_{j}\frac{P_j}{N_j}(\mb{k}_j+\mb{l}_j-[\mb{k}_j+\mb{l}_j]),
	\label{}
\end{equation}
where
\begin{equation}
	P_j=\sum_l\frac{N_0 p_{lj}\mb{h}_l}{N_{0l}}.
	\label{}
\end{equation}
This form of $n$ is the same as in Eq.~\eqref{eq:abelian-n}, with $p_i$ there replaced by $P_i$. Taking $\mb{h}=e_k$ and substituting the expression of $P_i$ into  Eqs.~\eqref{2dtheta1} and \eqref{2dtheta2}, we obtain the three-loop braiding statistics:
\begin{equation}
	\begin{gathered}
	\theta_{i,k}=\frac{\pi N_0^3 p_{ki}(N_0p_{ki}+N_{0k}q_i)}{N_i^2N_{0i}^2N_{0k}^2},\\
	\theta_{ij,k}=\frac{\pi N_0^3 (2N_0p_{ki}p_{kj}+N_{0k}p_{ki}q_j+N_{0k}p_{kj}q_i)}{N_iN_jN_{0i}N_{0j}N_{0k}^2},
	\end{gathered}
	\label{3dtheta1}
\end{equation}
and 
\begin{equation}
\begin{gathered}
\theta_{ei,k} = \frac{2\pi N_0 p_{ki}}{N_iN_{0i}N_{0k}}, \\
\theta_{vi,k} = \frac{\pi N_0(N_0p_{ki}+N_{0k}q_i)}{N_iN_{0i}N_{0k}}.
 \label{3dtheta2}
\end{gathered}
\end{equation}
Previously, we have identified $e^{N_0/2}v$ as the fermion particle in the 3D bulk. Since $e^{N_0/2}$ and $v$ both have a $\pi$ mutual statistics with respect to $e^{N_0/2}v$, we now should understand both of them as fermion parity loops. Accordingly, we understand $v^2$ as a bosonic particle in the 3D bulk, and $e$ is a loop excitation that has a $\frac{2\pi}{N_0}$ mutual statistics with respect to the fermion particle $e^{N_0/2}v$. 

Finally, we make the following comment. Throughout our computation, we do not keep track of the dependence of three-loop braiding statistics on the 4-cochain $\omega$. A consequence is that the three-loop braiding statistics given in Eqs.~\eqref{3dtheta1} and \eqref{3dtheta2} are not the complete result, in particular violating the constraints in Eqs.~\eqref{eqn:const5} and \eqref{eqn:const6}.  In order for the constraints to be obeyed, we have to keep track of the $\omega$ dependence carefully, which however is very complicated. Nevertheless, the indicator $\Theta_{\mathbb{f}i,k}$ of intrinsic FSPT phases does not change after attaching BSPT layers, therefore we can safely ignore the issue for the purpose of extracting the indicators.

\subsection{Realizations of intrinsic FSPT phases}
We now show that the state-sum model realizes all intrinsic FSPT phases that we found in Sec. \ref{sec:constraint}, completing the argument that the physical constraints Eqs.~\eqref{eqn:const1}-\eqref{eqn:const9} are complete. In Sec.~\ref{sec:constraint}, we find two kinds of intrinsic FSPT phases, supported by the representative groups $\mathbb{Z}_{2m}^f\times\mathbb{Z}_{N_1}$ and $\mathbb{Z}_{2m}^f\times\mathbb{Z}_{N_1}\times\mathbb{Z}_{N_2}$ respectively. Without loss of generality, we assume $m=2^{r_m}$, $N_1=2^{r_1}$ and $N_2=2^{r_2}$. According to Sec.~\ref{sec_solving}, existence of the two kinds of intrinsic FSPT phases requires $r_m \ge r_1+1\ge 2$ and $r_2 \ge r_1+1\ge 2$ respectively. Since the second kind of intrinsic FSPT does not put requirements on $m$, we assume $m=1$ for simplicity. One can easily extend the following discussion to general $m$ for the second kind of intrinsic FSPT phases.

Let us now take a unified view on the groups $\mathbb{Z}_{2m}^f\times\mathbb{Z}_{N_1}$ and $\mathbb{Z}_{2}^f\times\mathbb{Z}_{N_1}\times\mathbb{Z}_{N_2}$: both of them arise as central extensions of $\mathbb{Z}_{N_1}\times\mathbb{Z}_{N_2}$ by $\mathbb{Z}_2^f$. More specifically, the former is a nontrivial central extension of $\mathbb{Z}_{N_1}\times\mathbb{Z}_{N_2}$ by $\mathbb{Z}_2^f$, associated with $N_2=m$ and $q_2=1,q_1=0$ in the $2$-cocycle $\lambda$, while the latter is the trivial extension $\mathbb{Z}_{N_1}\times\mathbb{Z}_{N_2}$ by $\mathbb{Z}_2^f$ associated with $q_2=q_1=0$ in $\lambda$. Therefore, in the twisted Crane-Yetter state-sum model, we set $\mathcal{A}=\mathbb{Z}_2^f$ and choose $\lambda$ accordingly.\footnote{We note that the intrinsic FSPT phase for $\mathbb{Z}_{2m}^f\times\mathbb{Z}_{N_1}$ can not be realized directly with $\mathcal{A}=\mathbb{Z}_{2m}^f$ and $G=\mathbb{Z}_{N_1}$. The reason is that after dimensional reduction to a unit flux of $\mathbb{Z}_{N_1}$, the state-sum model reduces to a untwisted $\mathbb{Z}_{2m}$ gauge theory (in the SET version). However, for the intrinsic FSPT phase, such a dimensional reduction yields a twisted $\mathbb{Z}_{2m}$ gauge theory. } We will see below that through this choice, the state-sum model indeed can be viewed as $\mathbb{Z}_{2m}^f\times\mathbb{Z}_{N_1}$ and $\mathbb{Z}_{2}^f\times\mathbb{Z}_{N_1}\times\mathbb{Z}_{N_2}$   gauge theories coupled to fermionic matter respectively. In this notation, the condition $r_m\ge r_1+1 \ge 2$ on existence of intrinsic FSPT phases for $\mathbb{Z}_{2m}^f\times\mathbb{Z}_{N_1}$ translates to $r_2\ge r_1+1 \ge 2$, the same as that for $\mathbb{Z}_{2}^f\times\mathbb{Z}_{N_1}\times\mathbb{Z}_{N_2}$.

Let us specify the input data to the state-sum model. As discussed above, for the $2$-cocycle $\lambda$ in Eq.~\eqref{eqn:lambdaexpr}, we set $q_1=0$, and $q_2=0,1$ for  $\mathbb{Z}_{2}^f\times\mathbb{Z}_{N_1}\times\mathbb{Z}_{N_2}$ and $\mathbb{Z}_{2m}^f\times\mathbb{Z}_{N_1}$ respectively. For both groups, we set the 3-cocycle $\beta$ in Eq.~\eqref{eqn:betaexpr} by the following parameters:
\begin{equation}
	p_{11}=p_{12}=p_{22}=0, \quad p_{21}=1.
	\label{eqn:therealbeta}
\end{equation}
The non-Abelian part of $\beta$ is set to 0.


Before we discuss the loop braiding statistics, we need to check that with these choices of $\lambda$ and $\beta$ the topological state-sum model is well-defined, i.e. the obstruction class \eqref{obstruction-class} vanishes. In Appendix \ref{sec:obstruction_invariants}, we provide a complete set of invariants to distinguish all cohomology classes in $\mathcal{H}^5[G, \U(1)]$ when $G$ is a finite Abelian group. Applying these invariants to the present case, we find that the obstruction class vanishes, when the following equations hold:
\begin{equation}
	\begin{split}
		0&=\pi\frac{N_0}{N_{0i}}p_{ii}(1+q_i),\\
		0&=\pi \frac{N_0}{N_{0j}}\frac{N^{ij}}{N_i}p_{ji}(1+q_i)  + \pi \frac{N_0}{N_{0i}}\frac{N^{ij}}{N_j}(q_ip_{ij}+q_jp_{ii}),
	\end{split}
	\label{obfreecondition}
\end{equation}
where the second equation should hold for $i\neq j$, and $N_0=2$. These equations are defined modulo $2\pi$.
With our choice of the parameters $q_i$, $p_i$ and $p_{ij}$ in $\lambda$ and $\beta$, we find that the first equation in Eq.~\eqref{obfreecondition} is automatically satisfied, while the second one puts the following conditions on $r_1$ and $r_2$:
\begin{equation}
	\begin{gathered}
	0=\pi 2^{1-\min(1,r_2)+\max(r_1,r_2)-r_1},
	\end{gathered}
	\label{}
\end{equation}
which does not depend on $q_2$. For $r_2\ge r_1 +1 \ge 2$, we see that the above equation indeed holds modulo $2\pi$. Hence, the twisted Crane-Yetter state-sum model is obstruction-free.

The loop braiding statistics for the two groups are given by Eqs.~\eqref{3dtheta1} and \eqref{3dtheta2}, under the current choice of $q_i,p_i$ and $p_{ij}$. Let us check that the braiding statistics imply that they are indeed $\mathbb{Z}_{2m}^f \times\mathbb{Z}_{N_1}$ and $\mathbb{Z}_2^f\times\mathbb{Z}_{N_1}\times \mathbb{Z}_{N_2}$ gauge theories. Since $N_0=2$, the fermionic particle is $ev$, and the fermion parity loops are $e$ and $v$. According to Eq.~\eqref{3dtheta2}, the mutual braiding between $ev$ and the $\mathbb Z_{N_2}$ unit flux on any base loop is given by $q_2\pi/N_2$. The mutual braiding statistics between $ev$ and the $\mathbb{Z}_{N_1}$ unit flux is always $0$. Hence, it is indeed a $\mathbb{Z}_{2m}^f \times\mathbb{Z}_{N_1}$ gauge theory for $q_2=1$, and a $\mathbb{Z}_2^f\times\mathbb{Z}_{N_1}\times \mathbb{Z}_{N_2}$ for $q_2=0$. 

With this understanding,  we now calculate the indicator $\Theta_{\textbf{f}i,k}$ for intrinsic FSPT phases:  
\begin{equation}
	\begin{gathered}
	\Theta_{\mb{f}1,2}=N_1\theta_{e1,2}=\frac{\pi N_0p_{21}}{N_{10}N_{20}}=\pi,\\
	\Theta_{\mb{f}2,1}=wN_2\theta_{e2,1}=w\frac{\pi N_0p_{12}}{N_{10}N_{20}}=0,\\
	w=
	\begin{cases}
		1 & q_2=0\\
		2 & q_2=1
	\end{cases}.
	\end{gathered}
	\label{}
\end{equation}
where we understand that $e$ is the fermion parity loop, and $N_0=2$, $N_1=2^{r_1}$, $N_{2}=2^{r_2}$ with $r_2\ge r_1+1\ge 2$.  One may use the $v$ fermion parity loop to do the computation, which leads to the same result. This agrees with the results in Sec.~\ref{sec_solving} (the index ``2'' should be understood as ``0'' for the group $\mathbb{Z}_{2m}^f\times\mathbb{Z}_{N_1}$). Therefore, all intrinsic FSPT phases identified in Sec. \ref{sec:constraint} are realized in the twisted Crane-Yetter model.

\section{Anomalous SETs in \spd{3}}
In Sec. \ref{sec:constraint} we derived a set of physical constraints for Abelian loop braiding statistics. We now demonstrate that these constraints can be used to show that certain 3D symmetry-enriched gauge theories are anomalous. 

Let us discuss a simple example: $G=\mathbb{Z}_{2}$ and  $\mathcal{A}=\mathbb{Z}_{N_0}^f$ with $N_0=2$ and $R^{a,b}=(-1)^{ab}$.  In a $\mathbb{Z}_2^f$ gauge theory enriched by $\mathbb{Z}_{2}$ symmetry, if the fermion parity flux loop carries gapless modes whose symmetry transformations are identical to those of a nontrivial \spd{2} $\mathbb{Z}_{2}$  BSPT phase\cite{LevinGu2012}, such a SET is anomalous. To see the anomaly, gauging the $\mathbb{Z}_{2}$ symmetry we would obtain $\Theta_{1,0}=\pi$ forbidden by the constraints. More specifically, $\Theta_{1,0}=\pi$ implies $\Theta_{01,1}=\pi$, contradicting Eq. \eqref{eqn:const5}.

However, such a SET can actually be realized consistently on the surface of a \spd{4} bosonic $\mathbb{Z}_{2}$ SPT phase, for instance, by a coupled-``layer'' construction as presented in Ref. \onlinecite{Cheng3D}. 
Based on heuristic field theory arguments, Ref. \onlinecite{Cheng3D} also proposed that the bulk \spd{4} SPT state is the one obtained from group-cohomology classification ($\mathcal{H}^5[\mathbb{Z}_2, \mathrm{U}(1)]=\mathbb{Z}_2$). We can provide a more rigorous justification with the topological state-sum models. Let us set $\lambda=0$ for the moment, and the topological action has a variation
\begin{equation}
	\delta S\sim\beta\cup_1\beta,
	\label{}
\end{equation}
where $\sim$ means up to a $5$-coboundary. In order for the model to be a well-defined topological gauge theory in \spd{3}, $\delta S$ has to vanish cohomologically. When the obstruction class does not vanish, we have to couple the model to a \spd{4} theory. The fields in the bulk are just $G$ spins on vertices, and the topological action is given precisely by $\beta\cup_1\beta$. Therefore the bulk is essentially a group-cohomology model of a bosonic SPT phase. 

Back to the example, let us take $G=\mathbb{Z}_2,\mathcal{A}=\mathbb{Z}_2^f$, and a nontrivial $3$-cocycle given by $\beta(\mb{g,g,g})=[1]$ (we represent $\mathcal{A}=\{[0],[1]\}$). One can easily check that the obstruction class is nontrivial. If we naively apply the dimensional reduction method to compute loop braiding statistics, we would find $\Theta_{01,1}=\pi$.

We also notice that the same obstruction appears in the gauge theory with all bosonic charges Eq. \eqref{eqn:simpler} if we have $\lambda(\mb{g,g})=[1]$ and the same $\beta$.

\section{Discussions}
\label{sec:discussion}

\subsection{Relation to group-supercohomology models}
In \Ref{GuWen}, Gu and Wen proposed a systematic construction of fermionic SPT phases with a symmetry group $\mathbb{Z}_2^f\times G$. Let us summarize the mathematical structure of their construction: in \spd{$d$}, for each cohomology class $[\beta]\in \mathcal{H}^{d}[G, \mathbb{Z}_2]$, one can associate an obstruction class defined as the Steenrod square $\text{Sq}^2[\beta]$,  and viewed as an element of $\mathcal{H}^{d+2}[G, \mathrm{U}(1)]$. If the obstruction class vanishes, an FSPT phase can be constructed corresponding to $[\beta]$. \Ref{GuWen} proposed that the obstruction-free subgroup of $\mathcal{H}^{d}[G, \mathbb{Z}_2]$ gives a partial classification of \spd{$d$} FSPT phases.

For $d=3$, the mathematical structure of the Gu-Wen construction is completely identical to the twisted Crane-Yetter TQFT with $\mathcal{A}=\mathbb{Z}_2^f$ and a trivial $\lambda$. We believe that the state-sum model discussed in this work with $\lambda=0$ is indeed a gauged Gu-Wen model, where fermions are coupled to $\mathbb{Z}_2$ gauge fields. Our results also clarify the physical meaning of $[\beta]\in \mathcal{H}^3[G, \mathbb{Z}_2]$ for Abelian unitary $G$, that is, the cocycle $\beta$ encodes information about the three-loop braiding statistics. 

With a nontrivial $\lambda$ the state-sum model generalizes the Gu-Wen supercohomology constructions, by allowing gauge fermions to carry projective representations of the symmetry group. We have considered ``Abelian'' cocycles for $\lambda$. It will be interesting to explore the physics of ``non-Abelian'' 2-cocycles, corresponding to fermions carrying higher-dimensional projective representation of the symmetry group. A recent discussion on such terms in continuum field theories can be found in \Ref{Chan2017}.

\subsection{Non-abelian loop braiding statistics}

We have exclusively focused on Abelian loop braiding statistics in this work. Loops can also exhibit non-Abelian braiding statistics. This can happen even when the gauge group is Abelian, if we choose a ``non-Abelian'' 3-cocycle $\beta$ in \eqref{eqn:betaexpr}. We will present one such example, for $\mathcal{A}=\mathbb{Z}_2^f$, $G=\mathbb{Z}_4\times\mathbb{Z}_4\times\mathbb{Z}_4$, $\lambda=0$ and the 3-cocycle $\beta$ is parametrized by $p_{123}=1$ with all other components of $p$ set to 0. Using the invariants given in Appendix \ref{sec:obstruction_invariants}, it is easy to show that the state-sum model is obstruction-free. 

To see the non-Abelian loop braiding, consider a base loop $\phi_1$. From the dimensional reduction, the $e$ and $v$ anyons in the \spd{2} theory both carry two-dimensional projective representations of $G$. After $G$ is gauged, they become non-Abelian anyons and exhibit non-Abelian braiding statistics, similar to what has been found in certain Dijkgraaf-Witten gauge theories~\cite{WangLevinPRB, Wang_DW, GuPRB2016}.  

Recent works have constructed exactly-solvable lattice models for putative non-Abelian \spd{3} topological phases~\cite{ThorngrenKapustin2017,WangGu2017}, in which the dimensionally reduced theories may support non-Abelian Ising excitations. It will be interesting to extend the dimensional reduction approach to these models.

\begin{acknowledgments}
We acknowledge Xie Chen, Shawn X. Cui, Davide Gaiotto, Zheng-Cheng Gu, Yichen Huang, Anton Kapustin, Zhenghan Wang and Dominic Williamson for helpful discussions.  M.C. is particularly grateful to Shawn X. Cui for sharing his unpublished thesis. N.T. and C.W. are particularly grateful to Davide Gaiotto for insightful discussions on evaluating $\mathcal{H}^5[G,\U(1)]$ obstructions. M.C. and C.W. thank Aspen Center of Physics for hospitality and support under the National Science Foundation grant PHY-1066293, where the work was initiated. N.T. acknowledges support for the Perimeter Scholars International Master's program from the Marsland family through an Honorary PSI Scholarship Award.  Research at Perimeter Institute is supported by the Government of Canada through the Department of Innovation, Science and Economic Development Canada and by the Province of Ontario through the Ministry of Research, Innovation and Science.

\end{acknowledgments}

\appendix

\section{Equivalence between Topological Invariants and Loop Braiding Statistics}
\label{app:reconstruction}

We treat the topological invariants and three-loop braiding statistics interchangeably throughout the paper. Here, we show that they are indeed equivalent in the case of Abelian braiding statistics. The following argument is a simple generalization of that for BSPT phases given in Ref.~\onlinecite{WangLevinPRB}.

To show the equivalence, it is enough to reconstruct the full set of three-loop braiding statistics out of the topological invariants.  Consider an arbitrary set of vortices $\{v_i\}$ with $v_i$ carrying unit flux $\frac{2\pi}{N_i}e_i$. All of them are linked to a base loop that carries unit flux  $\frac{2\pi}{N_k}e_k$. According to the definitions of $\Theta_{ij,k}$ and $\Theta_{i,k}$, we have
\begin{align}
\theta_{v_iv_j,e_k} & = \frac{\Theta_{ij,k}}{N^{ij}} + \frac{2\pi y_{ijk}}{N^{ij}} \nonumber \\
\theta_{v_i,e_k}  & = \frac{\Theta_{i,k}}{\tilde N_i} + \frac{2\pi x_{ik}}{\tilde N_i} \label{equiv1}
\end{align}
where  $x_{ijk},y_{ik}$ are some integers that satisfy the relations $y_{iik}=2x_{ik}$ and $y_{ijk}=y_{jik}$. These relations follow from the properties $\theta_{\alpha\alpha,\gamma} = 2\theta_{\alpha,\gamma}$ and $\theta_{\alpha\beta,\gamma}  = \theta_{\beta\alpha,\gamma}$. We take $\{\Theta_{ij,k},\Theta_{i,k}\}$ in the interval $[0,2\pi)$, but in certain cases we set some of them in the interval $[2\pi,4\pi)$, which will be discuss below.

Then, we attach a charge $q^{ik}$ to the loop $v_i$ when it is linked to $\frac{2\pi}{N_k}e_k$ unit flux, for each $i$ and $k$. The new vortex loops $\{\hat v_i\}$ have the following mutual and self three-loop braiding: 
\begin{align}
\theta_{\hat v_i\hat v_j,e_k} & = \theta_{v_iv_j,e_k}  + \frac{2\pi q^{ik}_j}{N_j} + \frac{2\pi q^{jk}_i}{N_i}\nonumber \\
\theta_{\hat v_i,e_k}  & = \theta_{v_i,e_k}   + \frac{2\pi q^{ik}_i}{ N_i} + \pi q^{ik}_0
\end{align}
where $q^{ik}_j$ is the $j$th component of $q^{ik}$. We choose the charge $\{q^{ik}\}$ properly such that they satisfy the following relations:
\begin{align}
\frac{1}{N_{ij}}(N_iq^{ik}_j + N_jq^{jk}_i) & = - y_{ijk}, \quad (\text{mod} \ N^{ij}) \nonumber\\ 
\frac{\tilde{N}_i}{2N_i}(2q^{ik}_i+N_i q^{ik}_0) & = - x_{ik}, \quad (\text{mod} \ \tilde{N}_i) \label{equiv5}
\end{align}
One can show that for even $N_i$, such $\{q^{ik}\}$ always exist. For odd $N_i$ ($i\ge 1$), the existence of such $\{q^{ik}\}$ requires $y_{0ik} = x_{ik} \ (\text{mod } 2)$. Interestingly, it is actually a physical requirement for properly chosen $\Theta_{i,k}$ and $\Theta_{i0,k}$. Before we explain the case of odd $N_i$, we conclude that if Eq.~\eqref{equiv5} holds, we obtain a set of vortex loops $\{\hat{v}_i\}$ such that 
\begin{align}
\theta_{\hat v_i \hat v_j,e_k} & = \frac{\Theta_{ij,k}}{N^{ij}}  \nonumber \\
\theta_{\hat v_i,e_k}  & = \frac{\Theta_{i,k}}{\tilde N_i}
\end{align}
That is, these braiding statistics are determined by the topological invariants.

We now explain the case when $N_i$ is odd for $i\ge 1$. Consider $N_i$ copies of $v_i$ vortices linked to $\frac{2\pi}{N_k}e_k$ flux. Fusing the $v_i$ vortices together gives a pure charge $q$. Using the linearity relations \eqref{linearity3d1} and \eqref{linearity3d_ex1} which we will explain shortly,  we obtain 
\begin{equation}
N_i^2\theta_{v_i,e_k} = \pi q_0 = \frac{N_iN_0}{2}\theta_{v_iv_0,e_k} \label{equiv2}
\end{equation}
(More detailed discussion can be found in Appendix \ref{appendix_proof} in the proof of Eq.~\eqref{eqn:const5}.) In addition, using the constraints Eqs.~\eqref{eqn:const1}, \eqref{eqn:const3} and \eqref{eqn:const4} for odd $N_i$, we can write the topological invariants as follows
\begin{equation}
\Theta_{0i,k} = \frac{2\pi}{N_{0ik}}b_{ik}, \quad \Theta_{i,k}=\frac{2\pi}{N_{ik}}a_{ik} \label{equiv3}
\end{equation}
with $0\le a_{ik},b_{ik} < N_i$. Inserting Eqs.~\eqref{equiv1} and \eqref{equiv3} into Eq.~\eqref{equiv2}, we find that
\begin{equation}
\pi(a_{ik}+x_{ik})=\pi(b_{ik}+y_{0ik}), \quad (\text{mod } 2\pi)\label{equiv4}
\end{equation}
Now if $a_{ik}$ and $b_{ik}$ have the same parity, so do $x_{ik}$ and $y_{0ik}$. If $a_{ik}$ and $b_{ik}$ have opposite parity, we can replace $a_{ik}$ by $a_{ik}+N_{ik}$ in Eq.~\eqref{equiv3}. This only means we choose $2\pi\le \Theta_{i,k}<4\pi$. Now that $a_{ik}+N_{ik}$ and $b_{ik}$ have the same parity, so do $x_{ik}$ and $y_{0ik}$. This proves the claim that  $y_{0ik} = x_{ik} \ (\text{mod } 2)$ is a physical requirement for properly chosen $\Theta_{i,k}$ and $\Theta_{i0,k}$.

\begin{figure}

\begin{tikzpicture}[>=stealth, scale=0.9]
\def \lw {0.7 pt};

\begin{scope}[scale=1]
\node at (1.1, 1.2) {(a)};
\draw [line width = \lw] (1.7, -0.5) .. controls (2, -0.5) and (2, 0.5) ..(1.7,0.5 );
\draw [line width = \lw] (2, -0.5) .. controls (2.3, -0.5) and (2.3, 0.5) ..(2,0.5 );

\draw [white, line width = 3 pt](1, 0)--(3, 0);
\draw [-stealth, line width = \lw](1, 0)--(2.7, 0);

\draw [white, line width = 3pt] (1.7, -0.5) .. controls (1.4, -0.5) and (1.4, 0.5) ..(1.7,0.5 );
\draw [white, line width = 3pt] (2, -0.5) .. controls (1.7, -0.5) and (1.7, 0.5) ..(2,0.5 );
\draw [line width = \lw] (1.7, -0.5) .. controls (1.4, -0.5) and (1.4, 0.5) ..(1.7,0.5 );
\draw [line width = \lw] (2, -0.5) .. controls (1.7, -0.5) and (1.7, 0.5) ..(2,0.5 );

\node at (1.7, -0.7)[scale=0.8]{$\beta_1$};
\node at (2, -0.7)[scale=0.8]{$\beta_2$};
\node at (2.6, 0.2)[scale=0.8]{$\gamma$};

\draw [ -> ] (1.789, 0.2) --(1.775, 0.13);
\draw [ -> ] (1.489, 0.2) --(1.475, 0.13);
\draw [gray!60, line width=2 pt, -latex] (2.8, 0)--(3.3,0);

\end{scope}

\begin{scope}[xshift=2.4cm, scale=1]
\draw [line width = \lw] (1.7, -0.5) .. controls (2, -0.5) and (2, 0.5) ..(1.7,0.5 );
\draw [white, line width = 3 pt](1, 0)--(2.4, 0);
\draw [-stealth, line width = \lw](1, 0)--(2.4, 0);

\draw [white, line width = 3pt] (1.7, -0.5) .. controls (1.4, -0.5) and (1.4, 0.5) ..(1.7,0.5 );
\draw [line width = \lw] (1.7, -0.5) .. controls (1.4, -0.5) and (1.4, 0.5) ..(1.7,0.5 );

\node at (1.7, -0.7)[scale=0.8]{$\beta_1\!\times\!\beta_2$};
\node at (2.3, 0.2)[scale=0.8]{$\gamma$};
\draw [ -> ] (1.489, 0.2) --(1.475, 0.13);
\end{scope}

\begin{scope}[xshift=5cm, yshift=0.6cm, scale=1]
\node at (0.7, 0.6) {(b)};

\draw [line width = \lw] (1.7, -0.5) .. controls (2, -0.5) and (2, 0.5) ..(1.7,0.5 );

\draw [white, line width = 3 pt](1, 0)--(2.4, 0);
\draw [-stealth, line width = \lw](1, 0)--(2.4, 0);

\draw [white, line width = 3pt] (1.7, -0.5) .. controls (1.4, -0.5) and (1.4, 0.5) ..(1.7,0.5 );
\draw [line width = \lw] (1.7, -0.5) .. controls (1.4, -0.5) and (1.4, 0.5) ..(1.7,0.5 );

\draw [gray!60, line width=2 pt, -latex] (2.5, -0.6)--(3,-0.6);

\node at (1.4, -0.5)[scale=0.8]{$\beta_1$};
\node at (2.3, 0.2)[scale=0.8]{$\gamma_1$};
\draw [ -> ] (1.489, 0.2) --(1.475, 0.13);

\end{scope}

\begin{scope}[xshift=5cm, yshift=-0.6cm, scale=1]
\draw [line width = \lw] (1.7, -0.5) .. controls (2, -0.5) and (2, 0.5) ..(1.7,0.5 );

\draw [white, line width = 3 pt](1, 0)--(2.4, 0);
\draw [-stealth, line width = \lw](1, 0)--(2.4, 0);

\draw [white, line width = 3pt] (1.7, -0.5) .. controls (1.4, -0.5) and (1.4, 0.5) ..(1.7,0.5 );
\draw [line width = \lw] (1.7, -0.5) .. controls (1.4, -0.5) and (1.4, 0.5) ..(1.7,0.5 );

\node at (1.4, -0.5)[scale=0.8]{$\beta_2$};
\node at (2.3, 0.2)[scale=0.8]{$\gamma_2$};
\draw [ -> ] (1.489, 0.2) --(1.475, 0.13);
\end{scope}

\begin{scope}[xshift = 7.1cm, yshift=0 cm, scale=1]
\draw [line width = \lw] (1.7, -0.5) .. controls (2, -0.5) and (2, 0.5) ..(1.7,0.5 );

\draw [white, line width = 3 pt](1, 0.1)--(2.4, 0.1);
\draw [-stealth, line width = \lw](1, 0.1)--(2.4, 0.1);

\draw [white, line width = 3 pt](1, -0.1)--(2.4, -0.1);
\draw [-stealth, line width = \lw](1, -0.1)--(2.4,-0.1);

\draw [white, line width = 3pt] (1.7, -0.5) .. controls (1.4, -0.5) and (1.4, 0.5) ..(1.7,0.5 );
\draw [line width = \lw] (1.7, -0.5) .. controls (1.4, -0.5) and (1.4, 0.5) ..(1.7,0.5 );

\node at (1.7, -0.7)[scale=0.8]{$\beta_1\!\circ\!\beta_2$};
\node at (2.6, 0.15)[scale=0.8]{$\gamma_1$};
\node at (2.6, -0.15)[scale=0.8]{$\gamma_2$};
\draw [ -> ] (1.489, 0.2) --(1.475, 0.13);
\end{scope}
\end{tikzpicture}
\caption{Two ways of fusing loops: (a) Fusing $\beta_1$ and $\beta_2$, that are linked to the same base $\gamma$, into a new loop, denoted as $\beta_1\times\beta_2$; (b) Fusing $\beta_1$ and $\beta_2$, that are linked to different bases $\gamma_1$ and $\gamma_2$ and that carry the same amount of flux $\phi_{\beta_1}=\phi_{\beta_2}$, into a new loop, denoted as $\beta_1\circ\beta_2$. }
\label{fig_fusion}
\end{figure}
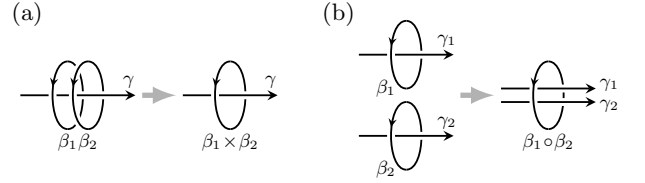

With the set $\{\hat{v}_i\}$ that are linked to unit flux $\frac{2\pi}{N_k}e_k$, the remaining three-loop braiding statistics are easy to reconstruct. To do that, we use the following general properties of Abelian three-loop braiding statistics:
\begin{subequations}
\begin{align}
\theta_{\alpha\alpha,\gamma} & = 2\theta_{\alpha,\gamma}, \label{exch3d}\\
\theta_{\alpha\beta,\gamma} & = \theta_{\beta\alpha,\gamma}, \label{symmetry3d} \\
\theta_{\alpha(\beta_1 \times \beta_2),\gamma} & = \theta_{\alpha\beta_1, \gamma} + \theta_{\alpha\beta_2, \gamma}, \label{linearity3d1} \\
\theta_{(\alpha \times \beta),\gamma} & = \theta_{\alpha,\gamma} + \theta_{\beta, \gamma} + \theta_{\alpha\beta,\gamma},  \label{linearity3d_ex1}\\
\theta_{(\alpha_1\circ\alpha_2)(\beta_1\circ\beta_2), (\gamma_1\times\gamma_2)} & = \theta_{\alpha_1\beta_1,\gamma_1} + \theta_{\alpha_2\beta_2,\gamma_2}, \label{linearity3d2}\\
\theta_{\alpha\circ\beta,\gamma_1\times\gamma_2} & = \theta_{\alpha,\gamma_1} + \theta_{\beta,\gamma_2}, \label{linearity3d_ex2}
\end{align}
\end{subequations}
The relation (\ref{exch3d}) follows immediately from the definition of exchange statistics. The relation (\ref{symmetry3d}) comes from the fact that braiding $\alpha$ around $\beta$ is topologically equivalent to braiding $\beta$ around $\alpha$, while both are linked to $\gamma$. Equations (\ref{linearity3d1})-(\ref{linearity3d_ex2}) will be referred to as {\it linearity reations}. They follow from the fact that braiding and exchanging of loops commute with the two fusion processes of loops, depicted in Fig.~\ref{fig_fusion}. More discussions on these linearity relations can be found in Refs.~\onlinecite{Wang_PRL2014} and \onlinecite{WangLevinPRB}. These works only consider the case that charge excitations are bosonic. Nevertheless, the linearity relations hold regardless of the exchange statistics of charge excitations.

We can now use the two types of fusions in Fig.~\ref{fig_fusion} and the linearity relations (\ref{linearity3d1})-(\ref{linearity3d_ex2}) to obtain braiding statistics between vortices that carry general gauge flux. Also, one can attach charge the vortices to exhaust vortices that carry the same gauge flux. Accordingly, the full set of three-loop braiding statistics indeed can be reconstructed out of the topological invariants. Hence, they are equivalent. 

\section{Proofs of the Constraints}
\label{appendix_proof}

In this appendix, we prove the constraints (\ref{eqn:const1})-(\ref{eqn:const9}) through various physical arguments, except that Eq.~(\ref{eqn:const8}) remains a conjecture. The proofs heavily rely on the general properties of Abelian loop braiding statistics Eqs.~\eqref{exch3d}-\eqref{linearity3d_ex2}.


\emph{Proofs of Eqs.~\eqref{eqn:const1}, \eqref{eqn:const2}, \eqref{eqn:const3} and \eqref{eqn:const7}.}---First of all, Eq. \eqref{eqn:const1} follows immediately from the relation (\ref{exch3d}) and the definitions of $\Theta_{ii,k}$ and $\Theta_{i,k}$. The constraint Eq. \eqref{eqn:const2} follows from the relation \eqref{symmetry3d}. The constraints Eqs. \eqref{eqn:const3} and \eqref{eqn:const7} only involve mutual braiding statistics between loops. The fact that there exist fermionic charge excitations does not matter for mutual statistics. Accordingly, they can be established using exactly the same arguments as those given in \Ref{WangLevinPRB}.

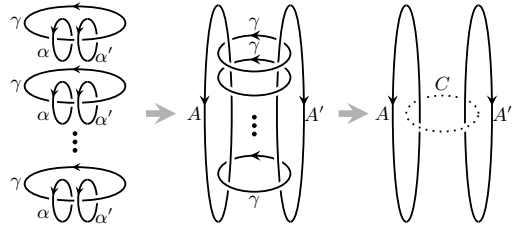
\begin{figure}
\begin{tikzpicture}[>=stealth]
\def \lw {0.7 pt};

\begin{scope}[yshift=-0.02cm, scale=0.93]
\begin{scope}[scale=0.6]

\draw [line width = \lw] (1.7, -0.5) .. controls (2, -0.5) and (2, 0.5) ..(1.7,0.5 );
\draw [line width = \lw] (2.3, -0.5) .. controls (2.6, -0.5) and (2.6, 0.5) ..(2.3,0.5 );

\draw [white, line width = 3pt ] (2, 0) .. controls (3.6,0) and (3.6, 0.8) ..(2, 0.8);
\draw [white, line width = 3pt ] (2, 0) .. controls (0.4,0) and (0.4, 0.8) ..(2, 0.8);
\draw [line width = \lw ] (2, 0) .. controls (3.6,0) and (3.6, 0.8) ..(2, 0.8);
\draw [line width = \lw ] (2, 0) .. controls (0.4,0) and (0.4, 0.8) ..(2, 0.8);

\draw [white, line width = 3pt] (1.7, -0.5) .. controls (1.4, -0.5) and (1.4, 0.5) ..(1.7,0.5 );
\draw [white, line width = 3pt] (2.3, -0.5) .. controls (2, -0.5) and (2, 0.5) ..(2.3,0.5 );
\draw [line width = \lw] (1.7, -0.5) .. controls (1.4, -0.5) and (1.4, 0.5) ..(1.7,0.5 );
\draw [line width = \lw] (2.3, -0.5) .. controls (2, -0.5) and (2, 0.5) ..(2.3,0.5 );

\draw [ ->](2, 0.8) --(1.93, 0.8);
\draw [ -> ] (2.089, 0.2) --(2.075, 0.13);
\draw [ -> ] (1.489, 0.2) --(1.475, 0.13);

\node at (1.25, -0.05)[anchor=north, scale=0.8]{$\alpha$};
\node at (2.7, 0)[anchor=north, scale=0.8]{$\alpha'$};
\node at (0.6,0.4)[scale=0.8]{$\gamma$};

\end{scope}

\begin{scope}[yshift=-0.9cm, scale=0.6]
\draw [line width = \lw] (1.7, -0.5) .. controls (2, -0.5) and (2, 0.5) ..(1.7,0.5 );
\draw [line width = \lw] (2.3, -0.5) .. controls (2.6, -0.5) and (2.6, 0.5) ..(2.3,0.5 );

\draw [white, line width = 3pt ] (2, 0) .. controls (3.6,0) and (3.6, 0.8) ..(2, 0.8);
\draw [white, line width = 3pt ] (2, 0) .. controls (0.4,0) and (0.4, 0.8) ..(2, 0.8);
\draw [line width = \lw ] (2, 0) .. controls (3.6,0) and (3.6, 0.8) ..(2, 0.8);
\draw [line width = \lw ] (2, 0) .. controls (0.4,0) and (0.4, 0.8) ..(2, 0.8);

\draw [white, line width = 3pt] (1.7, -0.5) .. controls (1.4, -0.5) and (1.4, 0.5) ..(1.7,0.5 );
\draw [white, line width = 3pt] (2.3, -0.5) .. controls (2, -0.5) and (2, 0.5) ..(2.3,0.5 );
\draw [line width = \lw] (1.7, -0.5) .. controls (1.4, -0.5) and (1.4, 0.5) ..(1.7,0.5 );
\draw [line width = \lw] (2.3, -0.5) .. controls (2, -0.5) and (2, 0.5) ..(2.3,0.5 );

\draw [ ->](2, 0.8) --(1.93, 0.8);
\draw [ -> ] (2.089, 0.2) --(2.075, 0.13);
\draw [ -> ] (1.489, 0.2) --(1.475, 0.13);

\node at (1.25, -0.05)[anchor=north, scale=0.8]{$\alpha$};
\node at (2.7, 0)[anchor=north, scale=0.8]{$\alpha'$};
\node at (0.6,0.4)[scale=0.8]{$\gamma$};

\end{scope}

\begin{scope}[yshift=-2.3cm, scale=0.6]

\draw [line width = \lw] (1.7, -0.5) .. controls (2, -0.5) and (2, 0.5) ..(1.7,0.5 );
\draw [line width = \lw] (2.3, -0.5) .. controls (2.6, -0.5) and (2.6, 0.5) ..(2.3,0.5 );

\draw [white, line width = 3pt ] (2, 0) .. controls (3.6,0) and (3.6, 0.8) ..(2, 0.8);
\draw [white, line width = 3pt ] (2, 0) .. controls (0.4,0) and (0.4, 0.8) ..(2, 0.8);
\draw [line width = \lw ] (2, 0) .. controls (3.6,0) and (3.6, 0.8) ..(2, 0.8);
\draw [line width = \lw ] (2, 0) .. controls (0.4,0) and (0.4, 0.8) ..(2, 0.8);

\draw [white, line width = 3pt] (1.7, -0.5) .. controls (1.4, -0.5) and (1.4, 0.5) ..(1.7,0.5 );
\draw [white, line width = 3pt] (2.3, -0.5) .. controls (2, -0.5) and (2, 0.5) ..(2.3,0.5 );
\draw [line width = \lw] (1.7, -0.5) .. controls (1.4, -0.5) and (1.4, 0.5) ..(1.7,0.5 );
\draw [line width = \lw] (2.3, -0.5) .. controls (2, -0.5) and (2, 0.5) ..(2.3,0.5 );

\draw [ ->](2, 0.8) --(1.93, 0.8);
\draw [ -> ] (2.089, 0.2) --(2.075, 0.13);
\draw [ -> ] (1.489, 0.2) --(1.475, 0.13);

\node at (1.25, -0.05)[anchor=north, scale=0.8]{$\alpha$};
\node at (2.7, 0)[anchor=north, scale=0.8]{$\alpha'$};
\node at (0.6,0.4)[scale=0.8]{$\gamma$};

\fill (2, 1.2) circle (0.057);
\fill (2, 1.4) circle (0.057);
\fill (2, 1.6) circle (0.057);

\end{scope}

\end{scope}




\begin{scope}[xshift=-0.1cm]

\begin{scope}[xshift=2 cm, yshift=-1 cm, scale=0.8]

\begin{scope}[yshift=0.6 cm]
\clip (0.5,0.4) rectangle +(3, 0.6);
\draw [white, line width = 3pt] (2, 0.4) ellipse (0.6 and 0.3);
\draw [line width = \lw] (2, 0.4) ellipse (0.6 and 0.3);
\draw [line width =\lw, -stealth] (2.0, 0.7)--(1.93, 0.7);
\node at (2, 0.9) [scale=0.8]{$\gamma$};
\end{scope}

\begin{scope}[yshift=0.2 cm]
\clip (0.5,0.4) rectangle +(3, 0.6);
\draw [white, line width = 3pt] (2, 0.4) ellipse (0.6 and 0.3);
\draw [line width = \lw] (2, 0.4) ellipse (0.6 and 0.3);
\draw [line width =\lw, -stealth] (2.0, 0.7)--(1.93, 0.7);
\node at (2, 0.9) [scale=0.8]{$\gamma$};
\end{scope}

\begin{scope}[yshift=-1.4 cm]
\clip (0.5,0.4) rectangle +(3, 0.6);
\draw [white, line width = 3pt] (2, 0.4) ellipse (0.6 and 0.3);
\draw [line width = \lw] (2, 0.4) ellipse (0.6 and 0.3);
\draw [line width =\lw, -stealth] (2.0, 0.7)--(1.93, 0.7);
\end{scope}

\begin{scope}[xshift=-0.3cm]
\draw [white, line width = 3pt] (1.7, -1.8) .. controls (2, -1.8) and (2, 1.8) ..(1.7, 1.8);
\draw [line width = \lw] (1.7, -1.8) .. controls (2, -1.8) and (2, 1.8) ..(1.7, 1.8);
\end{scope}

\begin{scope}[xshift=0.9cm]
\draw [white, line width = 3pt] (1.7, -1.8) .. controls (1.4, -1.8) and (1.4, 1.8) ..(1.7,1.8 );
\draw [line width = \lw] (1.7, -1.8) .. controls (1.4, -1.8) and (1.4, 1.8) ..(1.7,1.8 );
\end{scope}

\begin{scope}[yshift=0.6 cm]
\clip (0.5,0.4) rectangle +(3, -0.6);
\draw [white, line width = 3pt] (2, 0.4) ellipse (0.6 and 0.3);
\draw [line width = \lw] (2, 0.4) ellipse (0.6 and 0.3);
\end{scope}

\begin{scope}[yshift=0.2 cm]
\clip (0.5,0.4) rectangle +(3, -0.6);
\draw [white, line width = 3pt] (2, 0.4) ellipse (0.6 and 0.3);
\draw [line width = \lw] (2, 0.4) ellipse (0.6 and 0.3);

\end{scope}

\begin{scope}[yshift=-1.4cm]
\clip (0.5,0.4) rectangle +(3, -0.6);
\draw [white, line width = 3pt] (2, 0.4) ellipse (0.6 and 0.3);
\draw [line width = \lw] (2, 0.4) ellipse (0.6 and 0.3);
\node at (2, -0.07) [scale=0.8]{$\gamma$};
\end{scope}

\begin{scope}[yshift=-1.6cm]
\fill (2, 1.26) circle (0.04);
\fill (2, 1.4) circle (0.04);
\fill (2, 1.54) circle (0.04);
\end{scope}

\begin{scope}[xshift=-0.3cm]
\draw [white, line width = 3pt] (1.7, -1.8) .. controls (1.4, -1.8) and (1.4, 1.8) ..(1.7,1.8);
\draw [line width = \lw] (1.7, -1.8) .. controls (1.4, -1.8) and (1.4, 1.8) ..(1.7, 1.8);
\draw [line width = \lw, -stealth] (1.475, 0.2)--(1.475, 0.13);
\end{scope}

\begin{scope}[xshift=0.9cm]
\draw [white, line width = 3pt] (1.7, -1.8) .. controls (2, -1.8) and (2, 1.8) ..(1.7, 1.8);
\draw [line width = \lw] (1.7, -1.8) .. controls (2, -1.8) and (2, 1.8) ..(1.7,1.8);
\draw [line width = \lw, -stealth] (1.925, 0.2)--(1.925, 0.13);
\end{scope}

\node at (1  , 0)[scale=0.8]{$A$};
\node at (3, 0)[scale=0.8]{$A'$};

\draw [line width = 2.5pt, gray!60, -stealth] (3.4, 0) --(3.9,0);
\draw [line width = 2.5pt, gray!60, -stealth] (0.2, 0) --(0.7,0);

\end{scope}


\begin{scope}[xshift=4.5 cm, yshift=-1 cm, scale=0.8]

\begin{scope}[yshift=-0.4 cm]
\clip (0.5,0.4) rectangle +(3, 0.8);
\draw [white, line width = 3pt] (2, 0.4) ellipse (0.6 and 0.3);
\draw [dotted, line width = \lw] (2, 0.4) ellipse (0.6 and 0.3);
\node at (2, 0.9) [scale=0.8]{$C$};

\end{scope}

\begin{scope}[xshift=-0.3cm]
\draw [white, line width = 3pt] (1.7, -1.8) .. controls (2, -1.8) and (2, 1.8) ..(1.7, 1.8);
\draw [line width = \lw] (1.7, -1.8) .. controls (2, -1.8) and (2, 1.8) ..(1.7, 1.8);
\end{scope}

\begin{scope}[xshift=0.9cm]
\draw [white, line width = 3pt] (1.7, -1.8) .. controls (1.4, -1.8) and (1.4, 1.8) ..(1.7,1.8 );
\draw [line width = \lw] (1.7, -1.8) .. controls (1.4, -1.8) and (1.4, 1.8) ..(1.7,1.8 );
\end{scope}

\begin{scope}[xshift=-0.3cm]
\draw [white, line width = 3pt] (1.7, -1.8) .. controls (1.4, -1.8) and (1.4, 1.8) ..(1.7,1.8);
\draw [line width = \lw] (1.7, -1.8) .. controls (1.4, -1.8) and (1.4, 1.8) ..(1.7, 1.8);
\draw [line width = \lw, -stealth] (1.475, 0.2)--(1.475, 0.13);
\end{scope}

\begin{scope}[xshift=0.9cm]
\draw [white, line width = 3pt] (1.7, -1.8) .. controls (2, -1.8) and (2, 1.8) ..(1.7, 1.8);
\draw [line width = \lw] (1.7, -1.8) .. controls (2, -1.8) and (2, 1.8) ..(1.7,1.8);
\draw [line width = \lw, -stealth] (1.925, 0.2)--(1.925, 0.13);
\end{scope}

\begin{scope}[yshift=-0.4cm]
\clip (0.5,0.4) rectangle +(3, -0.6);
\draw [white, line width = 3pt] (2, 0.4) ellipse (0.6 and 0.3);
\draw [dotted, line width = \lw] (2, 0.4) ellipse (0.6 and 0.3);
\end{scope}

\node at (1  , 0)[scale=0.8]{$A$};
\node at (3, 0)[scale=0.8]{$A'$};

\end{scope}

\end{scope}

\end{tikzpicture}
\caption{The first thought experiment.}
\label{fig:exp1}
\end{figure}

\emph{Proof of Eq.~\eqref{eqn:const4}.}---Consider the thought experiment shown in Fig. \ref{fig:exp1}, where we have $N_k$ identical copies of a $\gamma$ base loop linked with $\alpha$ and $\alpha'$ loops. Here, $\alpha$ and $\alpha'$ are the same type of loops; we put a prime on the latter just for notational distinction. The loops carry gauge flux $\phi_\alpha = \phi_{\alpha'} = \frac{2\pi }{N_i}e_i$ and $\phi_\gamma = \frac{2\pi }{N_k}e_k$. Now imagine that we exchange $\alpha$ and $\alpha'$ in each copy. The total Berry phase is obviously given by $N_k \theta_{\alpha,\gamma}$. Then, we perform a  two-step deformation on the exchange process (Fig.~\ref{fig:exp1}): we first fuse the $\alpha$ ($\alpha'$) loops into a bigger $A$ ($A'$) loop that is linked with all the $\gamma$ base loops; then we fuse all the $\gamma$ loops together to a $C$ loop. It is not hard to see that $C$ carries no gauge flux. Therefore, the original exchange process is deformed to a process of exchanging two loops $A$ and $A'$ that are not linked to any base loop (Note that $A$ and $A'$ are the same type of loops). Applying the linearity relation \eqref{linearity3d_ex2} to the above deformation process, we obtain 
\begin{equation}
	N_k \theta_{\alpha,\gamma}=\frac{2\pi}{N_i}q_{A}\cdot e_i+\pi q_{A}\cdot e_0.
	\label{derivec4}
\end{equation}
where the right-hand side is the statistical phase associated with exchanging $A$ and $A'$, and $q_A$ is the gauge charge carried by the unlinked loop $A$. 

To obtain Eq.~\eqref{eqn:const4}, we multiply $\tilde N_i$ on both sides of \eqref{derivec4}. For $i=0$, the right-hand side reduces to $(2\tilde N_0/N_0 + \tilde N_0)\pi q_A\cdot e_0$. With Eq.~\eqref{tildeN0}, we find that $(2\tilde N_0/N_0 + \tilde N_0)$ is an even number. Accordingly, the right-hand side of \eqref{derivec4} equals 0 modulo  $2\pi$. If $i\neq 0$, we notice that $\tilde N_i/N_i$ is an integer and $\tilde N_i$ is even, thereby the right-hand side also equals 0 modulo $2\pi$. Hence, we prove Eq.~\eqref{eqn:const4}.


\emph{Proof of Eq.~\eqref{eqn:const5}.}---Consider a base loop $\gamma$ that is linked with $N_i$ copies of $\alpha$ loops. The loops carry gauge flux $\phi_\alpha=\frac{2\pi}{N_i}e_i$ and $\phi_\gamma=\frac{2\pi}{N_k}e_k$. Using the relations \eqref{linearity3d_ex1} and \eqref{exch3d}, it is not hard to see that the exchange statistics of the $\alpha$ loops as a whole is given by $N_i^2\theta_{\alpha,\gamma}$, which equals $N_i\Theta_{i,k}$. On the other hand, fusing the $\alpha$ loops together  gives a gauge charge $q$, whose exchange statistics is $\pi q_0$. Accordingly, we should have
\begin{equation}
N_i \Theta_{i,k} = \pi q_0
\label{derivec51}
\end{equation}
In addition, we consider $\frac{N_0}{2}$ copies of $\beta$ loops that are also linked to $\gamma$. The $\beta$ loops carry gauge flux $\phi_\beta = \frac{2\pi}{N_0}e_0$. According to the Aharonov-Bohm law, the mutual statistics between $q$ and the $\frac{N_0}{2}$ copies of $\beta$ is $\pi q_0$. That is, the mutual braiding statistics between $N_i$ copies of $\alpha$ as a whole and $\frac{N_0}{2}$ copies of $\beta$ as a whole is $\pi q_0$.  Using the linearity relations, we find the latter is also given by $\frac{N_0N_i}{2}\theta_{\alpha\beta,\gamma}$. Therefore, we have
\begin{equation}
	\pi q_0 = \frac{N_0N_i}{2}\theta_{\alpha\beta, \gamma}=\frac{N_{0i}}{2}\Theta_{0i,k}.
	\label{derivec52}
\end{equation}
where the second equality holds when $N_i$ is even. Combining Eqs.~\eqref{derivec51} and \eqref{derivec52}, we immediately obtain the constraint Eq.~\eqref{eqn:const5}.


\begin{figure}
\begin{tikzpicture}[>=stealth]
\def \lw {0.7 pt};

\begin{scope}[yshift=-0.02cm, scale=0.93]
\begin{scope}[scale=0.6]

\draw [line width = \lw] (1.7, -0.5) .. controls (2, -0.5) and (2, 0.5) ..(1.7,0.5 );
\draw [line width = \lw] (2.3, -0.5) .. controls (2.6, -0.5) and (2.6, 0.5) ..(2.3,0.5 );

\draw [white, line width = 3pt ] (2, 0) .. controls (3.6,0) and (3.6, 0.8) ..(2, 0.8);
\draw [white, line width = 3pt ] (2, 0) .. controls (0.4,0) and (0.4, 0.8) ..(2, 0.8);
\draw [line width = \lw ] (2, 0) .. controls (3.6,0) and (3.6, 0.8) ..(2, 0.8);
\draw [line width = \lw ] (2, 0) .. controls (0.4,0) and (0.4, 0.8) ..(2, 0.8);

\draw [white, line width = 3pt] (1.7, -0.5) .. controls (1.4, -0.5) and (1.4, 0.5) ..(1.7,0.5 );
\draw [white, line width = 3pt] (2.3, -0.5) .. controls (2, -0.5) and (2, 0.5) ..(2.3,0.5 );
\draw [line width = \lw] (1.7, -0.5) .. controls (1.4, -0.5) and (1.4, 0.5) ..(1.7,0.5 );
\draw [line width = \lw] (2.3, -0.5) .. controls (2, -0.5) and (2, 0.5) ..(2.3,0.5 );

\draw [ ->](2, 0.8) --(1.93, 0.8);
\draw [ -> ] (2.089, 0.2) --(2.075, 0.13);
\draw [ -> ] (1.489, 0.2) --(1.475, 0.13);

\node at (1.25, -0.05)[anchor=north, scale=0.8]{$\alpha$};
\node at (2.7, 0)[anchor=north, scale=0.8]{$\alpha'$};
\node at (0.6,0.4)[scale=0.8]{$\gamma$};

\end{scope}

\begin{scope}[yshift=-0.9cm, scale=0.6]
\draw [line width = \lw] (1.7, -0.5) .. controls (2, -0.5) and (2, 0.5) ..(1.7,0.5 );
\draw [line width = \lw] (2.3, -0.5) .. controls (2.6, -0.5) and (2.6, 0.5) ..(2.3,0.5 );

\draw [white, line width = 3pt ] (2, 0) .. controls (3.6,0) and (3.6, 0.8) ..(2, 0.8);
\draw [white, line width = 3pt ] (2, 0) .. controls (0.4,0) and (0.4, 0.8) ..(2, 0.8);
\draw [line width = \lw ] (2, 0) .. controls (3.6,0) and (3.6, 0.8) ..(2, 0.8);
\draw [line width = \lw ] (2, 0) .. controls (0.4,0) and (0.4, 0.8) ..(2, 0.8);

\draw [white, line width = 3pt] (1.7, -0.5) .. controls (1.4, -0.5) and (1.4, 0.5) ..(1.7,0.5 );
\draw [white, line width = 3pt] (2.3, -0.5) .. controls (2, -0.5) and (2, 0.5) ..(2.3,0.5 );
\draw [line width = \lw] (1.7, -0.5) .. controls (1.4, -0.5) and (1.4, 0.5) ..(1.7,0.5 );
\draw [line width = \lw] (2.3, -0.5) .. controls (2, -0.5) and (2, 0.5) ..(2.3,0.5 );

\draw [ ->](2, 0.8) --(1.93, 0.8);
\draw [ -> ] (2.089, 0.2) --(2.075, 0.13);
\draw [ -> ] (1.489, 0.2) --(1.475, 0.13);

\node at (1.25, -0.05)[anchor=north, scale=0.8]{$\alpha$};
\node at (2.7, 0)[anchor=north, scale=0.8]{$\alpha'$};
\node at (0.6,0.4)[scale=0.8]{$\gamma$};

\end{scope}

\begin{scope}[yshift=-2.3cm, scale=0.6]

\draw [line width = \lw] (1.7, -0.5) .. controls (2, -0.5) and (2, 0.5) ..(1.7,0.5 );
\draw [line width = \lw] (2.3, -0.5) .. controls (2.6, -0.5) and (2.6, 0.5) ..(2.3,0.5 );

\draw [white, line width = 3pt ] (2, 0) .. controls (3.6,0) and (3.6, 0.8) ..(2, 0.8);
\draw [white, line width = 3pt ] (2, 0) .. controls (0.4,0) and (0.4, 0.8) ..(2, 0.8);
\draw [line width = \lw ] (2, 0) .. controls (3.6,0) and (3.6, 0.8) ..(2, 0.8);
\draw [line width = \lw ] (2, 0) .. controls (0.4,0) and (0.4, 0.8) ..(2, 0.8);

\draw [white, line width = 3pt] (1.7, -0.5) .. controls (1.4, -0.5) and (1.4, 0.5) ..(1.7,0.5 );
\draw [white, line width = 3pt] (2.3, -0.5) .. controls (2, -0.5) and (2, 0.5) ..(2.3,0.5 );
\draw [line width = \lw] (1.7, -0.5) .. controls (1.4, -0.5) and (1.4, 0.5) ..(1.7,0.5 );
\draw [line width = \lw] (2.3, -0.5) .. controls (2, -0.5) and (2, 0.5) ..(2.3,0.5 );

\draw [ ->](2, 0.8) --(1.93, 0.8);
\draw [ -> ] (2.089, 0.2) --(2.075, 0.13);
\draw [ -> ] (1.489, 0.2) --(1.475, 0.13);

\node at (1.25, -0.05)[anchor=north, scale=0.8]{$\alpha$};
\node at (2.7, 0)[anchor=north, scale=0.8]{$\alpha'$};
\node at (0.6,0.4)[scale=0.8]{$\gamma$};

\fill (2, 1.2) circle (0.057);
\fill (2, 1.4) circle (0.057);
\fill (2, 1.6) circle (0.057);

\end{scope}
\end{scope}



\begin{scope}[xshift=2.8cm, yshift=4cm]

\begin{scope}[xshift=-1 cm, yshift=-5 cm, scale=0.8]

\begin{scope}[yshift=0.8 cm]
\begin{scope}
\clip (0.5,0.4) rectangle +(3, 0.6);
\draw [white, line width = 3pt] (2, 0.4) ellipse (0.6 and 0.3);
\draw [line width = \lw] (2, 0.4) ellipse (0.6 and 0.3);
\draw [-stealth](2, 0.7)--(1.93, 0.7);
\node at (2, 0.5)[scale=0.8]{$\gamma$};
\end{scope}
\draw [white, line width = 3pt] (2.6, 0.35) ellipse (0.15 and 0.25);
\draw [line width = \lw] (2.6, 0.35) ellipse (0.15 and 0.25);
\draw [-stealth](2.75, 0.35)--(2.75,0.28);
\node at (2.95, 0.35)[scale=0.8]{$\alpha'$};
\end{scope}

\begin{scope}[yshift=0.1 cm]
\begin{scope}
\clip (0.5,0.4) rectangle +(3, 0.8);
\draw [white, line width = 3pt] (2, 0.4) ellipse (0.6 and 0.3);
\draw [line width = \lw] (2, 0.4) ellipse (0.6 and 0.3);
\node at (2, 0.5)[scale=0.8]{$\gamma$};
\end{scope}
\draw [white, line width = 3pt] (2.6, 0.35) ellipse (0.15 and 0.25);
\draw [line width = \lw] (2.6, 0.35) ellipse (0.15 and 0.25);
\draw [-stealth](2.75, 0.35)--(2.75,0.28);
\node at (2.95, 0.35)[scale=0.8]{$\alpha'$};
\end{scope}

\begin{scope}[yshift=-1.6 cm]
\begin{scope}
\clip (0.5,0.4) rectangle +(3, 0.6);
\draw [white, line width = 3pt] (2, 0.4) ellipse (0.6 and 0.3);
\draw [line width = \lw] (2, 0.4) ellipse (0.6 and 0.3);
\draw [-stealth](2, 0.7)--(1.93, 0.7);
\node at (2, 0.5)[scale=0.8]{$\gamma$};
\end{scope}
\draw [white, line width = 3pt] (2.6, 0.35) ellipse (0.15 and 0.25);
\draw [line width = \lw] (2.6, 0.35) ellipse (0.15 and 0.25);
\draw [-stealth](2.75, 0.35)--(2.75,0.28);
\node at (2.95, 0.35)[scale=0.8]{$\alpha'$};
\end{scope}

\begin{scope}[xshift=-0.3cm]
\draw [white, line width = 3pt] (1.7, -1.8) .. controls (2, -1.8) and (2, 1.8) ..(1.7, 1.8);
\draw [line width = \lw] (1.7, -1.8) .. controls (2, -1.8) and (2, 1.8) ..(1.7, 1.8);
\end{scope}

\begin{scope}[yshift=0.8 cm]
\clip (0.5,0.4) rectangle +(3, -0.6);
\draw [white, line width = 3pt] (2, 0.4) ellipse (0.6 and 0.3);
\draw [line width = \lw] (2, 0.4) ellipse (0.6 and 0.3);
\end{scope}

\begin{scope}[yshift=0.1 cm]
\clip (0.5,0.4) rectangle +(3, -0.6);
\draw [white, line width = 3pt] (2, 0.4) ellipse (0.6 and 0.3);
\draw [line width = \lw] (2, 0.4) ellipse (0.6 and 0.3);

\end{scope}

\begin{scope}[yshift=-1.6cm]
\clip (0.5,0.4) rectangle +(3, -0.6);
\draw [white, line width = 3pt] (2, 0.4) ellipse (0.6 and 0.3);
\draw [line width = \lw] (2, 0.4) ellipse (0.6 and 0.3);
\end{scope}

\begin{scope}[yshift=-1.6cm]
\fill (2, 1.16) circle (0.04);
\fill (2, 1.3) circle (0.04);
\fill (2, 1.44) circle (0.04);
\end{scope}

\begin{scope}[xshift=-0.3cm]
\draw [white, line width = 3pt] (1.7, -1.8) .. controls (1.4, -1.8) and (1.4, 1.8) ..(1.7,1.8);
\draw [line width = \lw] (1.7, -1.8) .. controls (1.4, -1.8) and (1.4, 1.8) ..(1.7, 1.8);
\draw [line width = \lw, -stealth] (1.475, 0.2)--(1.475, 0.13);

\end{scope}

\node at (1  , 0)[scale=0.8]{$A$};
\draw [-stealth](2.0, 0.8)--(1.93, 0.8);

\draw [line width = 2.5pt, gray!60, -stealth] (3, 0) --(3.5,0);
\draw [line width = 2.5pt, gray!60, -stealth] (0.2, 0) --(0.7,0);

\end{scope}


\begin{scope}[xshift=1.2 cm, yshift=-5 cm, scale=0.8]

\begin{scope}[yshift=0.8 cm]
\begin{scope}
\clip (0.5,0.4) rectangle +(3, 0.6);
\draw [white, line width = 3pt] (2, 0.4) ellipse (0.6 and 0.3);
\draw [line width = \lw] (2, 0.4) ellipse (0.6 and 0.3);
\draw [-stealth](2, 0.7)--(1.93, 0.7);
\end{scope}
\node at (2.65, 0.7)[scale=0.8]{$\gamma'$};
\end{scope}

\begin{scope}[yshift=0.1 cm]
\begin{scope}
\clip (0.5,0.4) rectangle +(3, 0.8);
\draw [white, line width = 3pt] (2, 0.4) ellipse (0.6 and 0.3);
\draw [line width = \lw] (2, 0.4) ellipse (0.6 and 0.3);
\end{scope}
\node at (2.65, 0.7)[scale=0.8]{$\gamma'$};
\end{scope}

\begin{scope}[yshift=-1.6 cm]
\begin{scope}
\clip (0.5,0.4) rectangle +(3, 0.6);
\draw [white, line width = 3pt] (2, 0.4) ellipse (0.6 and 0.3);
\draw [line width = \lw] (2, 0.4) ellipse (0.6 and 0.3);
\draw [-stealth](2, 0.7)--(1.93, 0.7);
\end{scope}
\node at (2.65, 0.7)[scale=0.8]{$\gamma'$};
\end{scope}

\begin{scope}[xshift=-0.3cm]
\draw [white, line width = 3pt] (1.7, -1.8) .. controls (2, -1.8) and (2, 1.8) ..(1.7, 1.8);
\draw [line width = \lw] (1.7, -1.8) .. controls (2, -1.8) and (2, 1.8) ..(1.7, 1.8);
\end{scope}

\begin{scope}[yshift=0.8 cm]
\clip (0.5,0.4) rectangle +(3, -0.6);
\draw [white, line width = 3pt] (2, 0.4) ellipse (0.6 and 0.3);
\draw [line width = \lw] (2, 0.4) ellipse (0.6 and 0.3);
\end{scope}

\begin{scope}[yshift=0.1 cm]
\clip (0.5,0.4) rectangle +(3, -0.6);
\draw [white, line width = 3pt] (2, 0.4) ellipse (0.6 and 0.3);
\draw [line width = \lw] (2, 0.4) ellipse (0.6 and 0.3);

\end{scope}

\begin{scope}[yshift=-1.6cm]
\clip (0.5,0.4) rectangle +(3, -0.6);
\draw [white, line width = 3pt] (2, 0.4) ellipse (0.6 and 0.3);
\draw [line width = \lw] (2, 0.4) ellipse (0.6 and 0.3);
\end{scope}

\begin{scope}[yshift=-1.6cm]
\fill (2, 1.16) circle (0.04);
\fill (2, 1.3) circle (0.04);
\fill (2, 1.44) circle (0.04);
\end{scope}

\begin{scope}[xshift=-0.3cm]
\draw [white, line width = 3pt] (1.7, -1.8) .. controls (1.4, -1.8) and (1.4, 1.8) ..(1.7,1.8);
\draw [line width = \lw] (1.7, -1.8) .. controls (1.4, -1.8) and (1.4, 1.8) ..(1.7, 1.8);
\draw [line width = \lw, -stealth] (1.475, 0.2)--(1.475, 0.13);

\end{scope}

\node at (1  , 0)[scale=0.8]{$A$};
\draw [-stealth](2.0, 0.8)--(1.93, 0.8);

\draw [line width = 2.5pt, gray!60, -stealth] (3, 0) --(3.5,0);

\end{scope}


\begin{scope}[xshift=3.4 cm, yshift=-5 cm, scale=0.8]

\begin{scope}[yshift=0.8 cm]
\begin{scope}
\clip (0.5,0.4) rectangle +(3, 0.6);
\draw [white, line width = 3pt] (1.8, 0.4) ellipse (0.4 and 0.2);
\draw [line width = \lw] (1.8, 0.4) ellipse (0.4 and 0.2);
\draw [-stealth](1.81, 0.6)--(1.88, 0.6);
\end{scope}
\node at (2.4, 0.5)[scale=0.8]{$\bar \alpha''$};
\end{scope}

\begin{scope}[yshift=0.6 cm]
\begin{scope}
\clip (0.5,0.4) rectangle +(3, 0.6);
\draw [white, line width = 3pt] (1.8, 0.4) ellipse (0.4 and 0.2);
\draw [line width = \lw] (1.8, 0.4) ellipse (0.4 and 0.2);
\draw [-stealth](1.8, 0.6)--(1.73, 0.6);
\end{scope}
\node at (2.4, 0.3)[scale=0.8]{$\alpha''$};
\end{scope}

\begin{scope}[yshift=-0.4 cm]
\clip (0.5,0.4) rectangle +(3, 0.6);
\draw [white, line width = 3pt] (2, 0.4) ellipse (0.6 and 0.3);
\draw [dotted, line width = \lw] (2, 0.4) ellipse (0.6 and 0.3);
\end{scope}

\begin{scope}[xshift=-0.3cm]
\draw [white, line width = 3pt] (1.7, -1.8) .. controls (2, -1.8) and (2, 1.8) ..(1.7, 1.8);
\draw [line width = \lw] (1.7, -1.8) .. controls (2, -1.8) and (2, 1.8) ..(1.7, 1.8);
\end{scope}

\begin{scope}[yshift=-0.4 cm]
\clip (0.5,0.4) rectangle +(3, -0.6);
\draw [white, line width = 3pt] (2, 0.4) ellipse (0.6 and 0.3);
\draw [dotted, line width = \lw] (2, 0.4) ellipse (0.6 and 0.3);
\end{scope}


\begin{scope}[xshift=-0.3cm]
\draw [white, line width = 3pt] (1.7, -1.8) .. controls (1.4, -1.8) and (1.4, 1.8) ..(1.7,1.8);
\draw [line width = \lw] (1.7, -1.8) .. controls (1.4, -1.8) and (1.4, 1.8) ..(1.7, 1.8);
\draw [line width = \lw, -stealth] (1.475, 0.2)--(1.475, 0.13);
\end{scope}

\begin{scope}[yshift=0.8 cm]
\clip (0.5,0.4) rectangle +(3, -0.6);
\draw [white, line width = 3pt] (1.8, 0.4) ellipse (0.4 and 0.2);
\draw [line width = \lw] (1.8, 0.4) ellipse (0.4 and 0.2);
\end{scope}
\begin{scope}[yshift=0.6 cm]
\clip (0.5,0.4) rectangle +(3, -0.6);
\draw [white, line width = 3pt] (1.8, 0.4) ellipse (0.4 and 0.2);
\draw [line width = \lw] (1.8, 0.4) ellipse (0.4 and 0.2);
\end{scope}

\node at (1  , 0)[scale=0.8]{$A$};

\node at (2.5, -0.5)[scale=0.8]{$C'$};
\end{scope}

\end{scope}

\end{tikzpicture}
\caption{The second thought experiment. }
\label{fig:exp2}
\end{figure}
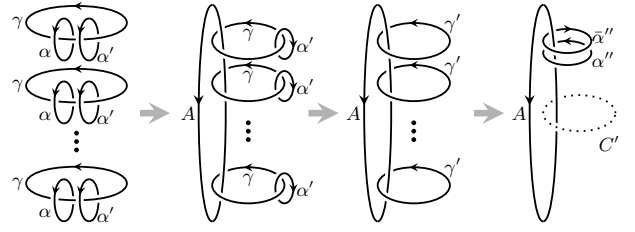

\emph{Proof of Eq.~\eqref{eqn:const6}.}---To show Eq.~\eqref{eqn:const6}, we perform several thought experiments. First, we consider again the thought experiment in Fig.~\ref{fig:exp1}, but now with $N^{ik}$ copies of the linked loops $\alpha,\alpha'$ and $\gamma$. Similarly to Eq.~\eqref{derivec4}, we obtain
\begin{equation}
	{N^{ik}}\theta_{\alpha,\gamma}=\frac{2\pi}{N_i}{q}_A\cdot {e}_i + \pi{q}_A\cdot e_0=\frac{N^{ik}}{\tilde N_i}\Theta_{i,k}.
	\label{c61}
\end{equation}
where we assume $N^{ik}$ is even. Note that this $q_A$ is different from that in Eq.~\eqref{derivec4}, since we start with a different number of copies of linked loops in Fig.~\ref{fig:exp1}.

Next, we consider another thought experiment, shown in Fig.~\ref{fig:exp2}. We start with the same $N^{ik}$ copies of  linked loops $\alpha,\alpha'$ and $\gamma$ as in Fig.~\ref{fig:exp1}. Then, we fuse all the $\alpha$ loops to a big loop $A$, but we shrink the $\alpha'$ loops and fuse them onto the $\gamma$ loops such that $\gamma$ turns to $\gamma'$. The gauge flux carried by $\gamma'$ is the same as that of $\gamma$, i.e., $\phi_{\gamma'}=\phi_{\gamma}$. Next, we fuse the $N^{ik}$ copies of $\gamma'$ together and obtain an excitation $C'$. It is not hard to see that $C'$ carries no gauge flux, hence it is actually a charge excitation. 

To proceed, we create
a pair of loops $\alpha''$ and $\bar{\alpha}''$, with the loop $\alpha''$ carrying unit flux of $\frac{2\pi}{N_i}e_i$. We imagine braiding $\alpha''$ around $C'$. Since $C'$ is a pure charge, the statistical phase is given by the Aharonov-Bohm phase $\frac{2\pi}{N_i}{q}_{C'}\cdot {e}_i$. On the other hand,  since $C'$ is composed of $N^{ik}$ copies of loops $\gamma'$, the linearity relations tell us that the braiding statistical phase is also given by $N^{ik}\theta_{\alpha''\gamma',A}=\Theta_{ik,i}$. Therefore we obtain
\begin{equation}
	\Theta_{ik,i}=\frac{2\pi}{N_i}{q}_{C'}\cdot {e}_i.
	\label{eqn:const6proof2}
\end{equation}
 
We now make use of Eqs.~\eqref{c61} and \eqref{eqn:const6proof2} to show the constraint Eq.~\eqref{eqn:const6}. Recall that we start with $N^{ik}$ identical $\alpha$-$\alpha'$-$\gamma$ links in both Figs.~\ref{fig:exp1} and \ref{fig:exp2}. Each link should carry a well-defined overall charge, which we denote as $q_{\rm link}$. Because of charge conversation, the overall gauge charge carried by the excitations does not vary in any step of the thought experiments. Accordingly, we have
\begin{equation}
q_A+q_{C'} = N^{ik}q_{\rm link} \label{c63}
\end{equation}
Adding together Eqs.~\eqref{c61} abd \eqref{eqn:const6proof2}, and using Eq.~\eqref{c63}, we arrive at
\begin{equation}
\frac{N^{ik}}{\tilde N_i}\Theta_{i,k} + \Theta_{ik,i} = \pi q_A\cdot e_0
\end{equation}
where we have used the fact that $q_{\rm link}$ is an integer vector.

Finally, we argue that $A$ must be a bosonic charge, i.e., $\pi q_A\cdot e_0=0$. To see that, consider the exchange statistics of $C'$ in Fig.~\ref{fig:exp2}:
\begin{align}
\theta_{C'} & =\pi q_{C'}\cdot e_0\nonumber\\ &
 =\left(N^{ik}\right)^2\theta_{\gamma',A} \nonumber\\
& =\frac{\left(N^{ik}\right)^2}{\tilde N_k } \Theta_{k,i} \nonumber \\
& =\frac{\left(N^{ik}\right)^2}{\tilde N_kN_i}\cdot N_i\Theta_{k,i}=0. 
\end{align}
Accordingly, $C'$ is bosonic. Then, using Eq.~\eqref{c63} with the assumption that $N^{ik}$ is even, we immediately conclude that $A$ is also bosonic. Hence, we prove the constraint Eq.~\eqref{eqn:const6}.





\emph{Proof of Eq.~\eqref{eqn:const9}.} --- Finally, we argue for the last constraint Eq.~\eqref{eqn:const9}. Physically, it follows from the requirement that the flux loops can not have any chiral modes. This is also new to fermionic theories (for bosonic ones, by condensing the bosonic gauge charges on the flux loops one can always make them gapped), and imposes a nontrivial constraint on the braiding statistics of the \spd{2} topological phase obtained from dimensional reduction. We recall the following bulk-boundary relation in a \spd{2} (bosonic) topological phase~\cite{Kitaev2006}:
\begin{equation}
	\frac{1}{\mathcal{D}}\sum_a d_a^2\theta_a = e^{\frac{\pi i}{4}c_-}.
	\label{eqn:2d}
\end{equation}
Here $\mathcal{D}=\sqrt{\sum_a d_a^2}$ is the total quantum dimension of the 2D theory, $a$ runs over all types of anyons, $d_a$ and $\theta_a$ are the quantum dimension and topological twist of an anyon of type $a$, respectively. The quantity $c_-$ is the chiral central charge, and  $c_-=0$ for nonchiral theories.

Let us apply Eq. \eqref{eqn:2d}. We fix a base loop, and dimensionally reduce the \spd{3} gauge theory. The anyon types in the \spd{2} theory can be labeled by a tuple $(q, m)$, where $q$ labels the gauge charges, and $\phi_i=\frac{2\pi}{N_i}m_i$ labels the gauge fluxes. These anyon excitations correspond to the vortex loops that are linked to the base loop in the original \spd{3} theory. Assuming all the \spd{2} anyons are Abelian, we have $d_a=1$ and $\mathcal{D}=|\mathcal{G}|=\prod_{i=0}^K N_i$. The topological twist of $(q, m)$ is given by
\begin{align}
	\theta_{(q,m)} = &(-1)^{q_0} e^{2\pi i\sum_j\frac{q_jm_j}{N_j}} \nonumber\\
	& \times e^{i\sum_j m_j^2\theta_{\alpha_j,e_k}+i\sum_{i<j}m_im_j\theta_{\alpha_i\alpha_j, e_k}},
	\label{}
\end{align}
where we have used $\alpha_i$ to denote the loop that corresponds to the anyon $(0,e_i)$ after dimensional reduction, and have taken the base loop to carry gauge flux $\frac{2\pi}{N_k}e_k$. 

Next we insert the expression of $\theta_{q,m}$ into Eq.~\eqref{eqn:2d} with the requirement $c_-=0$, and perform the summation over $q$ and $m$. We first sum over $q$, and the relevant part is  
\begin{equation}
	\begin{split}
		& \sum_{q_0=1}^{N_0}(-1)^{q_0}e^{\frac{2\pi i m_0}{N_0}q_0}\cdot \prod_{j>0} \sum_{q_j=1}^{N_j}e^{\frac{2\pi i m_j}{N_j}q_j}\\
		&=N_0\delta_{m_0, \frac{N_0}{2}}\prod_{j>0}N_j\delta_{m_j,0}.
	\end{split}
\end{equation}
Combining this part with the rest, we have
\begin{align}
1=\frac{1}{\mathcal{D}}\sum_{q,m}\theta_{q,m} = e^{i(\frac{N_0}{2})^2\theta_{\alpha_0\alpha_0,e_k}}.\label{c91}
\end{align}
According to the definitions of the topological invariants, the right-hand side of the above equation is given by
\begin{equation}
	e^{i(\frac{N_0}{2})^2\theta_{\alpha_0\alpha_0,e_k}}=
	\begin{cases}
		e^{i\frac{N_0}{4}\Theta_{0,k}} & \frac{N_0}{2}\equiv 0\,(\text{mod }2)\\	
		e^{i\frac{N_0}{2}\Theta_{0,k}} & \frac{N_0}{2}\equiv 1\,(\text{mod }2)	
	\end{cases}.
	\label{c92}
\end{equation}
Putting together Eqs.~\eqref{c91} and Eq.~\eqref{c92}, and properly rewriting the expressions, we obtain the constraint Eq.~\eqref{eqn:const9}.

\onecolumngrid

\section{Simplicial Calculus and Generalized Cup Product}
\label{sec:review}
We always work with a simplicial triangulation of a manifold $M$. A $p$-cochain is a function living on $p$-simplicies valued in an Abelian group $\mathcal{A}$. Denote the collection of all such cochains as $C^p(M, \mathcal{A})$, which naturally forms a group.

We define the coboundary operator $\delta$ that maps a $p$-cochain $f\in C^p(M, \mathcal{A})$ to a $(p+1)$-cochain:
\begin{equation}
	(\delta f)(i_0i_1\cdots i_{p+1})=\sum_{j=k}^{p+1} (-1)^k f(i_0i_1\cdots \hat{i}_k \cdots i_{p+1})
	\label{}
\end{equation}
where the variable $\hat{i}_k$ is omitted. $\delta$ can be considered as a discrete derivative, and satisfies $\delta^2=0$. If $\delta f=0$, $f$ is said to be closed, or a $p$-cocycle. If $f$ can be written as $f=\delta g$ with $g$ a $(p-1)$-cochain, $f$ is said to exact, or a $p$-coboundary. The group of $p$-cocycles is denoted as $Z^p(M, \mathcal{A})$ and the group of $p$-coboundaries $B^p(M, \mathcal{A})$. Clearly $B^p(M, \mathcal{A})\subset Z^p(M, \mathcal{A})$. The $p$-th cohomology group $H^p(M, \mathcal{A})=\frac{Z^p(M, \mathcal{A})}{B^p(M, \mathcal{A})}$.

The cup product of a $p$-cochain $f\in C^p$ and a $q$-cochain $g\in C^q$ is defined as
\begin{equation}
	[f\cup g]({i_0i_1\cdots i_{p+q}})=B[f(i_0\cdots i_p), g(i_p\cdots i_{p+q})].
	\label{}
\end{equation}
Here $B$ is a bilinear form on $\mathcal{A}$: $B(x+y, z)=B(x,z)+B(y,z)$ and $B(z,x+y)=B(z,x)+B(z,y)$. In our case, we have $R^{a,b}=e^{2\pi i B(a,b)}$, where $B(a,b)\in \mathbb{Q}/\mathbb{Z}$. Notice that we do not necessarily have $B(x,y)=B(y,x)$. For most of the calculations, we actually have $B(y,x)=-B(x,y) \:\text{mod }\mathbb{Z}$.

The most important property of the cup product is
\begin{equation}
	\delta(f\cup g)=\delta f\cup g+(-1)^p f\cup \delta g.
	\label{}
\end{equation}
Therefore, if $\delta f=\delta g=0$, $\delta (f\cup g)=0$. One can actually show that the cup product defines a product of cohomology classes. Cup product to some extent is the discrete version of wedge product of differential forms.

We also define higher cup product~\cite{Steenrod}. For our purpose, we mostly just need $\cup_1$:
\begin{equation}
	[f\cup_1 g](0,\dots, p+q-1)=\sum_{j=0}^{p-1}(-1)^{(p-j)(q+1)}B[f(0,\dots, j,j+q,\cdots,p+q-1), g(j,\dots, j+q)].
	\label{}
\end{equation}
They satisfy the property:
\begin{equation}
	 f\cup g + (-1)^{pq}g\cup f = (-1)^{p+q}\delta f\cup_{1} g + (-1)^{q} f\cup_{1}\delta g -(-1)^{p+q}\delta(f\cup_{1} g)
	\label{eqn:boundary_cup1}
\end{equation}
Notice that the sign on the LHS is reversed compared to the usual formula (e.g. see \Ref{Steenrod}) due to the skew symmetry of the bilinear form $B$.

Generally, higher cup products satisfy:
\begin{equation}
	f\cup_{a}g + (-1)^{pq+a}g\cup_{a}f = (-1)^{p+q+a}\delta f\cup_{a+1} g + (-1)^{q+a} f\cup_{a+1} \delta g-(-1)^{p+q+a}\delta(f\cup_{a+1} g).
	\label{eqn:steenrod}
\end{equation}

We list explicit expressions for the $\cup_1$ that we will use:
\begin{equation}
	\begin{split}
		q=1:\: & [f\cup_1 g](0,1,\cdots, p+1)=B[f(0,\dots, p), \sum_{i=0}^{p}g(i,i+1)]\\
		p=2,q=2:\: & [f\cup_1 g](0123)=B[f(023),g(012)]-B[f(013),g(123)]\\
		p=2,q=3:\: & [f\cup_1 g](01234)=B[f(034),g(0123)]+B[f(014),g(1234)] \\
		p=3,q=3:\: & [f\cup_1 g](012345)=B[f(0345), g(0123)] + B[f(0145), g(1234)] + B[f(0125), g(2345)]
	\end{split}
	\label{}
\end{equation}

\section{Invariance of the Partition Function under Pachner Moves}
\label{appendix_pachner}

We derive the expression for the obstruction class by checking the invariance of the partition function under Pachner moves. For a triangulated four manifold, there are essentially three kinds of Pachner moves: the 1-5, 2-4 and 3-3 moves. In the present example, all of them reduce to the following single condition:
\begin{equation}
	\mathcal{Z}^+(01234)\mathcal{Z}^+(01245)\mathcal{Z}^+(02345)=\mathcal{Z}^+(01235)\mathcal{Z}^+(01345)\mathcal{Z}^+(12345)
	\label{}
\end{equation}

Let us define
\begin{equation}
	\mathcal{Z}_0^+(01234)=R^{f_{012},f_{234}}, \mathcal{Z}_1^+(01234)=R^{f_{034},\beta_{0123}}R^{f_{014},\beta_{1234}}, \mathcal{Z}_2^+(01234)=R^{\lambda_{012}, f_{234}}.
	\label{}
\end{equation}
so that 
\begin{equation}
	\mathcal{Z}^+=\frac{\mathcal{Z}_0^+}{\mathcal{Z}_1^+\mathcal{Z}_2^+}.
	\label{}
\end{equation}

We simplify the Pachner move equation for $\mathcal{Z}_0^+$ and $\mathcal{Z}_1^+$ individually first:
\begin{equation}
	\begin{split}
	\frac{\mathcal{Z}_0^+(01234)\mathcal{Z}_0^+(01245)\mathcal{Z}_0^+(02345)}{\mathcal{Z}_0^+(01235)\mathcal{Z}_0^+(01345)\mathcal{Z}_0^+(12345)}&=\frac{R^{f_{012}, f_{234}+f_{245}-f_{235}}R^{f_{023},f_{345}}}{R^{f_{013}+f_{123},f_{345}}}\\
	&=R^{f_{012},-\beta_{2345}}\frac{R^{f_{012}, f_{345}}R^{f_{023},f_{345}}}{R^{f_{013}+f_{123},f_{345}}}\\
	&=R^{f_{012},-\beta_{2345}}R^{-\beta_{0123},f_{345}}
	\end{split}
	\label{}
\end{equation}

\begin{equation}
	\begin{split}
	&\frac{\mathcal{Z}_1^+(01234)\mathcal{Z}_1^+(01245)\mathcal{Z}_1^+(02345)}{\mathcal{Z}_1^+(01235)\mathcal{Z}_1^+(01345)\mathcal{Z}_1^+(12345)}\\
	&=R^{f_{015}, \beta_{1245}-\beta_{1235}-\beta_{1345}}R^{f_{045},\beta_{0124}+\beta_{0234}-\beta_{0134}}R^{f_{025}-f_{125},\beta_{2345}}R^{f_{014}-f_{145},\beta_{1234}}R^{f_{034}-f_{035},\beta_{0123}}\\
	&=R^{f_{015}, \beta_{1245}-\beta_{1235}-\beta_{1345}}R^{f_{045},\beta_{0124}+\beta_{0234}-\beta_{0134}}R^{f_{015}-f_{012},\beta_{2345}}R^{f_{015}-f_{045},\beta_{1234}}R^{f_{345}-f_{045},\beta_{0123}}\\
	&\quad\:R^{-\beta_{0125}, \beta_{2345}}R^{-\beta_{0145}, \beta_{1234}}R^{-\beta_{0345}, \beta_{0123}}\\
	&=R^{f_{015}, \beta_{1234}+\beta_{1245}+\beta_{2345}-\beta_{1235}-\beta_{1345}}R^{f_{045},\beta_{0124}+\beta_{0234}-\beta_{0134} - \beta_{0123}-\beta_{1234}}R^{-f_{012},\beta_{2345}}R^{f_{345},\beta_{0123}}\\
	&\quad\:R^{-\beta_{0125}, \beta_{2345}}R^{-\beta_{0145}, \beta_{1234}}R^{-\beta_{0345}, \beta_{0123}}\\
	&=R^{-f_{012},\beta_{2345}}R^{f_{345},\beta_{0123}}R^{-\beta_{0125}, \beta_{2345}}R^{-\beta_{0145}, \beta_{1234}}R^{-\beta_{0345}, \beta_{0123}}
\end{split}
	\label{}
\end{equation}
In the last step we use the $3$-cocycle condition of $\beta$.

\begin{equation}
	\begin{split}
	\frac{\mathcal{Z}_2^+(01234)\mathcal{Z}_2^+(01245)\mathcal{Z}_2^+(02345)}{\mathcal{Z}_2^+(01235)\mathcal{Z}_2^+(01345)\mathcal{Z}_2^+(12345)}&=R^{\lambda_{012},f_{234}+f_{245}-f_{235}}R^{\lambda_{023}-\lambda_{013}-\lambda_{123},f_{345}}\\
	&=R^{\lambda_{012}, f_{345}-\beta_{2345}}R^{-\lambda_{012},f_{345}}\\
	&=R^{\lambda_{012}, -\beta_{2345}} = (R^{\lambda_{012}, \beta_{2345}})^{-1}.
	\end{split}
	\label{}
\end{equation}

Combining the two pieces, we get
\begin{equation}
	\frac{\mathcal{Z}^+(01234)\mathcal{Z}^+(01245)\mathcal{Z}^+(02345)}{\mathcal{Z}^+(01235)\mathcal{Z}^+(01345)\mathcal{Z}^+(12345)} = \frac{R^{\beta_{0125}, \beta_{2345}}R^{\beta_{0145}, \beta_{1234}}R^{\beta_{0345}, \beta_{0123}}R^{\lambda_{012},\beta_{2345}}}{R^{\beta_{0123}, f_{345}}R^{f_{345},\beta_{0123}}}\frac{\omega_{01234}\omega_{01245}\omega_{02345}}{\omega_{01235}\omega_{01345}\omega_{12345}}.
	\label{}
\end{equation}
Since $\beta$ is valued in the transparent center, $R^{\beta_{0123}, f_{345}}R^{f_{345},\beta_{0123}}=1$ for all $f$. Therefore the invariance under Pachner moves requires that
\begin{equation}
	\frac{\omega_{12345}\omega_{01345}\omega_{01235}}{\omega_{01234}\omega_{01245}\omega_{02345}}=R^{\beta_{0125}, \beta_{2345}}R^{\beta_{0145}, \beta_{1234}}R^{\beta_{0345}, \beta_{0123}}R^{\lambda_{012},\beta_{2345}}.
	\label{}
\end{equation}

\section{Gauge Invariance of the Partition Function}
\label{sec:gauge-invariance}

We will show below that the action can be thought as a topological gauge theory for a $2$-form gauge field $f$ and a $1$-form gauge field $\mb{g}$. We will only consider the case with $\lambda=0$.

To show that the $f_{ijk}$'s can be regarded as a $2$-form gauge field, we need to show that the action is invariant under a $1$-form gauge transformation $f\rightarrow f+\delta \xi$ where $\xi$ is a $1$-cochain. The variation of the action is
\begin{equation}
	\delta\xi\cup f + f\cup \delta\xi + \delta\xi \cup \delta \xi - \delta\xi\cup_1\delta f
	=\delta (\delta\xi\cup_1 f)-\delta(\xi\cup\delta\xi).
	\label{}
\end{equation}
So the partition function does not change.

To establish that $\mb{g}_{ij}$'s are $1$-form gauge fields requires more work. Under gauge transformations $\mb{g}\rightarrow \mb{g}+\delta\mb{h}$ where $\mb{h}$ is a $0$-cochain.
First one needs to preserve the flatness condition, so the gauge transformations also affect the $2$-form gauge fields. Due to the $3$-cocycle condition of $\beta$, we can write
\begin{equation}
	 \beta(\mb{g}+\delta \mb{h}) - \beta(\mb{g}) = \delta\zeta .
	\label{}
\end{equation}
Here $\zeta$ is a $2$-cochain. 
Explicit expressions of $\zeta$ can be obtained, but extremely tedious in the general case. So we will illustrate by performing a gauge transformation on a single vertex $i$: now $\mb{g}_{ij}\rightarrow \mb{g}_{ij}+\mb{h}_i$ while the others remain unchanges. The $3$-cocycle $\beta(\mb{g}_{ij},\mb{g}_{jk},\mb{g}_{kl})$ transforms as
\begin{equation}
	\begin{split}
	\beta  (\mb{g}_{ij},\mb{g}_{jk},\mb{g}_{kl})\rightarrow \beta(\mb{g}_{ij},\mb{g}_{jk},\mb{g}_{kl}) + 
	 \beta(\mb{h}_i, \mb{g}_{ij},\mb{g}_{jk}) + \beta(\mb{h}_i, \mb{g}_{ik}, \mb{g}_{kl}) - \beta(\mb{h}_i, \mb{g}_{ij}, \mb{g}_{jl})
	\end{split}
	\label{}
\end{equation}
So we can set $\zeta_{imn}=-\beta(\mb{h}_i, \mb{g}_{im},\mb{g}_{jn})$. 

To preserve the flatness condition $\delta f=\beta$, $f$ has to transform as $f\rightarrow f+\zeta$. The change in $f\cup f-f\cup_1\beta$ is
\begin{equation}
	\begin{split}
	\zeta\cup f & + f\cup\zeta + \zeta\cup\zeta - f\cup_1\delta \zeta - \zeta \cup_1 \delta f - \zeta\cup_1\delta \zeta.
	\end{split}
	\label{eqn:gauge1}
\end{equation}
Using the formula Eq. \eqref{eqn:boundary_cup1}, we can write
\begin{equation}
	\zeta\cup f  + f\cup\zeta= \zeta\cup_1\delta f+ \delta \zeta\cup_1 f - \delta(\zeta\cup_1 f).
	\label{}
\end{equation}
Neglecting the boundary term, Eq. \eqref{eqn:gauge1} becomes
\begin{equation}
	 \zeta\cup\zeta - \zeta\cup_1\delta \zeta - f\cup_1\delta \zeta + \delta\zeta \cup_1 f. 
	\label{}
\end{equation}

Now applying the formula Eq. \eqref{eqn:steenrod}, we obtain
\begin{equation}
	f\cup_1\delta\zeta - \delta\zeta \cup_1 f = \beta\cup_2\delta\zeta - \delta(f\cup_2 \delta\zeta).
	\label{}
\end{equation}
So finally, the change of $f\cup f-f\cup_1\beta$ is
\begin{equation}
		f\cup f-f\cup_1\beta \rightarrow
	f\cup f-f\cup_1\beta + \zeta\cup\zeta - \zeta\cup_1\delta \zeta - \beta\cup_2\delta\zeta, 
	\label{eqn:ff1}
\end{equation}
up to a boundary term.

We also need to take into account the change in $\eta$. The change in $\beta\cup_1\beta$ is given by:
\begin{equation}
	\beta\cup_1\beta \rightarrow \beta\cup_1\beta + \beta\cup_1\delta\zeta+\delta\zeta\cup_1\beta + \delta \zeta\cup_1\delta\zeta.
	\label{}
\end{equation}
Using Eq. \eqref{eqn:steenrod}, we have
\begin{equation}
	\beta\cup_1\delta\zeta + \delta\zeta\cup_1\beta = \delta\beta\cup_2\delta\zeta + \beta\cup_2\delta^2\zeta + \delta(\beta\cup_2\delta\zeta),
	\label{}
\end{equation}
and
\begin{equation}
	\delta(\zeta\cup\zeta)=\zeta\cup\delta\zeta + \delta\zeta \cup \zeta = -\delta\zeta\cup_1\delta\zeta + \delta(\zeta\cup_1\delta\zeta).
	\label{}
\end{equation}
Because $\delta\eta=\beta\cup_1\beta$, $\eta$ must be modified in the following way:
\begin{equation}
	\eta\rightarrow \eta- \zeta\cup\zeta+\zeta\cup_1\delta\zeta + \beta\cup_2\delta\zeta.
	\label{}
\end{equation}
Combining with Eq. \eqref{eqn:ff1}, the action is indeed invariant.

\section{Dimensional Reduction}
\label{sec:dim-reduc}
First we need to choose a triangulation of $M_3\times S^1$. We start from an open manifold $M_3\times D^1$ (where $D^1$ stands for an interval), and triangulate the two boundaries into $3$-simplicies denoted by $[ijkl]$ and $[i'j'k'l']$, respectively. We will exploit an ordering in which $i<i'$ for all vertices $i$ in the simplicial triangulation of $M_3$.  We then identify the two boundaries, i.e. $i$ is identified with $i'$, to obtain $M_3\times S^1$. The $G$ flux through $S^1$ is measured by a Wilson loop along $S^1$, which is the $G$ label on $ii'$ (it is easy to see that all $jj'$ should have the same label by the flatness condition). We will set $\mb{g}_{ii'}=\mb{h}$ from now on.

\begin{figure}
	\begin{tikzpicture}[scale=0.80, baseline={([yshift=-.5ex]current  bounding  box.center)}]
		\def \lw {0.7 pt};
		\def \x {5.5}
		\def \y {-1}
		\draw [line width=\lw, middlearrow={stealth}] (0,0) node[below] {$0$} -- (2,1) node[right] {$1$};
		\draw [line width=\lw, middlearrow={stealth}] (2,1) -- (0.2, 3) node[above] {$2$};
		\draw [line width=\lw, middlearrow={stealth}] (0.2, 3) -- (-1.5,1) node[left] {$3$};
		\draw [line width=\lw, middlearrow={stealth}] (0,0) -- (-1.5, 1);
		\draw [line width=\lw, middlearrow={stealth}] (0,0) -- (0.2, 3);
		\draw [dotted, line width=\lw, middlearrow={stealth}] (2,1) -- (-1.5, 1);

	\draw [line width=\lw, middlearrow={stealth}] (\x,\y) node[below] {$0'$} -- (\x+2,\y+1) node[right] {$1'$};
	\draw [line width=\lw, middlearrow={stealth}] (\x+2,\y+1) -- (\x+0.2, \y+3) node[above] {$2'$};
	\draw [line width=\lw, middlearrow={stealth}] (\x+0.2,\y+3) -- (\x-1.5,\y+1) node[left] {$3'$};
	\draw [line width=\lw, middlearrow={stealth}] (\x,\y) -- (\x-1.5,\y+1);
	\draw [line width=\lw, middlearrow={stealth}] (\x,\y) -- (\x+0.2, \y+3);
	\draw [dotted, line width=\lw, middlearrow={stealth}] (\x+2,\y + 1) -- (\x-1.5, \y+1);

	\draw [blue, middlearrow={stealth}] (0,0) -- (\x, \y);
\draw [blue, middlearrow={stealth}] (2,1) -- (\x+2, \y+1);
\draw [blue, middlearrow={stealth}] (0.2,3) -- (\x+0.2, \y+3);
\draw [blue, middlearrow={stealth}] (-1.5,1) -- (\x-1.5, \y+1);
\draw (0.2+2.5, 2.85) node {$\mb{h}$};
\end{tikzpicture}
\caption{Illustration of a prism.}
\end{figure}
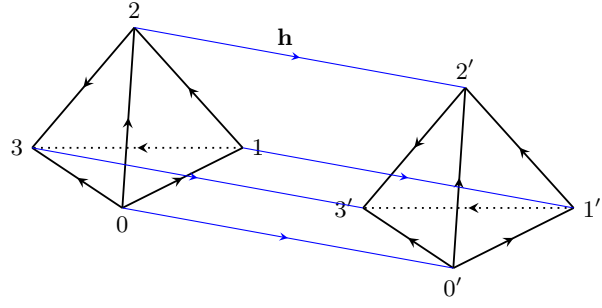

The basic building block of the triangulation is a prism $[ijkli'j'k'l']$.  Each 4D prism $[ijkli'j'k'l']$ is further triangulated into four $4$-simplices $[ijkll'], [ijkk'l'], [ijj'k'l'], [ii'j'k'l']$.

Next, we divide the 2-simplices into two types: the ``in-plane'' ones, which lie entirely inside the $3$-manifold $M_3$, i.e. $f_{ijk}\equiv f_{i'j'k'}$, and those going ``out of plane'', e.g. $f_{ii'j'}$. They need to be treated differently. This is already evident when we examine the twisted flatness conditions: the flatness conditions for the in-plane fields are essentially properties of $M_3$, and may depend on the topology of the manifold. The flatness conditions for the out-of-plane fields can be dealt with explicitly, which we will analyze now.

Let us consider the twisted flatness conditions on a 3D prism $[ijki'j'k']$, which is further triangulated into three $3$-simplices $[ijkk'], [ijj'k']$ and $[ii'j'k']$. Let us write the flatness conditions out:
\begin{subequations}
	\begin{align}
		&f_{jkk'}-f_{ikk'}+f_{ijk'}-f_{ijk}=\beta_{ijkk'} \label{eq:tflata}\\
		&f_{jj'k'}-f_{ij'k'}+f_{ijk'}-f_{ijj'}=\beta_{ijj'k'} \label{eq:tflatb}\\
		&f_{ijk}-f_{ij'k'}+f_{ii'k'}-f_{ii'j'}=\beta_{ii'j'k'} \label{eq:tflatc}.
	\end{align}
	\label{}
\end{subequations}
The meaning of these conditions is uncovered by considering \eqref{eq:tflata}$-$\eqref{eq:tflata}$+$\eqref{eq:tflatc}:
\begin{equation}
	m_{ik}=m_{ij}+m_{jk}+n(\mb{g}_{ij},\mb{g}_{jk}).
	\label{eqn:twisted-flatness-2d}
\end{equation}
Here $m_{ij}$ are defined as $m_{ij}=f_{ii'j'}-f_{ijj'}$, and $n(\mb{g}_{ij}, \mb{g}_{jk})$ is given by
\begin{equation}
	\begin{split}
	n(\mb{g}_{ij}, \mb{g}_{jk})&=\beta_{ijkk'}-\beta_{ijj'k'}+\beta_{ii'j'k'}\\
	&=\beta(\mb{g}_{ij}, \mb{g}_{jk}, \mb{h})-\beta(\mb{g}_{ij}, \mb{h}, \mb{g}_{jk})  +\beta(\mb{h}, \mb{g}_{ij}, \mb{g}_{jk}) \\
	& \equiv  (i_\mb{h}\beta)(\mb{g}_{ij}, \mb{g}_{jk}).
	\end{split}
	\label{}
\end{equation}
Here $i_\mb{h}\beta$ is called the slant product of $\beta$: $[i_\mb{h}\beta]\in \mathcal{H}^2[G, \mathcal{T}]$. Once we fix the fields involved in $m_{ij}, m_{jk}$ and $m_{ik}$, as well as the in-plane one $f_{ijk}$, the remaining two, $f_{ij'k'}$ and $f_{ijk'}$, are also fixed.

At this point, it is clear that $m$ should be thought as a $1$-form gauge field valued in $\mathcal{A}$ on $M_3$, satisfying a twisted flatness condition Eq. \eqref{eqn:twisted-flatness-2d}. What we have shown is that the twisted flatness conditions on $M_3\times S^1$ naturally decouple into the flatness conditions of the in-plane $2$-form fields $\{f_{ijk}\}$, and the flatness conditions of the 1-form gauge fields $\{m_{ij}\}$.

Now we need to evaluate the partition function. First we collect the contributions from $\ftj$'s on a single prism $[01230'1'2'3']$:
\begin{equation}
	\begin{split}
		\ftj([01230'1'2'3'])
		&=\frac{\ftj([01233'])\ftj([011'2'3'])}{\ftj([0122'3'])\ftj([00'1'2'3'])}\\
	&=\frac{R^{f_{012},f_{233'}}R^{f_{011'},f_{123}}}{R^{f_{012},f_{22'3'}}R^{f_{00'1'},f_{123}}}
	\frac{R^{f_{02'3'}, n_{012}}}{R^{f_{013'}, n_{123}}}\frac{R^{f_{00'3'}, \beta_{0123}}}{R^{f_{033'},\beta_{0123}}} \frac{R^{\lambda_{012}, f_{22'3'}}R^{\lambda_{00'1'},f_{1'2'3'}}}{R^{\lambda_{012}, f_{233'}}R^{\lambda_{011'}, f_{1'2'3'}}}
	i_\mb{h}\omega(\mb{g}_{01},\mb{g}_{12},\mb{g}_{23})\\
	&=\frac{R^{m_{03},\beta_{0123}}}{R^{f_{012},m_{23}}R^{m_{01},f_{123}}}
	\frac{R^{f_{02'3'}, n_{012}}}{R^{f_{013'}, n_{123}}} R^{\lambda_{012}, m_{23}}R^{\lambda(\mb{h},\mb{g}_{01})-\lambda(\mb{g}_{01}, \mb{h}), f_{123}}
	i_\mb{h}\omega(\mb{g}_{01},\mb{g}_{12},\mb{g}_{23})
	\end{split}
		\label{}
\end{equation}


We can use flatness conditions on $[00'2'3']$ and $[0133']$:
\begin{equation}
	\begin{gathered}
		f_{023}-f_{02'3'}+f_{00'3'}-f_{00'2'}=\beta(\mb{h},\mb{g}_{02},\mb{g}_{23}),\\
		f_{133'}-f_{033'}+f_{013'}-f_{013}=\beta(\mb{g}_{01},\mb{g}_{13},\mb{h}),
	\end{gathered}
	\label{}
\end{equation}
to rewrite
\begin{equation}
	\frac{R^{f_{02'3'}, n_{012}}}{R^{f_{013'}, n_{123}}}=\frac{R^{f_{023},n_{012}}}{R^{f_{013},n_{123}}}
	\frac{R^{f_{00'3'}-f_{00'2'}-\beta(\mb{h},\mb{g}_{02},\mb{g}_{23}),n_{012}}}{R^{f_{033'}-f_{133'}+\beta(\mb{g}_{01},\mb{g}_{13},\mb{h}),n_{123}}}
	\label{}
\end{equation}

Let us collect the following factors involving $f_{ijk}$'s:
\begin{equation}
	\frac{R^{f_{023},n_{012}}}{R^{f_{013},n_{123}}R^{f_{012},m_{23}}R^{m_{01},f_{123}}}
	\label{}
\end{equation}
To see that these terms do not contribute to the partition function, we recall the following property of the cup product:
\begin{equation}
	f\cup m + m\cup f = \delta(f\cup_1 m)-\delta f\cup_1 m - f\cup_1 \delta m.
	\label{eqn:cup1}
\end{equation}
Recall that $\delta f= \beta, \delta m=-n$, Eq. \eqref{eqn:cup1} implies
\begin{equation}
	\frac{R^{f_{023},n_{012}}}{R^{f_{012},m_{23}}R^{m_{01},f_{123}}R^{f_{013},n_{123}}}\sim R^{\beta_{0123},m_{01}+m_{12}+m_{23}}.
	\label{}
\end{equation}
Here $\sim$ means up to a boundary term.

To further simplify the expressions let us do the following gauge-fixing: we fix all $f_{ijj'}=0$, by using the gauge degrees of freedom on $ij'$, and then $m_{ij}=f_{ii'j'}$. So $f_{00'3'}-f_{00'2'}=m_{03}-m_{02}=m_{23}+n_{023}, f_{033'}-f_{133'}=0$. Notice that after the gauge-fixing $m_{ij}$, we also need to multiply the partition function by a factor of $|\mathcal{A}|$ to correctly normalize it. In total the partition function should be multiplied by $|\mathcal{A}|^{|\Delta_1(M_3)|}$.

Now we have obtained the following expression for the partition function on a prism:
\begin{equation}
	\begin{split}
		\ftj(01230'1'2'3') &={R^{m_{03},\beta_{0123}}}\frac{R^{f_{023},n_{012}}}{R^{f_{012},m_{23}}R^{m_{01},f_{123}}R^{f_{013},n_{123}}}\frac{R^{m_{23}+n_{023}-\beta(\mb{h},\mb{g}_{02},\mb{g}_{23}),n_{012}}}{R^{\beta(\mb{g}_{01},\mb{g}_{13},\mb{h}),n_{123}}} R^{\lambda_{012}, m_{23}}R^{\xi_{01}, f_{123}}
i_\mb{h}\omega(\mb{g}_{01},\mb{g}_{12},\mb{g}_{23})\\
		&=R^{m_{23}, n_{012}}R^{\lambda_{012}, m_{23}}R^{\xi_{01}, f_{123}} \frac{R^{n_{012}+n_{023}, \beta_{0123}}R^{n_{023}-\beta(\mb{h},\mb{g}_{02},\mb{g}_{23}),n_{012}}}{R^{\beta(\mb{g}_{01},\mb{g}_{13},\mb{h}),n_{123}}}i_\mb{h}\omega(\mb{g}_{01},\mb{g}_{12},\mb{g}_{23}).
	\end{split}
		\label{}
\end{equation}

We separate the weights associated to $m$ and $f$:
\begin{equation}
	\begin{split}
		\mathsf{S}_{2+1}([0123])&\equiv R^{\xi_{01},f_{012}}\\
		\ftj_{2+1}([0123])&\equiv R^{m_{23}, n_{012}}R^{\lambda_{012}, m_{23}}\alpha(\mb{g}_{01},\mb{g}_{12},\mb{g}_{23}),
	\end{split}
		\label{}
\end{equation}
where
\begin{equation}
\alpha(\mb{g}_{01},\mb{g}_{12},\mb{g}_{23})=\frac{R^{n_{012}+n_{023}, \beta_{0123}}R^{n_{023}-\beta(\mb{h},\mb{g}_{02},\mb{g}_{23}),n_{012}}}{R^{\beta(\mb{g}_{01},\mb{g}_{13},\mb{h}),n_{123}}}i_\mb{h}\omega(\mb{g}_{01},\mb{g}_{12},\mb{g}_{23}).
	\label{}
\end{equation}
The exact expression of $\alpha$ is not important for our analysis.

Let us now take care of the normalization factors. We have
\begin{equation}
	\begin{gathered}
	|\Delta_0(M_4)|=|\Delta_0(M_3)|,\\
	|\Delta_1(M_4)|=|\Delta_0(M_3)| + 2|\Delta_1(M_3)|.
	\end{gathered}
	\label{}
\end{equation}
Thus the normalization factor becomes
\begin{equation}
	\frac{|\mathcal{A}|^{|\Delta_1(M_3)|}}{|G|^{|\Delta_0(M_4)|}|\mathcal{A}|^{|\Delta_1(M_4)|-|\Delta_0(M_4)| }}=\frac{1}{|G|^{|\Delta_0(M_3)|}|\mathcal{A}|^{|\Delta_1(M_3)|}}
	\label{}
\end{equation}
To summarize, we have found that the partition function on $M_3\times S^1$ can be written as
\begin{equation}
	\mathcal{Z}_\mb{h}(M_3\times S^1)=\frac{1}{|G|^{|\Delta_0|}|\mathcal{A}|^{|\Delta_1|}} \sum_{\{\mb{g}\}\in l(\Delta_0)}\bigg(\sum_{\{f\} \in l(\Delta_2)}\delta_{\delta f=\beta}\mathsf{S}_{2+1}^{\varepsilon(\sigma_3)}(\sigma_3)\bigg) \bigg(  \sum_{\{m\}\in l(\Delta_1)}\delta_{\delta m=-n}\prod_{\sigma_3\in\Delta_3} \ftj_{2+1}^{\varepsilon(\sigma_3)}(\sigma_3)\bigg)
	\label{eq:Zh}
\end{equation}
It should be clear that in this expression $\Delta_k\equiv \Delta_k(M_3)$.

In the following we discuss the physical interpretation of the sum over the ``in-plane'' fields $f$. We will only consider the simple case $\xi=0$.  The sum over $f$ can be evaluated:
\begin{equation}
	\begin{gathered}
	\sum_{\{f\} \in l(\Delta_2)}\delta_{\delta f=\beta} = |Z^2(M_3, \mathcal{A})|\delta\Big(\sum_{\sigma_3\in \Delta_3} \varepsilon(\sigma_3)\beta(\sigma_3)\Big).
	\end{gathered}
		\label{}
\end{equation}
In the following we will write
\begin{equation}
	\int_{M_3}\beta\equiv \sum_{\sigma_3\in \Delta_3} \varepsilon(\sigma_3)\beta(\sigma_3).
	\label{}
\end{equation}
Let us calculate the number of $1$- and $2$-cocycles:
\begin{equation}
	\begin{split}
		|Z^1| &=|H^1|\cdot |B^1| = |H^1|\cdot\frac{|C^0|}{|Z^0|},\\
	|Z^2|&= |H^2|\cdot|B^2| =|H^2|\cdot\frac{|C^1|}{|Z^1|}\\
	&=|H^2|\cdot\frac{|C^1|}{|H^1|\cdot |B^1|}\\
	&=\frac{|H^2|}{|H^1|}\cdot \frac{|C^1|\cdot |Z^0|}{|C^0|}.
	\end{split}
	\label{}
\end{equation}
Now we can easily see $|C^k|=|\mathcal{A}|^{|\Delta_k|}, |Z^0|=|\mathcal{A}|$ (assuming $M_3$ is connected).  Note that due to Poincare duality $|H^2|=|H^1|$. Putting together we find
\begin{equation}
	|Z^1|=|H^1|\cdot |\mathcal{A}|^{|\Delta_0|-1}, |Z^2|=|\mathcal{A}|^{|\Delta_1|-|\Delta_0|+1}.
	\label{}
\end{equation}

Therefore we find that
\begin{equation}
	\mathcal{Z}_\mb{h}(M_3\times S^1)=\frac{1}{|G|^{|\Delta_0|}|\mathcal{A}|^{\Delta_0-1}}\sum_{\{\mb{g}\}}'\delta\Big(\int_{M_3}\beta\Big)\bigg(  \sum_{\{m\}\in l(\Delta_1)}\delta_{\delta m=-n}\ftj_{2+1}^{\varepsilon(\sigma_3)}(\sigma_3)\bigg).
	\label{eq:Zh2}
\end{equation}

To better understand the sum over $f$, we consider the following example of a Dijkgraaf-Witten theory with the gauge group given by $G\times\mathcal{A}$. Group elements are labeled by $(\mb{h},x)$. We further assume that the group $4$-cocycle takes the following form:
\begin{equation}
	\omega((\mb{g}_1,a_1), (\mb{g}_2,a_2), (\mb{g}_3,a_3), (\mb{g}_4,a_4))=\chi_{a_4}(\mb{g}_1,\mb{g}_2, \mb{g}_3).
	\label{}
\end{equation}
Let us compute the partition function of this DW theory on $M_3\times S^1$. Following the derivation in \Ref{Wang_PRL2014}, fixing the holonomy along $S^1$ to be $(\mb{h},x)$, the dimensionally reduced partition function becomes
	\begin{equation}
		\mathcal{Z}_{(\mb{h},x)}(M_3\times S^1)=\frac{1}{|G|^{|\Delta_0|}|\mathcal{A}|^{|\Delta_0|}}\sum_{\{\mb{g},a\}}'\prod_{\sigma_3\in \Delta_3}\ftj_1^{\varepsilon(\sigma_3)}(\sigma_3)\ftj_2^{\varepsilon(\sigma_3)}(\sigma_3).
	\label{}
\end{equation}
Here we define
\begin{equation}
	\begin{gathered}
	\ftj^+_1(\sigma_3)=\chi_x(\mb{g}_i, \mb{g}_j,\mb{g}_k),\\
	\ftj^+_2(\sigma_3)=(i_\mb{h}\chi_{a_l})(\mb{g}_i, \mb{g}_j)^{-1}
	\end{gathered}
	\label{}
\end{equation}

Notice that this DW theory can be related to the theory Eq. \eqref{eqn:simpler} discussed in Sec. \ref{sec:particles}. More precisely, after the duality transformation, we arrive at exactly such a DW theory with
\begin{equation}
	\chi_x(\mb{g}_1,\mb{g}_2,\mb{g}_3)=R^{\beta(\mb{g}_1,\mb{g}_2,\mb{g}_3),x}.
	\label{}
\end{equation}

Let us consider $\mathcal{Z}_\mb{h}\equiv\sum_{x\in\mathcal{A}} \mathcal{Z}_{(\mb{h},x)}$. The only dependence on $x$ comes from $\prod_{\sigma_3\in \Delta_3}\ftj_1^{\varepsilon(\sigma_3)}(\sigma_3)$. For a given $x$,
\begin{equation}
	\prod_{\sigma_3\in \Delta_3}\ftj_1^{\varepsilon(\sigma_3)}(\sigma_3)=R^{\int_{M_3}{\beta},x}.
	\label{}
\end{equation}
We now carry out the sum over $x$:
\begin{equation}
	\frac{1}{|\mathcal{A}|}\sum_{x\in\mathcal{A}}R^{\int_{M_3}{\beta},x}=\delta\Big(\int_{M_3}{\beta}\Big).
	\label{}
\end{equation}
Therefore
\begin{equation}
	\mathcal{Z}_{\mb{h}}(M_3\times S^1)=\frac{1}{|G|^{|\Delta_0|}|\mathcal{A}|^{|\Delta_0|-1}}\sum_{\{\mb{g}\}}'\delta\Big(\int_{M_3}{\beta}\Big)\sum_{\{a\}}'\prod_{\sigma_3\in \Delta_3}\ftj_2^{\varepsilon(\sigma_3)}(\sigma_3).
	\label{eq:ZDW}
\end{equation}
The similarity between Eq. \eqref{eq:Zh2} and Eq. \eqref{eq:ZDW} is quite obvious. 
 Motivated by this computation, we believe the $\delta$ function in Eq. \eqref{eq:Zh2} can be in general understood as the result of summing over the flux of the $\mathcal{A}$ gauge field, dual to ``2-form gauge charges''.

\section{Evaluating Obstructions in the Twisted Crane-Yetter Models}
\label{sec:obstruction_invariants}

Recall that the twisted Crane-Yetter models are well defined only if the following 5-cocycle $\nu \in \mathcal{H}^5[G, \U(1)]$ is a 5-coboundary:
\begin{align}
\nu(\mathbf{g}_1,\mathbf{g}_2,\mathbf{g}_3,\mathbf{g}_4,\mathbf{g}_5) = & R^{\beta(\mathbf{g}_1\mathbf{g}_2\mathbf{g}_3,\mathbf{g}_4,\mathbf{g}_5), \beta(\mathbf{g}_1,\mathbf{g}_2,\mathbf{g}_3)} R^{\beta(\mathbf{g}_1,\mathbf{g}_2\mathbf{g}_3\mathbf{g}_4,\mathbf{g}_5),\beta(\mathbf{g}_2,\mathbf{g}_3,\mathbf{g}_4)} \nonumber\\
& R^{\beta(\mathbf{g}_1,\mathbf{g}_2,\mathbf{g}_3\mathbf{g}_4\mathbf{g}_5),\beta(\mathbf{g}_3,\mathbf{g}_4,\mathbf{g}_5)} R^{\lambda(\mathbf{g}_1, \mathbf{g}_2), \beta(\mathbf{g}_3, \mathbf{g}_4, \mathbf{g}_5)} \label{5-obstruction}
\end{align}
where  $R^{a,b} = (-1)^{ab}$, $\beta$ is a 3-cocycle in $\mathcal{H}^3[G, \mathcal{T}]$, and $\lambda$ is a 2-cocycle in $\mathcal H^2[G, \mathcal{T}]$. When $\nu$ is not a 5-coboundary, we say that the corresponding model has an $\mathcal{H}^5[G,\U(1)]$ obstruction. The main purpose of this appendix is to determine when the twisted Crane-Yetter models are obstruction-free, i.e. when the $\nu$ in Eq.~\eqref{5-obstruction} is a 5-coboundary. We only discuss the case that $G$ is an Abelian group and  $\mathcal{T}=\mathbb{Z}_{N_0}$. 

To do that, we define six quantities $\{\Theta_{i,l,m}, \Theta_{ij,l,m}, \Theta_{ijk,l,m}, \Omega_i, \Omega_{ik}, \Omega_{ijk}\}$ for a general 5-cocycle $\nu\in\mathcal{H}^5(G,\U(1))$, where $G$ is any finite Abelian group $\prod\mathbb Z_{N_i}$. These quantities have the important property that they are defined in a way such that they are \emph{invariant} under a coboundary transformation $\nu \rightarrow \nu\delta\mu$. Hence, we call these quantities invariants for $\mathcal{H}^5[G,\U(1)]$. We claim that a 5-cocycle  is a 5-coboundary \emph{if} and \emph{only if} all the six corresponding invariants vanish. We define these invariants for general 5-cocycles in appendix \ref{app_obstruction1}, and we  prove the claim in appendix \ref{app_obstruction2}. Finally, we apply the invariants to the specific 5-cocycle given in Eq.~\eqref{5-obstruction}.

\subsection{Defining the invariants}
\label{app_obstruction1}

To define the invariants, let us first define the following functions:
\begin{align}
i_{\mathbf{a}}\nu(\mathbf{g},\mathbf{h},\mathbf{k},\mathbf{l}) & = \frac{\nu(\mathbf{a}, \mathbf{g},\mathbf{h},\mathbf{k},\mathbf{l})\nu( \mathbf{g},\mathbf{h},\mathbf{a},\mathbf{k},\mathbf{l})\nu(\mathbf{g},\mathbf{h},\mathbf{k},\mathbf{l},\mathbf{a})}{\nu(\mathbf{g},\mathbf{a},\mathbf{h},\mathbf{k},\mathbf{l})\nu(\mathbf{g},\mathbf{h},\mathbf{k},\mathbf{a},\mathbf{l})} 
\end{align}
The function $i_\mathbf{a}\nu$ is usually called the ``slant product'' of $\nu$. It is actually a 4-cocycle in $\mathcal{H}^4[G, \U(1)]$, when $\mathbf{a}$ is treated a parameter. One may continue to apply the slant product on $i_{\mathbf{a}}\nu$:
\begin{align}
i_{\mathbf{b}}i_{\mathbf{a}}\nu(\mathbf{g},\mathbf{h},\mathbf{k}) & = \frac{i_\mathbf{a}\nu(\mathbf{g}, \mathbf{b},\mathbf{h},\mathbf{k})i_\mathbf{a}\nu(\mathbf{g}, \mathbf{h},\mathbf{k},\mathbf{b})}{i_\mathbf{a}\nu(\mathbf{b},\mathbf{g}, \mathbf{h},\mathbf{k})i_\mathbf{a}\nu(\mathbf{g},\mathbf{h}, \mathbf{b},\mathbf{k})} \nonumber\\
i_\mathbf{c} i_{\mathbf{b}}i_{\mathbf{a}}\nu(\mathbf{g},\mathbf{h})  & = \frac{i_{\mathbf{b}}i_{\mathbf{a}}\nu(\mathbf{c},\mathbf{g},\mathbf{h})i_{\mathbf{b}}i_{\mathbf{a}}\nu(\mathbf{g},\mathbf{h},\mathbf{c})}{i_{\mathbf{b}}i_{\mathbf{a}}\nu(\mathbf{g},\mathbf{c},\mathbf{h})} 
\end{align}
where $i_{\mathbf{b}}i_{\mathbf{a}}\nu$ is a 3-cocycle and $i_\mathbf{c}
i_{\mathbf{b}}i_{\mathbf{a}}\nu$ is a 2-cocycle. In addition, we also define the following function
\begin{align}
i_{\mathbf{a},\mathbf{b}}\nu(\mathbf{g},\mathbf{h},\mathbf{k})  &  =  \frac{\nu(\mathbf{g},\mathbf{h},\mathbf{k},\mathbf{a},\mathbf{b})\nu(\mathbf{g},\mathbf{a},\mathbf{h},\mathbf{k},\mathbf{b})}{\nu(\mathbf{g},\mathbf{h},\mathbf{a},\mathbf{k},\mathbf{b})\nu(\mathbf{a},\mathbf{g},\mathbf{h},\mathbf{k},\mathbf{b})}\frac{\nu(\mathbf{g},\mathbf{h},\mathbf{a},\mathbf{b},\mathbf{k})\nu(\mathbf{a},\mathbf{g},\mathbf{h},\mathbf{b},\mathbf{k})}{\nu(\mathbf{g},\mathbf{a},\mathbf{h},\mathbf{b},\mathbf{k})}\frac{\nu(\mathbf{g},\mathbf{a},\mathbf{b},\mathbf{h},\mathbf{k})}{\nu(\mathbf{a},\mathbf{g},\mathbf{b},\mathbf{h},\mathbf{k})}\nu(\mathbf{a},\mathbf{b},\mathbf{g},\mathbf{h},\mathbf{k})
\end{align}
The function $i_{\mathbf{a},\mathbf{b}}\nu$, however, is not a 3-cocycle.

With these functions, we now define the invariants for $\mathcal{H}^5[G,\U(1)]$ for Abelian group $G=\prod_i\mathbb{Z}_{N_i}$. Let $\mathbf{e}_i$ be the generator associated with the $\mathbb{Z}_{N_i}$ subgroup of $G$. First, we  define the following invariants for a given 5-cocycle $\nu$:
\begin{align}
e^{i\Theta_{ijk,l,m}} & = \frac{i_{\mathbf e_{i}} i_{\mathbf e_{l}} i_{\mathbf e_{m}} \nu(\mathbf e_{k},\mathbf e_{j})}{i_{\mathbf e_{i}} i_{\mathbf e_{l}} i_{\mathbf e_{m}} \nu(\mathbf e_{j}, \mathbf e_{k})}\label{i1}\\
e^{i\Theta_{ij,l,m}} &=\prod_{n=1}^{N^{ij}}i_{\mathbf e_{i}} i_{\mathbf e_{l}} i_{\mathbf e_{m}} \nu(\mathbf e_{j},n \mathbf e_{j}) i_{\mathbf e_{j}} i_{\mathbf e_{l}} i_{\mathbf e_{m}} \nu(\mathbf e_{i},n\mathbf e_{i})\label{i2}\\
e^{i\Theta_{i,l,m}}  &=\prod_{n=1}^{N_{i}}i_{\mathbf e_{i}} i_{\mathbf e_{l}} i_{\mathbf e_{m}} \nu(\mathbf e_{i}, n\mathbf e_{i})\label{i3}
\end{align}
One may check that $\Theta_{i,l,m}, \Theta_{ij,l,m}$ and $\Theta_{ijk,l,m}$ are ideed invariant under a coboundary transformation $\nu\rightarrow \nu \delta \mu$.

Next, we define the following invariants
\begin{align}
e^{i\Omega_i} & = \prod_{m,n=1}^{N_i} i_{\mathbf{e}_i}(\mathbf{e}_i, m\mathbf{e}_i, \mathbf{e}_i, n\mathbf{e}_i)\label{in1} \\
e^{i\Omega_{ik}} & =\left(\prod_{m=1}^{N^{ik}} \prod_{n=1}^{N_i} i_{\mathbf{e}_k, m\mathbf{e}_k}\nu(\mathbf{e}_i,n\mathbf{e}_i, \mathbf{e}_i)i_{\mathbf{e}_k}\nu(\mathbf{e}_i, n \mathbf{e}_i, \mathbf{e}_i, m \mathbf{e}_i) \right)
\end{align}
Again these two quantities are invariant under a coboundary transformation $\nu\rightarrow \nu \delta \mu$.

Finally, we define the invariant $\Omega_{ijk}$. To do that, we write 
\begin{equation}
N_i = \prod_p p^{r_p}, \quad N_j = \prod_p p^{s_p}, \quad N_k = \prod_p p^{t_p} \nonumber
\end{equation}
where the products are taken over all prime numbers $p$. Then, we have the following group isomorphisms
\begin{align}
\mathbb Z_{N_i} &=  \mathbb{Z}_{2^{r_2}}\times  \mathbb{Z}_{3^{r_3}} \times  \mathbb{Z}_{5^{r_5}} \times \cdots \nonumber \\
\mathbb Z_{N_j} &=  \mathbb{Z}_{2^{s_2}}\times  \mathbb{Z}_{3^{s_3}} \times  \mathbb{Z}_{5^{s_5}} \times \cdots \nonumber\\
\mathbb Z_{N_k} &=  \mathbb{Z}_{2^{t_2}}\times  \mathbb{Z}_{3^{t_3}} \times  \mathbb{Z}_{5^{t_5}} \times \cdots  \nonumber
\end{align}
Let $\mathbf{e}_i^p\equiv \frac{N_i}{p^{r_p}}\mathbf{e}_i$ be the generator associated with the $\mathbb{Z}_{p^{r_p}}$ subgroup in $\mathbb{Z}_{N_i}$, and $\mathbf{e}_j^p$, $\mathbf{e}_k^p$ are similarly defined. In the case that $r_p\le s_p \le t_p$, we define 
\begin{equation}
e^{ i\Omega_{ijk}^p } =\prod_{m=1}^{p^{t_p}} \prod_{n=1}^{p^{s_p}}  \frac{i_{\mathbf{e}^p_k,m\mathbf{e}^p_k}\nu(\mathbf{e}_i^p+\mathbf{e}^p_j, n\mathbf{e}^p_i + n\mathbf{e}^p_j, \mathbf{e}^p_i+\mathbf{e}^p_j)}{i_{\mathbf{e}_k^p,m\mathbf{e}^p_k}\nu(\mathbf{e}^p_i, n \mathbf{e}^p_i, \mathbf{e}^p_i)i_{\mathbf{e}^p_k,m\mathbf{e}^p_k}\nu(\mathbf{e}^p_j, n\mathbf{e}^p_j, \mathbf{e}^p_j)} \frac{i_{\mathbf{e}^p_k}(\mathbf{e}^p_i+\mathbf{e}^p_j, n\mathbf{e}^p_i+n\mathbf{e}^p_j, \mathbf{e}^p_i+\mathbf{e}^p_j, m\mathbf{e}^p_i+m\mathbf{e}^p_j)}{i_{\mathbf{e}^p_k}(\mathbf{e}^p_i, n\mathbf{e}^p_i, \mathbf{e}^p_i, m\mathbf{e}^p_i) i_{\mathbf{e}^p_k}(\mathbf{e}^p_j, n\mathbf{e}^p_j, \mathbf{e}^p_j, m\mathbf{e}^p_j)} \label{omegaijk} 
\end{equation}
If $r_p, s_p, t_p$ are in different orders, $\Omega_{ijk}^p$ are defined similarly with a corresponding permutation of indices $i,j,k$ in Eq.~\eqref{omegaijk}. At the end, we define the total invariant
\begin{equation}
\Omega_{ijk} = \sum_p \Omega_{ijk}^p
\end{equation}
Again, one can show that $\Omega_{ijk}$ is invariant under a coboundary transformation $\nu\rightarrow \nu \delta \mu$.

\subsection{Completeness of the invariants}
\label{app_obstruction2}

Let us now show that the invariants $\{\Theta_{i,l,m}, \Theta_{ij,l,m}, \Theta_{ijk,l,m}, \Omega_i, \Omega_{ik}, \Omega_{ijk}\}$  are complete, in the sense that they have the resolution to distinguish every cohomology class in $\mathcal{H}^5[G,\U(1)]$. To do that, we perform a counting argument. First of all, for Abelian group $G = \prod_{i} \mathbb Z_{N_i}$, the cohomology group is given by
\begin{align}
\mathcal{H}^5[G,\U(1)] = \prod_i \Z_{N_i} \prod_{i<j} \Z_{N_{ij}}^2 \prod_{i<j<k} \Z_{N_{ijk}}^4 \prod_{i<j<k<l} \Z_{N_{ijkl}}^3 \prod_{i<j<k<l<p} \Z_{N_{ijklp}}
\end{align}
That means the invariants can take \emph{at most} $|\mathcal{H}^5[G,\U(1)]|$ distinct values. If we are able to show that the invariants can take exactly $|\mathcal{H}^5[G, \U(1)]|$ distinct values, we prove the invariants are complete.

To do that, we evaluate the values of the invariants for the following  explicit 5-cocycles
\begin{align}
\nu_1(\mathbf{a},\mathbf{b},\mathbf{c},\mathbf{d},\mathbf{e}) & = \exp\left\{ i2\pi \sum_{ijk} \frac{P_{ijk}}{N_iN_jN_k}a_i(b_j+c_j-[b_j+c_j])(d_k+e_k-[d_k+e_k])\right\}\nonumber\\
\nu_2(\mathbf{a},\mathbf{b},\mathbf{c},\mathbf{d},\mathbf{e}) &= \exp\left\{i2\pi\sum_{ijkl} \frac{Q_{ijkl}}{N_{ijk}N_l} a_ib_jc_k(d_l+e_l-[d_l+e_l])\right\} \cdot \exp\left\{i2\pi\sum_{ijklp} \frac{R_{ijklm}}{N_{ijklm}} a_ib_jc_kd_le_m\right\}
\end{align}
where $P_{ijk}, Q_{ijkl}, R_{ijklp}$ are integer parameters. For simplicity, we assume $R_{ijklm}=0$ if any of its two indices are equal.  We have used integer vectors $\mathbf{a}=(a_1,a_2,\dots)$ to denote the group elements of $G$ with $0\le a_i < N_i$, and  $[b_j+c_j]$ is defined as $b_j+c_j$ modulo $N_j$. We use additive convention for group multiplication of the Abelian group $G$.  One may check that for $\nu_1$ and $\nu_2$ are indeed  5-cocycles.

Inserting the expression of $\nu_2$ into the definition of $\Theta_{ijk,l,m}, \Theta_{ij,l,m}, \Theta_{i,l,m}$, we find that
\begin{align}
	\Theta_{ijk,l,m} & = -\frac{2\pi}{N_{ijklm}}\sum_{\sigma} \text{sgn}\,(\sigma) R_{\sigma(i)\sigma(j)\sigma(k)\sigma(l)\sigma(m)} \nonumber\\
\Theta_{ij,l,m} & = \frac{2\pi N^{ij}}{N_{ilm}N_j}(Q_{milj}+Q_{ilmj}+Q_{lmij}-Q_{imlj}-Q_{mlij}-Q_{limj}) +(i\leftrightarrow j) \nonumber\\
\Theta_{i,l,m} & = \frac{2\pi}{N_{ilm}}(Q_{mili}+Q_{ilmi}+Q_{lmii}-Q_{imli}-Q_{mlii}-Q_{limi})
\end{align}
(Note that $\Omega_i,\Omega_{ij},\Omega_{ijk}$ can be evaluated but the expressions are complicated, so we do not list them here. This does not affect the counting argument below. ) Insertiging the expression of $\nu_1$ in to the definitions of $\Omega_i, \Omega_{ik}, \Omega_{ijk}$, we find that $\Theta_{ijk,l,m}=\Theta_{ij,l,m}=\Theta_{i,l,m}=0$ and
\begin{align}
\Omega_i & = 2\pi \frac{P_{iii}}{N_i} \nonumber\\
\Omega_{ik} & = 2\pi  \frac{P_{iik}+P_{iki}+P_{kii}}{N_{ik}} \nonumber\\
\Omega_{ijk} &  = 2\pi  \frac{P_{ijk}+P_{ikj}+P_{kij}+P_{jik}+P_{jki}+P_{kji}}{N_{ijk}} X 
\end{align}
where $X$ is an integer with the property that $X$ and $N_{ijk}$ are coprime. 

Let us count how many distinct values these  invariants can take. First, $\Theta_{ijk,l,p}$ is fully antisymmetric, and it can take $N_{ijklp}$ different values. One can show that $N_{ijlp}\Theta_{ij,l,p}=0$, and it is symmetric in $i,j$ and antisymmetric in $l,p$. A more careful calculation shows that for fixed indices $i\neq j\neq l\neq p$, $\Theta_{ij,l,p}$ and those related to  $\Theta_{ij,l,p}$ by index permutations can take $N_{ijlp}^3$ distinct values. The invariant $\Theta_{i,l,p}$ is antisymmetric in $l,p$. For fixed indices $i\neq j\neq p$, $\Theta_{i,l,p}$ and those related by index permutations can take $N_{ilp}^3$ distinct values. Hence, $\Theta_{ijk,l,p}, \Theta_{ij,l,p},\Theta_{i,l,p}$ together can take $\mathcal{N}_\Theta$ different values with
\begin{equation}
\mathcal{N}_\Theta= \prod_{i<j<k}N_{ijk}^3 \prod_{i<j<k<l}N_{ijkl}^3\prod_{i<j<k<l<m}N_{ijklm}
\end{equation}
Next, we count the possible number of values for $\Omega_i,\Omega_{ij},\Omega_{ijk}$ from $\nu_1$. Since for $\nu_1$ we have $\Theta_{ijk,l,m}=\Theta_{ij,l,m}=\Theta_{i,l,m}=0$, these possible values of invariants are distinct from those from $\nu_2$. The invariant $\Omega_i$ obviously can take $N_i$ distinct values. For given $i
\neq k$, $\Omega_{ik}$ and $\Omega_{ki}$ are independent. They together can take $N_{ik}^2$ distinct values. The invariant $\Omega_{ijk}$ is fully symmetric and it can take $N_{ijk}$ distinct values (note that in the definition of $\Omega_{ijk}$, it is only symmetric in $i$ and $j$. This full symmetry in all three indices is only a consequence of the specific cocycle $\nu_1$). Accordingly, we have that the invariants $\Omega_{i},\Omega_{ij},\Omega_{ijk}$ can take $\mathcal{N}_\Omega$ distinct values with
\begin{equation}
\mathcal{N}_\Omega = \prod_i N_i \prod_{i<j} N_{ij}^2 \prod_{i<j<k} N_{ijk}
\end{equation}
Putting all together, the invariants can take $\mathcal{N}_\Theta \mathcal{N}_\Omega = |\mathcal{H}^5[G,\U(1)]|$ distinct values in total. Hence, they are complete.

\subsection{Evaluating obstruction}
\label{app_obstruction3}

We now evaluate the values of the invariants $\{\Theta_{i,l,m}, \Theta_{ij,l,m}, \Theta_{ijk,l,m}, \Omega_i, \Omega_{ik}, \Omega_{ijk}\}$ for the 5-cocycle given in Eq.~\eqref{5-obstruction}. That 5-cocycle depends on a 3-cocycle $\beta$ in $\mathcal{H}^3[G, \mathbb{Z}_{N_0}]$ and a 2-cocycle $\lambda$ in $\mathcal{H}^2[G, \mathbb{Z}_{N_0}]$. Below, we work for the following explicit $\beta$ and $\lambda$: 
\begin{align}
\beta(\mathbf{a},\mathbf{b},\mathbf{c}) & = \sum_{ij} \frac{N_0 p_{ij}}{ N_{i0}N_j} a_i (b_j+c_j -[b_j+c_j]) + \sum_{ijk} \frac{N_0 p_{ijk}}{N_{ijk0}} a_ib_jc_k, \quad (\text{mod } N_0) \nonumber\\
\lambda(\mathbf{a},\mathbf{b}) & = \sum_i \frac{q_{i}}{N_i}(a_i+b_i - [a_i+b_i]), \quad (\text{mod } N_0)
\end{align}
where $p_{ij}, p_{ijk}, q_i$ are integer parameters, and for simplicity, we assume that $p_{ijk}=0$ if any two of the indices are the same. 



We now insert the above expressions of $\beta$ and $\lambda$ into the expression of $\nu$ in Eq.~\eqref{5-obstruction}, and further insert $\nu$ into the definitions of the invariants $\{\Theta_{i,l,m}, \Theta_{ij,l,m}, \Theta_{ijk,l,m}, \Omega_i, \Omega_{ik}, \Omega_{ijk}\}$. After a long tedious calculation, we find that
\begin{align}
\Omega_i &  = \pi \frac{N_0}{N_{0i}} p_{ii} (1+q_i) \nonumber\\
\Omega_{ik} &  = \pi \frac{N_0}{N_{0k}}\frac{N^{ik}}{N_i}p_{ki}(1+q_i) + \pi \frac{N_0}{N_{0i}}\frac{N^{ik}}{N_k}(q_ip_{ik}+q_kp_{ii}), \quad \quad (\text{when } i \neq k) \nonumber\\
\Omega_{ijk} & =\pi \frac{N^{ijk}(N^{ijk}-1)}{2}\frac{N_0}{N_{ijk0}}(1+q_i+q_j)\hat p_{ijk}+ \pi\frac{N_0}{N_{ijk0}}(q_i\hat p_{jk} + q_j \hat p_{ki}+ q_k \hat p_{ij}), \quad (\text{when }  i \neq j\neq k) \nonumber\\
\Theta_{i,l,m} & = \pi\frac{N_0}{N_{ilm0}}q_i\hat{p}_{ilm}\nonumber\\
\Theta_{ij,l,m} & = \pi\frac{N^{ij}}{N_i}\frac{N_0}{N_{jlm0}}q_i\hat{p}_{jlm} + (i\leftrightarrow j) \nonumber\\
\Theta_{ijk,l,m} & =0
\end{align}
where $\hat{p}_{ijk}\equiv p_{ijk}+p_{jki}+p_{kij}-p_{kji}-p_{jik}-p_{ikj}$ and $\hat{p}_{ij}=p_{ij}+p_{ji}$. (Note that we only calculated $\Omega_{ik}$ and $\Omega_{ijk}$ with $i\neq j\neq k$ for simplicity. For $\Omega_{ijk}$, $k$ is the index such that $N_k$ contains the most the factor 2 among $N_i,N_j,N_k$.) The twisted Crane-Yetter models are obstruction free if and only if all the above expressions vanish.


\twocolumngrid

\bibliography{spt}
\end{document}